\documentclass[aps,prl,amssymb,groupedaddress,nofootinbib]{revtex4}

\usepackage[english]{babel}
\usepackage{amsfonts}
\usepackage{amssymb}
\usepackage{epsfig}
\usepackage{amsmath}
\usepackage{fancyhdr}
\usepackage{textcomp}
\usepackage{setspace}

\def\a{\alpha}
\def\b{\beta}
\def\D{\Delta}
\def\d{\delta}
\def\e{\epsilon}
\def\G{\Gamma}

\def\H{\Theta}

\def\n{\eta}
\def\O{\Phi}
\def\o{\phi}

\def\s{\sigma}

\def\t{\tau}
\def\u{\mu}

\def\W{\Omega}
\def\w{\omega}

\def\z{\zeta}

\def\hf{\frac{1}{2}}
\def\pl{\parallel}

\def\px{\approx}
\def\pr{\prime}

\def\={\nonumber &=}
\def\nn{\nonumber}
\def\&{{}&}

\def\({\left(}
\def\){\right)}
\def\[{\left[}
\def\]{\right]}
\def\<{\left\langle}
\def\>{\right\rangle}

\def\uk{{\bf \hat{k}}}

\def\ux{{\bf \hat{x}}}
\def\bk{{\bf k}}
\def\bn{{\bf n}}

\def\bx{{\bf x}}
\def\bl{{\bf l}}

\def\curl{\mathcal}
\def\it{\textit}

\def\and{\quad \mbox{and} \quad}
\def\imp{\quad \Rightarrow \quad}

\begin{document}

\title{Shape of primordial non-Gaussianity and the CMB bispectrum}

\author{J.R.~Fergusson}

\author{E.P.S.~Shellard}

\affiliation{Centre for Theoretical Cosmology,\\
Department of Applied Mathematics and Theoretical Physics,\\
University of Cambridge,
Wilberforce Road, Cambridge CB3 0WA, United Kingdom}

\date{\today}

\begin{abstract}

\noindent We present a set of formalisms for comparing, evolving and 
constraining primordial non-Gaussian models through the CMB bispectrum. 
We describe improved methods for efficient computation of the full CMB 
bispectrum  for any general (non-separable) primordial bispectrum, incorporating
a flat sky approximation and a new cubic interpolation.   We 
review all the primordial non-Gaussian models in the present literature and 
calculate the CMB bispectrum up to $l <2000$ for each different model.   This allows
us to determine the observational independence of these models by calculating 
the cross-correlation of their CMB bispectra.  We are able to identify several 
distinct classes of primordial shapes - including equilateral, local, warm, flat 
and feature (non-scale invariant) - which should 
be distinguishable given a significant detection of CMB non-Gaussianity. 
We demonstrate that a  simple shape correlator provides a fast and reliable 
method for determining whether or not CMB shapes are well correlated.  We 
use an eigenmode decomposition of the primordial shape to 
characterise and understand model independence.   Finally, we 
advocate a standardised normalisation method for $f_{NL}$ based 
on the shape autocorrelator, so that observational limits and errors 
$\Delta f_{NL}$ can be consistently compared for different models.

\end{abstract}

\pacs{1}

\maketitle

%%%%%%%%%%%%%%%%%%%%%%%%%%%%%%%%% INTRODUCTION %%%%%%%%%%%%%%%%%%%%%%%%%%%%%%%%

\section{Introduction}

Constraints on non-Gaussianity arguably provide the most stringent observational tests of
the simplest inflationary paradigm and, in the near future, these limits are set 
to tighten substantially.   Single-field slow-roll inflation predicts to high precision that 
the CMB will be a Gaussian random field and hence can be completely described by its 
angular power spectrum. However, if there was mechanism for generating some 
non-Gaussianity in the initial perturbations then its measurement would open 
up a wealth of extra information about the physics governing the early universe.  
Motivated by the first discussions of the local case which is dominated by squeezed states, non-Gaussianity is 
usually parametrised by $f_{NL}$, a quantity that can be extracted from CMB observations. 
The purpose of this paper is to apply and improve the methods detailed in ref.~\cite{0612713} for 
calculating the CMB bispectrum for any general (non-separable) primordial non-Gaussian model 
and, on this basis, to determine the extent to which competing models are independent and 
can be constrained by present limits on $f_{NL}$.

At present, CMB observations directly constrain only three separable primordial non-Gaussian models because of the 
calculational difficulties of using bispectrum estimators in the general case.   These are the local model and the 
equilateral model   (a separable approximation to the DBI inflation) and warm inflation with the latest (model-dependent) CMB constraints on $f_{NL}$ becoming  \cite{09012572,08030547,07071647}
\begin{align} \label{eq:locallim}
-4 &< f^{local}_{NL} < 80\,, \\ \label{eq:equilim}
-151 &< f^{equi}_{NL} < 253\,, \\ \label{eq:warmlim}
-375 &< f^{warm}_{NL} < 37\,
\end{align}
The above are 2$\sigma$-limits on non-Gaussianity, though we note that there is a tentative claim
 \cite{07121148} of a detection of $f^{local}_{NL}$ at almost $3\s$, with limits $27 < f^{local}_{NL} < 147 \;(2\s)$.  Given the growing range of theoretical 
possibilities for other (non-separable) non-Gaussian bispectra, an important goal is the development of  methods 
which can be used to place direct constraints in the general case \cite{inprep}.  However, our purpose
here is to note the models which are tightly constrained by the present limits and, conversely, to 
identify those for which further investigation is warranted.  

The bispectrum has been shown to be optimal for detecting primordial non-Gaussianity \cite{0503375}, but
alternative approaches seem to be able to produce comparable limits.  A wavelet analysis of the WMAP 5 year
data yields   $-8 < f_{NL} < 111 \;(2\s)$ \cite{08070231} and positional information from wavelets can be used to 
examine the likelihood of specific features.  Minkowski functionals provide a geometric characterisation of 
the temperature fluctuations in the CMB, yielding a slightly weaker constraint $-70 < f_{NL} < 91 \;(2\s)$ \cite{08023677}.
Of course, the effects of non-Gaussianity are not only felt in the CMB but could be detectable in a wide range of astrophysical measurements, such as cluster abundances and the large scale clustering of highly biased tracers.
However, given the imminent launch of the Planck satellite with a projected constraint $|f_{NL}| \le 5 \;(2\s)$  \cite{0005036} (almost at the level of cosmic variance, $|f_{NL}| \le 3 \;(2\s)$), the focus of this paper remains on
the CMB bispectrum.

In section 2,  we review the relationship between the primordial bispectrum and its counterpart in the CMB, 
noting the importance of separability for present estimators.   We also present a new analytic solution for the 
CMB bispectrum from a \it{constant} primordial model, against which it is useful to normalise other models. 
In section 3, we introduce the shape function relevant for nearly scale-invariant bispectra, and we review 
the current literature classifying current non-Gaussian models into centre-, corner- and 
edge-weighted categories, as well as those which are not scale-invariant.    We use a shape correlator to forecast the  
cross-correlation of the CMB bispectra for all these different models, identifying five distinct classes of shapes.   
This work takes forward earlier discussions of the non-Gaussian shape (see, for example, \cite{0405356,0509029,0612571}) but here we are able to directly compare to the actual CMB bispectra for all the models.   Moreover, we propose a specific eigenfunction decomposition of the shape function which offers
insight as to why particular shapes are related or otherwise. 

In section 4, we describe important improvements to the numerical methods which we use to make accurate 
computations of the CMB bispectrum for all the models surveyed \cite{0612713}.  The most important of these are 
the flat sky approximation and a cubic interpolation scheme for the tetrahedral domain of allowed multipoles.  What was previously regarded as an insurmountable computational problem has now become tractable, irrespective
of separability. 
In section 5, we present the main results detailing the cross-correlations for all the different models, while
confirming the different classes of independent shapes previously identified, some of which remain to be fully constrained (even by present data).   We compare the results of the shape and CMB bispectra correlators, noting the 
efficacy of the shape approach in identifying models for which quantitative CMB analysis is  required.   
Finally, in section 6, we propose an alternative normalisation procedure for $f_{NL}$ which brings the 
constraints into a more consistent pattern, allowing for a model-independent comparison of the true level of 
non-Gaussianity.

\section{Relating the primordial and CMB bispectrum}\label{se:estimator}

The primordial gravitational potential $\Phi({\bf k})$ induces CMB temperature anisotropies 
which we represent using $a_{lm}$'s, that is, 
$$ 
\frac{\Delta T}{T}(\hat {\bf n}) = \sum_{lm} a_{lm} Y_{lm}(\hat {\bf n})\,.
$$
The linear evolution which relates them is mediated by the transfer functions $\Delta_l(k)$ through the
integral, 
\begin{align} \label{eq:alm}
a_{lm} = 4\pi (-i)^l \int \frac{d^3 k}{(2\pi)^3} \D_l(k) \O(\bk) Y_{lm}(\uk)\,.
\end{align}
The CMB bispectrum is the three point correlator of the $a_{lm}$,
\begin{align}
B^{l_1 l_2 l_3}_{m_1 m_2 m_3} = \<a_{l_1 m_1} a_{l_2 m_2} a_{l_3 m_3}\>\,,
\end{align}
and so, substituting (\ref{eq:alm}), we obtain
\begin{align}\label{eq:bispectrum1}
\nn B^{l_1 l_2 l_3}_{m_1 m_2 m_3} =\& (4\pi)^3 (-i)^{l_1+l_2+l_3} \int \frac{d^3 k_1}{(2\pi)^3} \frac{d^3 k_2}{(2\pi)^3} \frac{d^3 k_3}{(2\pi)^3} \D_{l_1}(k_1) \D_{l_2}(k_2) \D_{l_3}(k_3) \\
& \<\O(\bk_1)\O(\bk_2)\O(\bk_3)\> Y_{l_1 m_1}(\uk_1) Y_{l_2 m_2}(\uk_2) Y_{l_3 m_3}(\uk_3)\,.
\end{align}
The primordial bispectrum is defined as
\begin{align}\label{eq:primbispect}
\<\O(\bk_1)\O(\bk_2)\O(\bk_3)\> = (2\pi)^3 B_\O(k_1,k_2,k_3)\, \d(\bk_1+\bk_2+\bk_3)\,,
\end{align}
where the delta function enforces the triangle condition. We replace the delta function with its integral form and expand the exponential into spherical harmonics. If we substitute this into equation (\ref{eq:bispectrum1}) and integrate out the angular parts of the three $k_i$ integrals, which yield delta functions, then we can remove the summation to obtain,
\begin{align}\label{eq:bispectrum3}
\nn B^{l_1 l_2 l_3}_{m_1 m_2 m_3} = \(\frac{2}{\pi}\)^3 \int \& d x d k_1 d k_2 d k_3 (x k_1 k_2 k_3)^2 B_\O(k_1,k_2,k_3) \D_{l_1}(k_1) \D_{l_2}(k_2) \D_{l_3}(k_3)\\
& j_{l_1}(k_1 x) j_{l_2}(k_2 x) j_{l_3}(k_3 x) \int d\W_x Y_{l_1 m_1}(\ux) Y_{l_2 m_2}(\ux) Y_{l_3 m_3}(\ux) \,.
\end{align}
The integral over the angular part of $x$ is known as the Gaunt integral and has a geometric solution,
\begin{align}
\nn \curl{G}^{l_1 l_2 l_3}_{m_1 m_2 m_3} &\equiv \int d\W_x Y_{l_1 m_1}(\ux) Y_{l_2 m_2}(\ux) Y_{l_3 m_3}(\ux) \\
&= \sqrt{\frac{(2l_1+1)(2l_2+1)(2l_3+1)}{4\pi}} \( \begin{array}{ccc} l_1 & l_2 & l_3 \\ 0 & 0 & 0 \end{array} \) \( \begin{array}{ccc} l_1 & l_2 & l_3 \\ m_1 & m_2 & m_3 \end{array} \)\,,
\end{align}
where $\textstyle {\( \begin{array}{ccc} l_1 & l_2 & l_3 \\ m_1 & m_2 & m_3 \end{array} \)}$
is the Wigner 3j symbol. First, as we expect the bispectrum to be isotropic, it is common to work with the angle averaged bispectrum,
\begin{align}
B_{l_1 l_2 l_3} = \sum_{m_i} \( \begin{array}{ccc} l_1 & l_2 & l_3 \\ m_1 & m_2 & m_3 \end{array} \)\<a_{l_1 m_1} a_{l_2 m_2} a_{l_3 m_3}\>\,.
\end{align}
We will find it more convenient to work with the reduced bispectrum where we drop the geometric factors associated with the Gaunt integral,
\begin{align}
B^{l_1 l_2 l_3}_{m_1 m_2 m_3} = \curl{G}^{l_1 l_2 l_3}_{m_1 m_2 m_3} b_{l_1 l_2 l_3}\,.
\end{align}
In removing the 3j symbols it is important to remember some of the constraints that they place on the l values. The first constraint is that the sum of the three $l_i$ must be even. The other is that if we treat the three $l_i$ as lengths, they must be able to form a triangle. This is analogous to the constraint that the three $k_i$ are able to form a triangle, which is demanded by the delta function in (\ref{eq:primbispect}),
and in $l$-space is enforced through the $x$ integral over the three spherical Bessel functions, which evaluates to zero if the corresponding triangle condition is violated. The reduced bispectrum is of the form,
\begin{align} \label{eq:redbispect}
\nn b_{l_1 l_2 l_3}= \(\frac{2}{\pi}\)^3 \int & dx d k_1 d k_2 d k_3\, \(x k_1 k_2 k_3\)^2\, B_\O(k_1,k_2,k_3)\\
& \D_{l_1}(k_1) \D_{l_2}(k_2) \D_{l_3}(k_3)\, j_{l_1}(k_1 x) j_{l_2}(k_2 x) j_{l_3}(k_3 x)\,.
\end{align}

In looking for a way in which we can simplify the integral associated with $b_{l_1 l_2 l_3}$, the most obvious place to start is to try to find simplifications for the primordial bispectrum $B_\O$. The most common, and simplest, approach is that pioneered by Komatsu and Spergel in \cite{0005036}. Here they expand the primordial gravitational potential perturbation as a Taylor expansion around the Gaussian part,
\begin{align}
\O(\bx) = \O_{G}(\bx) + f_{NL} \(\O_{G}^2(\bx) - \<\O_{G}(\bx)\>^2\)\,,
\end{align}
where $f_{NL}$ parametrises the level of non-Gaussianity. Simple calculation results in a primordial bispectrum of the form,
\begin{align}\label{eq:localbi}
B_\O(k_1,k_2,k_3) = 2 f_{NL} \( P_\O(k_1) P_\O(k_2) + P_\O(k_2) P_\O(k_3) + P_\O(k_3) P_\O(k_1) \)\,.
\end{align}
This is known as the \textit{local} model as the non-Gaussianity of the gravitational potential is local in space. Substituting this into equation (\ref{eq:redbispect}) for the reduced bispectrum we see that, as the primordial bispectrum only consists of products of functions of a single $k_i$, the integral can be separated into an integral over $x$ of the products of integrals over $k$,
\begin{align}\label{eq:localredbispectrum}
b_{l_1 l_2 l_3} = 2 f^{local}_{NL} \int x^2 dx \(\a_{l_1}(x)\b_{l_2}(x)\b_{l_3}(x) + 2\,\mbox{permutations}\)\,,
\end{align} 
where,
\begin{align}
\a_l(x) &= \frac{2}{\pi} \int k^2 dk \D_l(k) j_l(kx)\,, \\
\b_l(x) &= \frac{2}{\pi} \int k^2 dk P_\O(k) \D_l(k) j_l(kx)\,.
\end{align}
This reduces the dimension of integration from four to two, and the separation allows us to calculate the reduced bispectrum easily.

If we choose to work in the Sachs-Wolfe approximation, where we replace the transfer function with a Bessel function,
\begin{align}\label{eq:largeangle}
\D_{l}\(k\) = \frac{1}{3} j_{l}\((\t_o - \t_{dec}) \, k\)\,,
\end{align}
the integral for the reduced bispectrum can be expressed in closed form and it is possible to derive an analytic solution in simple cases. The exact analytic result for the local model on large angles is
\begin{align} \label{eq:analy}
B_{l_1l_2l_3} =  f_{NL}\(\frac{2 A_\O^2}{27\pi^2}\) G(l_1,l_2,l_3)\,,
\end{align}
where $G(l_1,l_2,l_3)$ is shorthand for,
\begin{align} \label{eq:gfactor}
G(l_1,l_2,l_3) = \frac{1}{l_1(l_1+1)l_2(l_2+1)} + \frac{1}{l_2(l_2+1)l_3(l_3+1)} + \frac{1}{l_3(l_3+1)l_1(l_1+1)}\,.
\end{align}
Here, we present a second analytic solution, that of the simplest primordial bispectrum that scales in the correct manner, 
\begin{align} \label{eq:constmodel}
B_\O(k_1,k_2,k_3) = 1 / (k_1 k_2 k_3)^2\,,
\end{align}
which we will denote as the \it{constant} model.
 Here the reduced bispectrum integral becomes,
\begin{align}
b_{l_1 l_2 l_3}= f_{NL} \(\frac{2}{3\pi}\)^3 \int x^2 dx\, \Pi_{i=1}^3 I_{l_i}(0,x)\,,
\end{align}
where
\begin{align}
I_{l}(p,x) = \int k^p dk j_{l}(k)j_{l} (x k)\,.
\end{align}
The $I_{l}(0,x)$ integral  has a nice solution (see ref~\cite{watson} p405),
\begin{align}\label{eq:zerobesselsolution}
x>1 \imp \frac{\pi}{2}\frac{x^{-(l+1)}}{2l+1} \,,&& x<1 \imp \frac{\pi}{2}\frac{x^{l}}{2l+1}\,,
\end{align}
from which we can obtain an exact large-angle solution for the constant model  bispectrum,
\begin{align}\label{eq:constbispectrum}
\nn b_{l_1 l_2 l_3} &= \frac{f_{NL}}{27} \frac{1}{(2l_1+1)(2l_2+1)(2l_3+1)}\[\int^1_0 x^{l_1+l_2+l_3+2} dx + \int_1^\infty x^{-l_1-l_2-l_3-1} dx \]\\
&= f_{NL} \(\frac{1}{3}\)^3 D(l_1,l_2,l_3)\,,
\end{align}
where we have defined,
\begin{align}\label{eq:dfactor}
D(l_1,l_2,l_3) = \frac{1}{(2l_1+1)(2l_2+1)(2l_3+1)}\[\frac{1}{l_1+l_2+l_3+3} + \frac{1}{l_1+l_2+l_3}\]\,.
\end{align}

Unfortunately the bispectrum signal is too weak for us to be able to measure individual multipoles directly from data, so to compare theory with observations we must use an estimator which sums over all multipoles. At the most basic level estimators can be thought of as performing a least squares fit of the bispectrum predicted by theory, $\<a_{l_1 m_1} a_{l_2 m_2} a_{l_3 m_3}\>$, to the bispectrum obtained from observations, $a^{obs}_{l_1 m_1} a^{obs}_{l_2 m_2} a^{obs}_{l_3 m_3}$. If we ignore the effects of sky cuts and inhomogeneous noise the estimator can be written,
\begin{align} \label{eq:estimator}
\curl{E} = \frac{1}{\curl{N}^2} \sum_{l_i m_i} \frac{\<a_{l_1 m_1} a_{l_2 m_2} a_{l_3 m_3}\>}{C_{l_1}C_{l_2}C_{l_3}} a^{obs}_{l_1 m_1} a^{obs}_{l_2 m_2} a^{obs}_{l_3 m_3}\,,
\end{align}
where
\begin{align} \label{eq:normalisation}
\curl{N}(B) = \sqrt{\sum_{l_i} \frac{B^2_{l_1 l_2 l_3}}{C_{l_1}C_{l_2}C_{l_3}}}\,.
\end{align}
The above estimator has been shown to be optimal \cite{0503375} for general bispectra in the limit where the non-Gaussianity is small and the observed map is free of instrument noise and foreground contamination. This is of course an idealised case and we need to consider taking into account the effect of sky cuts, inhomogeneous noise, and beam effects which were considered in some detail in refs~\cite{0509029,07114933}. 

Here, for clarity we consider only the simple form of the estimator (\ref{eq:estimator}), since all statements made about it can be extended to the more complete version. If we consider the local model we can rewrite it as,
\begin{align}
\curl{E} = \frac{1}{\curl{N}^2} \sum_{l_i m_i} \curl{G}^{l_1 l_2 l_3}_{m_1 m_2 m_3} b_{l_1 l_2 l_3} \frac{a^{obs}_{l_1 m_1} a^{obs}_{l_2 m_2} a^{obs}_{l_3 m_3}}{C_{l_1} C_{l_2} C_{l_3}}\,.
\end{align}
From this we see that we only need to be able to calculate the reduced bispectrum, $b_{l_1 l_2 l_3}$, from theory rather than the full bispectrum, $\<a_{l_1 m_1} a_{l_2 m_2} a_{l_3 m_3}\>$, which would be much more challenging. For the local model we can use the separation of $b_{l_1 l_2 l_3}$, from equation (\ref{eq:localredbispectrum}), to separate the sums in the estimator, so,
\begin{align}
\curl{E} = \frac{1}{\curl{N}^2}\int d^3x A(\bx) \(B(\bx)\)^2\,,
\end{align}
where,
\begin{align}
A(\bx) &\equiv \sum_{lm} \a_l(x) \frac{a^{obs}_{lm}}{C_l} Y_{lm}(\ux) \\
B(\bx) &\equiv \sum_{lm} \b_l(x) \frac{a^{obs}_{lm}}{C_l} Y_{lm}(\ux)\,.
\end{align}
From this we can conclude that if the primordial bispectrum is separable then we can overcome both the issue with the multi-dimensional integration, and the calculation of the 3j symbols. Thus separability has become the cornerstone of all non-Gaussian analysis.

\section{The shape of primordial bispectra} \label{se:shapes}

\subsection{Shape function}

Here we will introduce the shape function for the primordial bispectrum. As the power spectrum is constrained to be very close to scale invariant we expect that the bispectrum will show similar behaviour. Exact scale invariance for the local model results in an equal $k$ primordial bispectrum of the form,
\begin{align}
B^{local}_\O(k,k,k) = 6 f_{NL} \frac{A_\O^2}{ k^6}\,.
\end{align}
This equal-$k$ behaviour with  $B_\O(k,k,k) \propto k^{-6}$ turns out to be the expected scaling of a large number of non-Gaussian models, and so the difference between these models is only due to the dependence of the primordial bispectrum on the ratios $k_1:k_2$ and $k_1:k_3$ \cite{0405356}. 
As we already have the factor, $(k_1 k_2 k_3)^2$, in the integral for the reduced bispectrum (\ref{eq:redbispect}), it is natural to use it to divide out the scale dependence of the primordial bispectrum. Thus, we can define the momentum dependence through a shape function $S$ as,
\begin{align} \label{eq:shapefn}
S(k_1,k_2,k_3) = \frac{1}{N} (k_1 k_2 k_3)^2 B_\O(k_1,k_2,k_3)\,,
\end{align}
where $N$ is an appropriate normalisation, often taken to be $N= 1/f_{NL}$. (Inconsistent definitions for $f_{NL}$ for different models mean that it is difficult to compare their observational limits; we will discuss this further in section 6, but here we will generally factor out $N$).

The two most commonly discussed models are the local model,
\begin{align} \label{eq:local}
S^{local}(k_1,k_2,k_3) \propto \frac{K_3}{K_{111}}\,,
\end{align}
and the equilateral model
\begin{align} \label{eq:equi}
S^{equi}(k_1,k_2,k_3) \propto \frac{\tilde k_1\tilde k_2\tilde k_3}{K_{111}}\,.
\end{align}
However, we should also keep in mind the constant model $S^{const} (k_1,k_2,k_3) =1$, for 
which we have a large-angle analytic solution $D(l_1,l_2,l_3)$.
Here in eqns (\ref{eq:local}--\ref{eq:equi}), and throughout this section, we will adopt a shorthand notation for the possible combinations of wavenumbers that can contribute to the bispectrum (i.e.\ the simplest terms consistent with its symmetries):
\begin{align}\label{eq:compactnotn}
K_p &= \sum_i (k_i)^p \quad\mbox{with}\quad K = K_1 \\
K_{pq} &= \frac{1}{\D_{pq}}\sum_{i\not=j} (k_i)^p (k_j)^q\\
K_{pqr} &= \frac{1}{\D_{pqr}}\sum_{i\not=j\not=l} (k_i)^p (k_j)^q (k_l)^r\\
\tilde{k}_{ip} &= K_p - 2 (k_i)^p \quad\mbox{with}\quad \tilde{k}_{i} = \tilde{k}_{i1}\,.
\end{align}
where $\Delta_{pq} = 1 +\delta_{pq}$ and $\Delta_{pqr} = \Delta_{pq}(\Delta_{qr} + \delta_{pr})$ (no summation) . This notation significantly compresses the increasingly complex bispectrum expressions quoted in the literature.

One of the first models discussed with a specific shape was Maldacena's calculation for single field slow-roll inflation \cite{0210603}:
\begin{align}\label{eq:Maldshape}
\nn S^{Mald}(k_1,k_2,k_3) &\propto (3\e-2\n)\frac{K_3}{K_{111}} + \e(K_{12} + 8\frac{K_{22}}{K}) \\
&\px (6\e-2\n)S^{local}(k_1,k_2,k_3) + \frac{5}{3}\e S^{equi}(k_1,k_2,k_3))\,,
\end{align}
where $S^{local}$ and $S^{equi}$ are normalised so that $S^{local}(k,k,k) = S^{equi}(k,k,k)$. While we know the predicted non-Gaussianity in this case is negligible, there are more recent models which yield similar combinations of equilateral and local terms which are measurable (e.g. non-local inflation \cite{08023218}). We need to know, therefore, the extent to which we can distinguish between the relative contributions from these different shapes and the degree to which they are observationally independent.

\subsection{Shape correlator}

One obvious way to distinguish between models is to use the estimator discussed previously (\ref{eq:estimator}), replacing the observed bispectrum with one calculated from a competing theory,
\begin{align}\label{eq:cmbcor}
\curl{C}(B,B^\pr) = \frac{1}{\curl{N}(B)\curl{N}(B^\pr)}\sum_{l_i} \frac{B_{l_1 l_2 l_3}B^\pr_{l_1 l_2 l_3}}{C_{l_1} C_{l_2} C_{l_3}}\,.
\end{align}
If the observational data contained a bispectrum of the form $B^\pr_{l_1 l_2 l_3}$ then $\curl{C}(B,B^\pr)$ is an estimate of the proportion of the correct $f^{\pr}_{NL}$ we would recover by using an estimator based on $B_{l_1 l_2 l_3}$. However, this Fisher matrix approach is extremely computationally demanding as we must calculate the full bispectrum for each model before we can make any comparison. What we would like in addition, therefore, is a simple method allowing us to predict the value of the correlator directly from the shape functions, thus indicating cases in which a full Fisher
matrix analysis is warranted. 

\begin{figure}[t]
\centering
\includegraphics[width=0.6\linewidth]{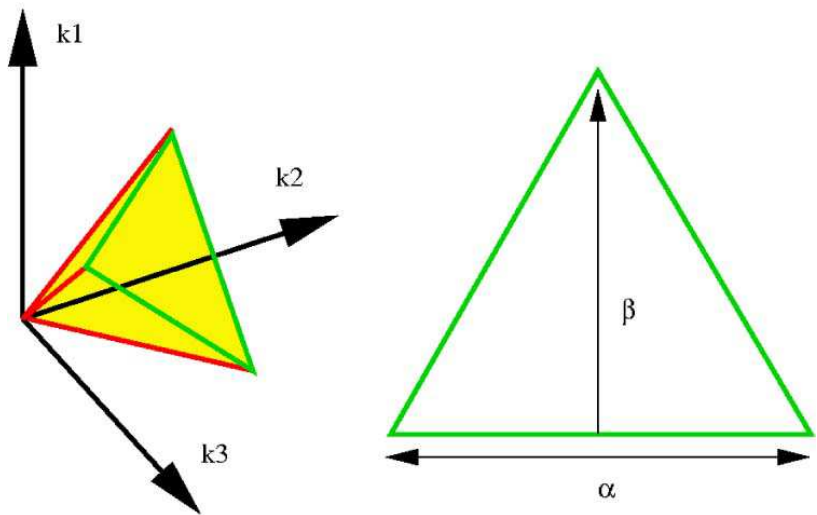}
\caption[The region of $k$-space allowed by the triangle inequality.]{\small (a) The region of $k$-space allowed by the triangle inequality, i.e., for which the primordial bispectrum is valid. The red lines are $k_1=k_2,\, k_3=0;\; k_2=k_3,\, k_1=0;\; k_3=k_1,\, k_2=0$ and the allowed region is in yellow. (b) This area can be parametrised into slices represented by the green triangle and the distance ${2}|\vec k|/{\sqrt{3}}$ of the centre of the triangle from the origin.}
\label{fig:region}
\end{figure}

If we return to equation (\ref{eq:redbispect}) for the reduced bispectrum and substitute the expression for the shape function we have,
\begin{align} \label{eq:biint}
b_{l_1 l_2 l_3} = f_{NL} \(\frac{2}{\pi}\)^3 \int_{\curl{V}_k} d\curl{V}_k S(k_1,k_2,k_3) \D_{l_1}(k_1) \D_{l_2}(k_2) \D_{l_3}(k_3) I^G_{l_1 l_2 l_3}(k_1,k_2,k_3)\,,
\end{align}
where $\curl{V}_k$ is the area inside the cube $[0,k_{max}]$ allowed by the triangle condition (refer to figure~\ref{fig:region}). The integral $I^G$ is given by,
\begin{align}
I^G_{l_1 l_2 l_3}(k_1,k_2,k_3) = \int x^2 dx j_{l_1}(k_1 x) j_{l_2}(k_2 x) j_{l_3}(k_3 x)\,.
\end{align}
So $S(k_1,k_2,k_3)$ is the signal that is evolved via the transfer functions to give the bispectrum today, with $I^G$ giving an additional, purely geometrical, factor. Essentially, $I^G$ acts like a window function on all the shapes as it projects from $k$ to $l$-space, that is, it tends to smear out their sharper distinguishing features, but only erasing significant differences in extreme cases (as we shall discuss later). This means that the shape function $S(k_1,k_2,k_3)$, especially in the scale-invariant case, can be thought of as the primordial counterpart of the reduced bispectrum $b_{l_1 l_2 l_3}$ before projection.

To construct a shape correlator that predicts the value of (\ref{eq:cmbcor}) correctly we then should consider something of the form
\begin{align}\label{eq:shapeint}
F(S,S^\pr) = \int_{\curl{V}_k} S(k_1,k_2,k_3) \,S^\pr(k_1,k_2,k_3) \,\w(k_1,k_2,k_3) \, d\curl{V}_k\,,
\end{align}
where $\w$ is an appropriate weight function. We note that authours in ref.~\cite{0405356} define a cosine between shape functions to be used as a correlator. However, we comment later on the quantitative differences of their definition, notably its breakdown in the non-scale invariant case.

The question now is what weight function should we choose? Our goal is to choose $S^2 \w$ in $k$-space such that it produces the same scaling as the estimator $B^2/C^3$ in $l$-space. Let us consider the simplest case where both $k_1=k_2=k_3=k$ and $l_1=l_2=l_3=l$. For primordial bispectra which are scale invariant, then,
\begin{align}
S^2(k,k,k) \w(k,k,k) \propto \w(k,k,k)\,.
\end{align}
If we work in the large angle approximation, and assume that $l+1 \px l$, then we know $C_l \propto 1 / l^2$ and from the analytic solutions $G$, $D$, (\ref{eq:gfactor}, \ref{eq:dfactor}) that $b_{lll} \propto 1 / l^4$. The angle averaged bispectrum is related to the reduced bispectrum by,
\begin{align}
B_{l_1 l_2 l_3} = \sqrt{\frac{(2l_1+1)(2l_2+1)(2l_3+1)}{4\pi}} \( \begin{array}{ccc} l_1 & l_2 & l_3 \\ 0 & 0 & 0 \end{array} \) b_{l_1 l_2 l_3}\,.
\end{align}
For equal $l$, we can deduce that,
\begin{align}
B_{l l l} \propto \( \begin{array}{ccc} l & l & l \\ 0 & 0 & 0 \end{array} \) l^{-5/2}\,,
\end{align}
The Wigner 3J symbol has an exact solution for which
\begin{align}
\( \begin{array}{ccc} l & l & l \\ 0 & 0 & 0 \end{array} \) \px (-1)^{3l/2} \frac{1}{\sqrt{3l+1}} \sqrt{\frac{l!^3}{3l!}}\frac{(3l/2)!}{(l/2)!^3} \px (-1)^{3l/2} \sqrt{\frac{2}{\sqrt{3}\pi}} \frac{1}{l} \,,
\end{align}
with the last expression using Stirling's approximation, $l! \px \sqrt{2\pi l} (l/e)^l$. Combining these results gives,
\begin{align}\label{eq:estimatorscale}
\frac{B^2_{lll}}{C^3_l} \propto l^{-1}\,,
\end{align}
and so we find that we should choose a weight function $w(k,k,k) \propto k^{-1}$. This is a very simple approximation which ignores the cross-sectional weighting inherent in (\ref{eq:cmbcor}). Our analysis of the behaviour in the $(\a,\b)$-slices with a constant primordial shape function ($S(\a,\b) = \mbox{const}$) shows that $B^2/C^3$ is flat in the interior and then grows to a finite value on the boundary. However this variation is confined to be very close to the boundary and we choose to 
neglect this effect. As a result we take the explicit flat $k^{-1}$ weighting:
\begin{align}\label{eq:weight}
w(k_1,k_2,k_3) = \frac{1}{k_1+k_2+k_3}\,,
\end{align}
Note that this weighting does not incorporate damping due to photon diffusion at large $l$, edge effects or smoothing due to the projection from $k$ to $l$-space. These could be included using phenomenological window functions, but our purpose here is simplicity.   In any case, 
the choice of the weight function may significantly improve forecasting accuracy, but it does not impact important qualitative insights. 

\begin{table}[t]
\centering
\begin{tabular}[t]{|@{\hspace{1mm}}c@{\hspace{1mm}}|@{\hspace{1mm}}c@{\hspace{1mm}}|@{\hspace{1mm}}c@{\hspace{1mm}}|@{\hspace{1mm}}c@{\hspace{1mm}}|@{\hspace{1mm}}c@{\hspace{1mm}}|@{\hspace{1mm}}c@{\hspace{1mm}}|@{\hspace{1mm}}c@{\hspace{1mm}}|@{\hspace{1mm}}c@{\hspace{1mm}}|@{\hspace{1mm}}c@{\hspace{1mm}}|@{\hspace{1mm}}c@{\hspace{1mm}}|}
\hline
 & DBI & Equi & Feat & FlatS & Ghost & Local & Single& Warm & WarmS \\
\hline
DBI & 1.00 & 0.99 & -0.41 & 0.39 & 0.94 & 0.50 & 0.98 & 0.38 & 0.55 \\
Equi & & 1.00 & -0.36 & 0.30 & 0.98 & 0.46 & 0.95 & 0.44 & 0.63 \\
Feat & & & 1.00 & -0.44 & -0.26 & -0.41 & -0.46 & -0.05 & -0.08 \\
FlatS & & & & 1.00 & 0.15 & 0.62 & 0.49 & 0.01 & -0.03 \\
Ghost & & & & & 1.00 & 0.37 & 0.86 & 0.50 & 0.71 \\
Local & & & & & & 1.00 & 0.55 & 0.30 & 0.27 \\
Single & & & & & & & 1.00 & 0.29 & 0.44 \\
Warm & & & & & & & & 1.00 & 0.80 \\
WarmS & & & & & & & & & 1.00 \\
\hline
\end{tabular}
\caption{Shape correlations} 
\label{tb:shapecorrelator}
\end{table}

With this choice of weight (\ref{eq:weight}), the primordial shape correlator from (\ref{eq:shapeint}) then takes the form
\begin{align}\label{eq:shapecor}
\bar{\curl{C}}(S,S^\pr) = \frac{F(S,S^\pr)}{\sqrt{F(S,S)F(S^\pr,S^\pr)}}\,.
\end{align}
This correlator is very simple to calculate and allows us to quickly categorise new models and their degree of independence.

Before surveying model shapes currently in the literature, we note the utility of this correlator (\ref{eq:shapecor}) with the simple shape examples discussed above. The local and equilateral shapes (\ref{eq:local}, \ref{eq:equi}) yield a 46\% correlation. Thus, supposing the universe to have local non-Gaussianity, a highly significant observation using a local estimator would be expected to produce a (less) significant result also for an equilateral estimator. Nevertheless, 46\% is a relatively low correlation and the equilateral and local shapes can be regarded as distinguishable in principle. What of Maldacena's shape with $\epsilon \approx \eta$ (\ref{eq:Maldshape}) (as, for example, in $m^2\phi^2$ inflation)? If this were indeed observable, it would yield the rather striking result that it is 99.7\% correlated with the local shape (\ref{eq:local}) (see also \cite{0405356}) and that it is only 53\% correlated with equilateral (\ref{eq:equi}). As we shall see, such strong correspondences between models with apparently different shapes are rather common. In this case, we need to go to a fine-tuned regime near $\eta \approx 2.84\epsilon$ in order for (\ref{eq:Maldshape}) to have an equal 86\% correlation to both local and equilateral shapes; it is clearly generically in the local class of models. Table~\ref{tb:shapecorrelator} provides a summary of the correlations between all the shapes we discuss in the next sections.

Finally we note that the definition in ref.~\cite{0405356} of a cosine correlator has three weaknesses. The first is that it is calculated only on a 2D slice, $k_1 = \mbox{const}$, through the tetrahedron (in contrast to our 3D integration over the $k=\mbox{const}$ slices in figure \ref{fig:region}). This choice may be tolerable for comparing shapes which scale in exactly the same way as each slice will give the same correlation. However, if the two models being compared have differing scale dependence, like for comparisons with the feature model, then it will produce poor results. Even in the case when comparing shapes which scale in exactly the same way there is a second more subtle issue which makes the cosine correlator unsuitable. The region covered in the bispectrum correlator is the intersection of the cube defined by $[0,l_{max}]$ and the tetrahedron defined by the triangle condition on the three $l_i$ and so we should cover a similar region in $k$-space. If we just look at the correlation on one slice then we miss the effect the shape of the region has on the result. If we think of the region as being composed of many parallel slices then some will be incomplete due to the effect of restricting the individual $k_i<k_{max}$. Different slices will give different correlations depending on how much they have been cut and so no slice is truly representative of the true correlation. The third and related problem if that the weight in (\ref{eq:weight}) is required to give accurate representation of the CMB correlation using shape functions in $k$-space.

\subsection{Shape decomposition}

Given strong observational limits on the scalar tilt we expect all shape functions to exhibit behaviour close to scale-invariance, so that $S(k_1,k_2, k_3)$ will only depend weakly on $|{\vec k}|$. Here, we choose to parametrise the magnitude of the $k_i$'s with both $|{\vec k}| = (k_1^2+ k_2^2 + k_3^2)^{1/2}$
and the semi-perimeter,
\begin{align}
k \equiv \frac{1}{2}(k_1+k_2+k_3)\,.
\end{align}
A consequence of this scaling behaviour is that the form of the shape function on a cross-section 
is essentially independent of $k$, so that for the models under consideration we can write 
\begin{align}\label{eq:shapesplit1}
S(k_1,k_2,k_3) = f(k)\bar S(\hat k_1,\hat k_2, \hat k_3)\,.
\end{align}
where 
\begin{align}\label{eq:hatparam}
\hat k_1 = \frac{{k}_1}{k} \,, \qquad \hat k_2 = \frac{{k}_2}{k} \,, \qquad \hat k_3 = \frac{{k}_3}{k}\,, 
\end{align}
and we note that $\hat{k}_1+\hat{k}_2+\hat{k}_3 = 2$.  Since we are restricted to the region where the three $k_i$ are able to form a triangle by momentum conservation, we will reparametrise the allowed region to separate out the overall scale $k$ from the behaviour on a cross-sectional slice $\curl{S}_k$. This two-dimensional slice is spanned by the remaining coordinates (see figure~\ref{fig:region}),
\begin{align}\label{eq:parameters}
\nonumber k_1 &= k \(1-\b\)\,, \\
\nonumber k_2 &= \hf k \(1 + \a + \b\)\,,\\
k_3  &= \hf k \(1 - \a + \b\)\,.
\end{align}
The surface $k = {\rm const.}$\ defines a plane with normal $(1,1,1)$ at a distance ${2}|{\vec k}|/\sqrt{3}$ from the origin. Our new parameters $\a,\,\b$, parametrise the position on the triangular domain formed by the intersection of the tetrahedral region and that plane \cite{0410486}. They have the following domains, $0 \le k < \infty$, $0 \le \b \le 1$ and $ -(1-\b)\le \a \le 1 - \b$.  In this parametrisation we can re-write shape function (\ref{eq:shapesplit1}) and the volume element respectively as
\begin{align}\label{eq:shapesplit2}
S(k_1,k_2,k_3) = f(k) \bar{S}(\a,\b)\,, \qquad d\curl{V}_k = dk_1 dk_2 dk_3 = k^2 dk d\a d\b\,.
\end{align}
Here, we note that $f(k)\approx {\rm const.}$\ for all the model shapes to be discussed in the next section, with the exception of the feature and oscillatory models.

\begin{figure}[t]
\centering
\includegraphics[width=0.9\linewidth]{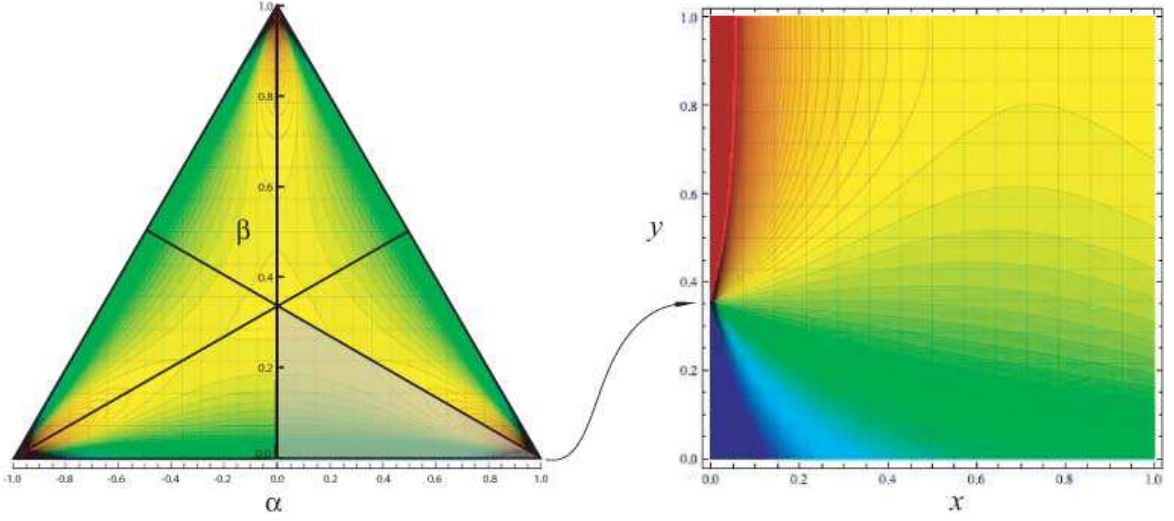}
\caption[Coordinate transformation required for eigenmode expansion on a uniform domain.]{\small Coordinate transformation from a shaded subtriangle on the equilateral $(\a,\,\b)$ slice to the uniform square $(x,\,y)$ domain suitable for an eigenmode expansion, i.e.$\a = 1 - x\,,\;\beta = yx/3\, $. Here, we illustrate the transformation with contour plots for the warm inflation shape function. Note the non-trivial behaviour in the corner region where the function diverges and the sign changes.} \label{fig:coordtransf}
\end{figure}

Having restricted our discussion to two-dimensional $(\a,\b)$-slices, we now note that bispectrum symmetries are such that we need only characterise the shape on one sixth of this domain (refer to figure \ref{fig:coordtransf}). This is a right-angled subtriangle with corners defined by the centre of the original triangle plus any corner together with the midpoint of an adjacent side. Here, we choose the bottom right triangle (containing $k_3 \rightarrow 0$), with corners ($\textstyle\frac{1}{ 3}$, 0), (1, 0) and (0, 0) (i.e. the shaded region in figure~\ref{fig:coordtransf}). In order to set up a straightforward eigenfunction decomposition, we make the following coordinate transformation to take our subtriangle to a unit square, 
\begin{align} \label{eq:abcoordxy}
\a = 1 - x,\qquad \beta = yx/3\,,
\end{align}
with $0\le x\le 1$ and $0\le y\le1$. Analogous to polar coordinates at $r=0$, this transformation expands
the $k_3=0$ corner (see figure~\ref{fig:coordtransf}) and our $\curl{S}_k$ volume element becomes
\begin{align} 
d\a \, d\b = x\, dx\, dy\,.
\end{align}
Here, we note that $x$ and $k_3$ share the same squeezed limit ($x \rightarrow 0$
as $k_3 \rightarrow 0$). 

With the simple weight function $w(x,y)= x$ we can now decompose an arbitrary shape function 
$\bar{S}(x,y)$ defined on $\curl{S}_k$ into a sum, 
\begin{align} \label{eq:eigenmode}
\bar{S}(x,y) = \sum_{m,n} c_{mn} X_m(x) Y_n(y)\,,
\end{align}
consisting of products of orthogonal eigenfunctions $X_m$ and $Y_n$ defined on the unit interval. One possible choice would be Bessel functions $X_n(x) = J_p(\lambda_{pn} x)$ (given the weight $w=x$) and trigonometric functions $Y_m(y) = A \sin(ny) + B \cos(ny)$. However, the shape function in general is neither periodic nor vanishing on the boundary, leading to an ill-conditioned problem with poor series convergence. Given these arbitrary boundary conditions, a better choice employs Legendre polynomials $P_n(y)$ in the $y$-direction and analogous radial polynomials $R_m(x)$ in the $x$-direction. The domain $0\le y\le 1$ requires shifted Legendre polynomials $\bar P_n(y)$ which, unit normalised, become
\begin{align} 
\bar P_0(y) = 1\,, \quad \bar P_1(y) = \sqrt{3} (-1 + 2y) \, \quad \bar P_2(y) = \sqrt{5} (1 - 6y + 6y^2) \,\quad ...
\end{align}
The eigenfunctions in the $x$-direction can be created using the generating function,
\begin{align} 
R_n (x) = \frac{1}{N} \left| \begin{array}{ccccc}
{1}/{2} & {1}/{3} & {1}/{4} & ... &{1}/{(n+2)} \\
{1}/{3} & {1}/{4} &{1}/{5} & ... & {1}/{(n+3)} \\
... & ... & ... & ... & ... \\
{1}/{n} & {1}/{n+1} &{1}/{n+2} & ... & {1}/{(2n+1)} \\
1& x & x^2 & ... & x^n \end{array} \right|\,,
\end{align}
with the first unit normalized polynomials given by,
\begin{align} 
R_0(x) = \sqrt{2}\,,\quad R_1(x) = - 4 + 6 x\,, \quad R_2(x) = \sqrt{6} (3 - 12x + 10 x^2) \,,\quad ...
\end{align}

\begin{figure}[t]
\centering
\includegraphics[width=.65\linewidth]{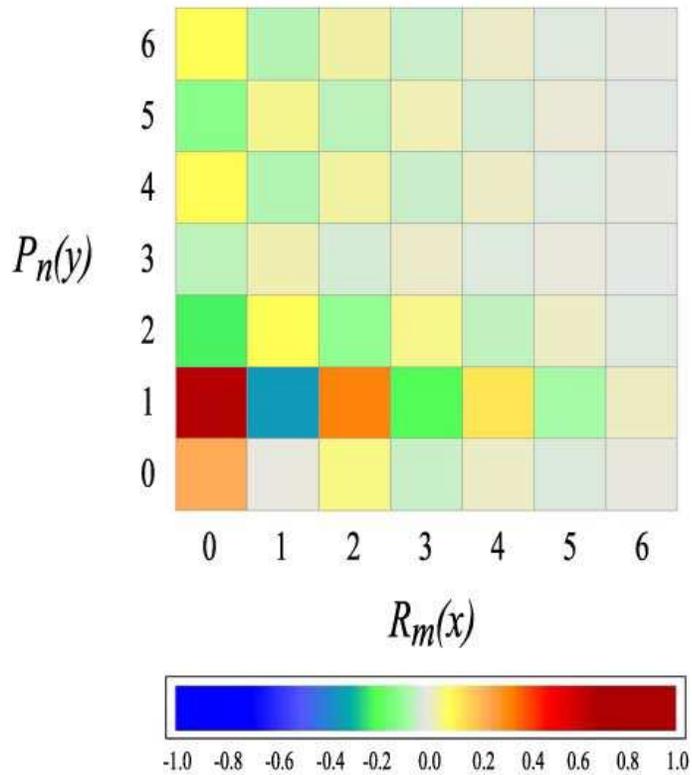}
\caption[Eigenvalues for warm inflation.]{\small Eigenvalues for the two-dimensional eigenmode expansion of the warm inflation shape function. Here, we denote the conventions with $m$ (for $R_m(x)$ incrementing in the horizontal $x$-direction and the $n$ (for $\bar P_n(y)$) in the vertical $y$-direction. Note the dominance of $R_m(x)P_1(y)$ modes. The colour coding (used also in figure~\ref{fig:eigen_all}) is such that only blue and red colours can contribute at above the 10\% level to the autocorrelator $C_k(S, S')$, with yellow and pale green below 1\%. }
\label{fig:eigen_warm}
\end{figure}

We can now find an eigenmode decomposition (\ref{eq:eigenmode}) on the domain $\curl{S}_k$ for any given shape function $S(x,y)$ with the expansion coefficients given by 
\begin{align} \label{eq:shapedec}
c_{mn} = \int_0^1\int_0^1 R_m(x) \, \bar P_n(y)\, \bar{S}(x,y) \, x\, dx\,dy\,.
\end{align}
Exploiting eigenmode orthogonality, the counterpart of the correlator (\ref{eq:shapecor}) between two shapes
$S$, $S'$ on the $\curl{S} _k$ slice then is
\begin{align} \label{eq:slicecor}
\bar C_{\curl{S}_k} (\bar{S}, \bar{S}') = \int_{\curl{S}_k} \bar{S}(\a,\b)\,\bar{S}'(\a,\b)\,d\a\,d\b = \sum_{n.m} c_{nm} c'_{nm}\,,
\end{align}
where we have assumed unit normalised $\bar{S}$, $\bar{S}'$ on $\curl{S}_k$. Even for scale-invariant shapes, the slice correlator (\ref{eq:slicecor}) is not identical to the overall shape correlator (\ref{eq:shapecor}) because the integration domain for the latter includes the cubic region $[0, k_{max}]$, that is, for large $k>k_{max}$ it integrates over interior regions of the slices which weights the centre more heavily. (The integration domain is illustrated explicitly in figure~\ref{fig:transform}.) Nevertheless, $\bar{C}_{\curl{S}_k}(\bar{S},\bar{S}')$ is in close agreement with $\bar{C}(S,S')$ for highly correlated shapes and is able to reliably distinguish between independent shapes. 

\begin{figure}[t]
\centering
\includegraphics[width=.85\linewidth]{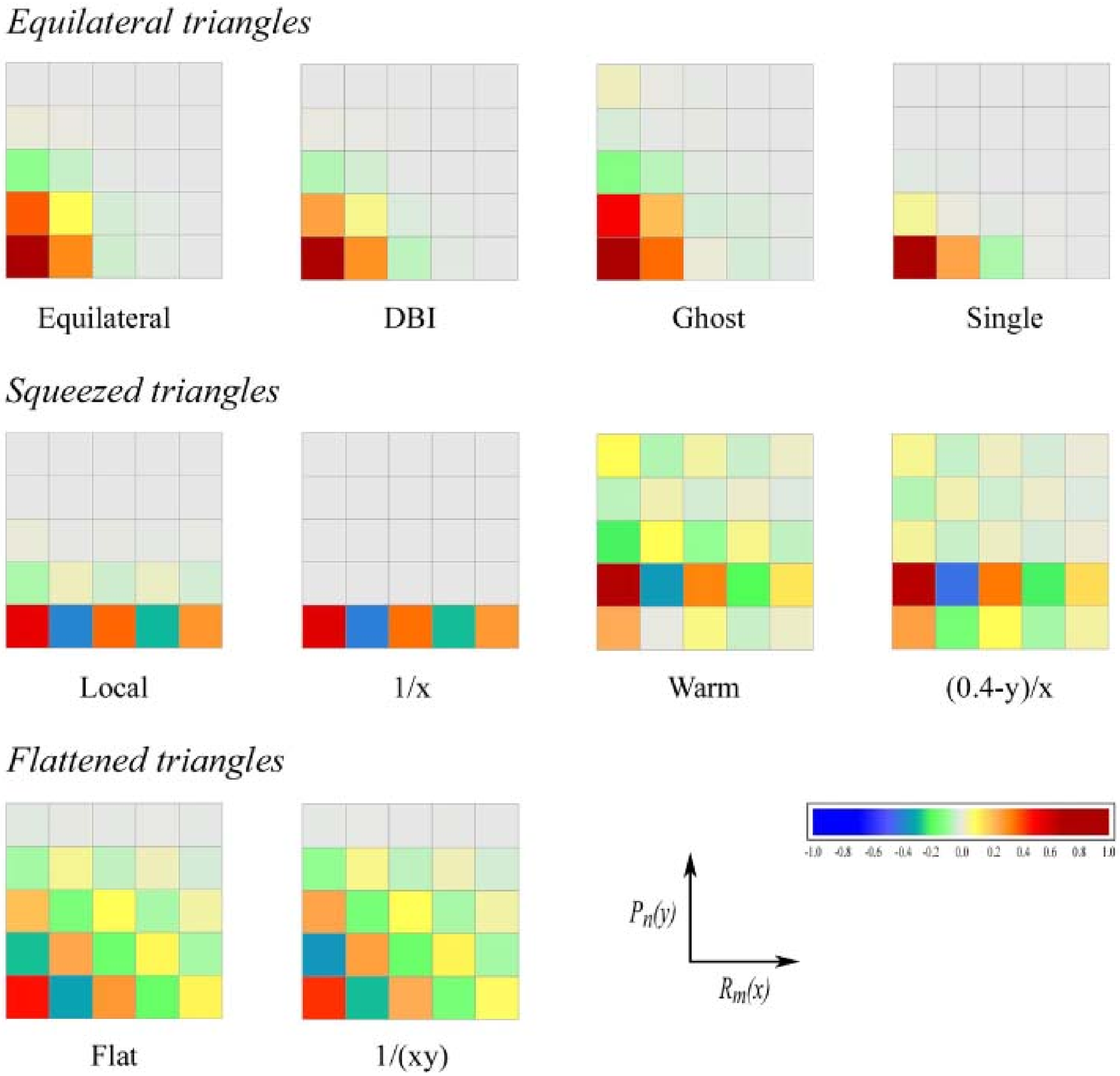}
\caption[Eigenvalues for different shape functions.]{\small Eigenmodes for different shape functions using the conventions and colour scale defined in figure~\ref{fig:eigen_warm}. Strong similarities are apparent between the equilateral family of models which are all highly correlated and would prove very difficult to distinguish observationally. The independence of the local and warm models is also apparent from the orthogonality of the dominant eigenmodes. } 
\label{fig:eigen_all}
\end{figure}

Eigenmode expansion matrices $(c_{mn})$ illustrating the two qualitatively different results for the equilateral $S_{equil}$ and local $S_{local}$ shapes are, respectively,
\begin{align} 
\left ( \begin{array}{cccccccc}
 .88 & .35 & -.10 &.01 \\
 .27 & .10 &- .04 &.01 \\
 -.02 & -.02 & .00 &.00 \\
 -.01 & -.01 & .00 &.00 \end{array} \right)\quad \hbox{and} \quad
 \left ( \begin{array}{cccc}
 .55 & -.07 & .01 &.00 \\
- .39 & .03 & .00 &.00 \\
 .34 & -.03 & .00 &.00 \\
- .28 & .02 & .00 &.00 \end{array} \right)\,.
\end{align}
The eigenmode coefficients converge rapidly for the equilateral shape (left) and it can be very well approximated by just three linear terms (with which it is 98\% correlated):
\begin{align}\label{eq:equillinear}
S_{equil} (x,y) \approx 0.88 + 0.35 R_1(x) + 0.27 P_1(y) = -0.74+ 1.65x + 1.76 y\,.
\end{align}
The local shape, on the other hand, which is divergent except for a cut-off at $k_{min}/k_{max} \approx 2/l_{max}$, oscillates as $\sim \pm 1/\sqrt{m}$ in the $R_m(x)$ modes (for $n=0$), although it converges rapidly for the higher $P_n(y)$ $(n>1)$. It is immediately apparent that the 46\% correlation between local and equilateral shapes discussed earlier arises primarily from the dominant constant term $c_{00}c'_{00}$ in the expansion (\ref{eq:slicecor}). Removing the constant mean from the local shape, then yields a small -15\% anticorrelation between the models, thus suggesting possible strategies for distinguishing them. 

In the discussion that follows we shall illustrate the eigenvalues graphically as shown in figure~\ref{fig:eigen_warm} for the warm inflation shape. Like the local model, the warm shape is corner-weighted, however, the dominant presence of the $R_m(x) P_1(y)$ modes (see also figure~\ref{fig:coordtransf}), which are orthogonal to the $R_m(x)$ modes in the local case, implies that the two shapes exhibit little correlation (only 30\%) and can be regarded as independent. Figure~\ref{fig:eigen_all} provides a summary of the largest eigenvalues for all the shapes we discuss in the next sections.

\begin{figure}[t]
\centering
\begin{tabular}{@{}c@{}c@{}}
\includegraphics[width=0.5\linewidth]{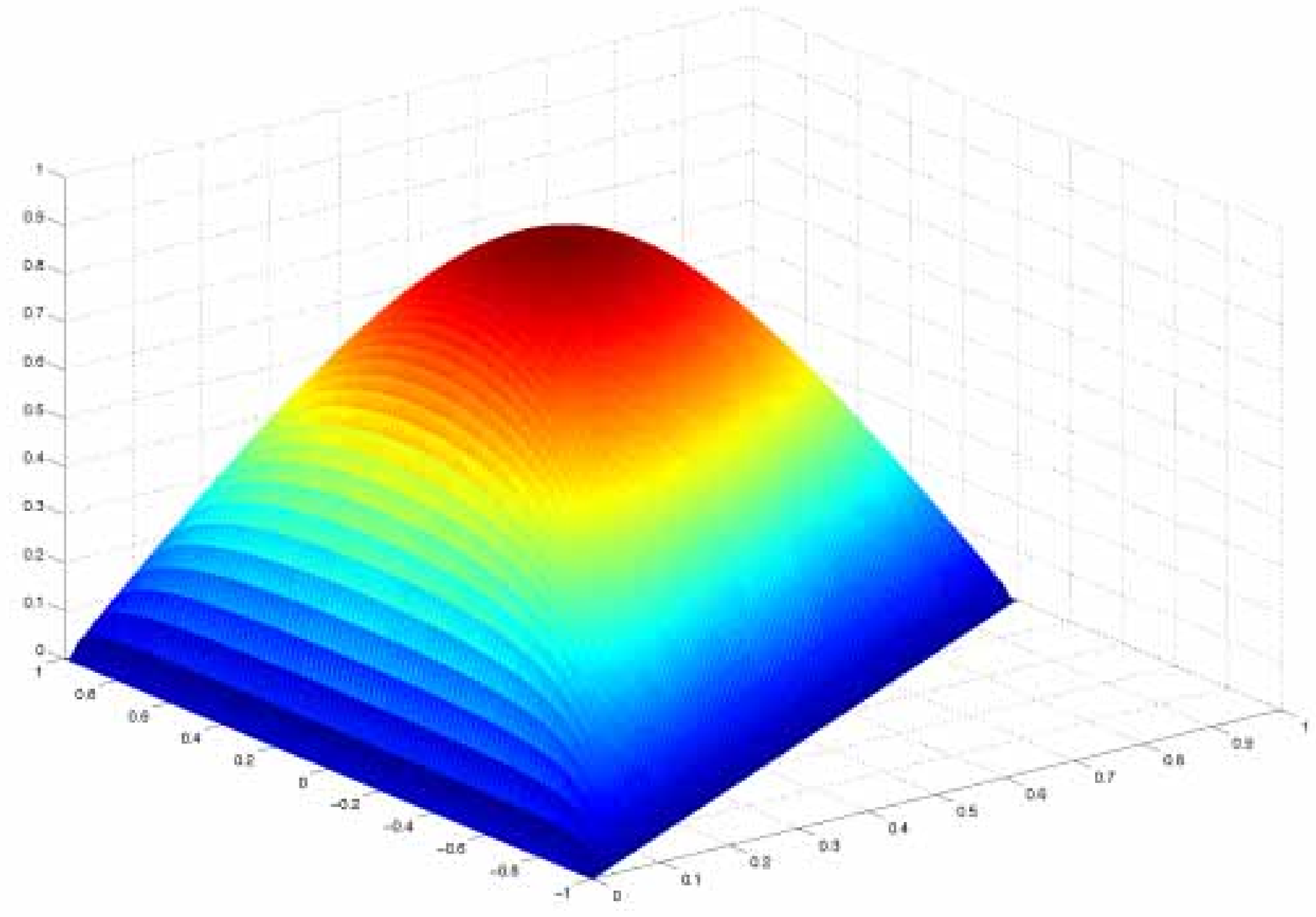} &
\includegraphics[width=0.5\linewidth]{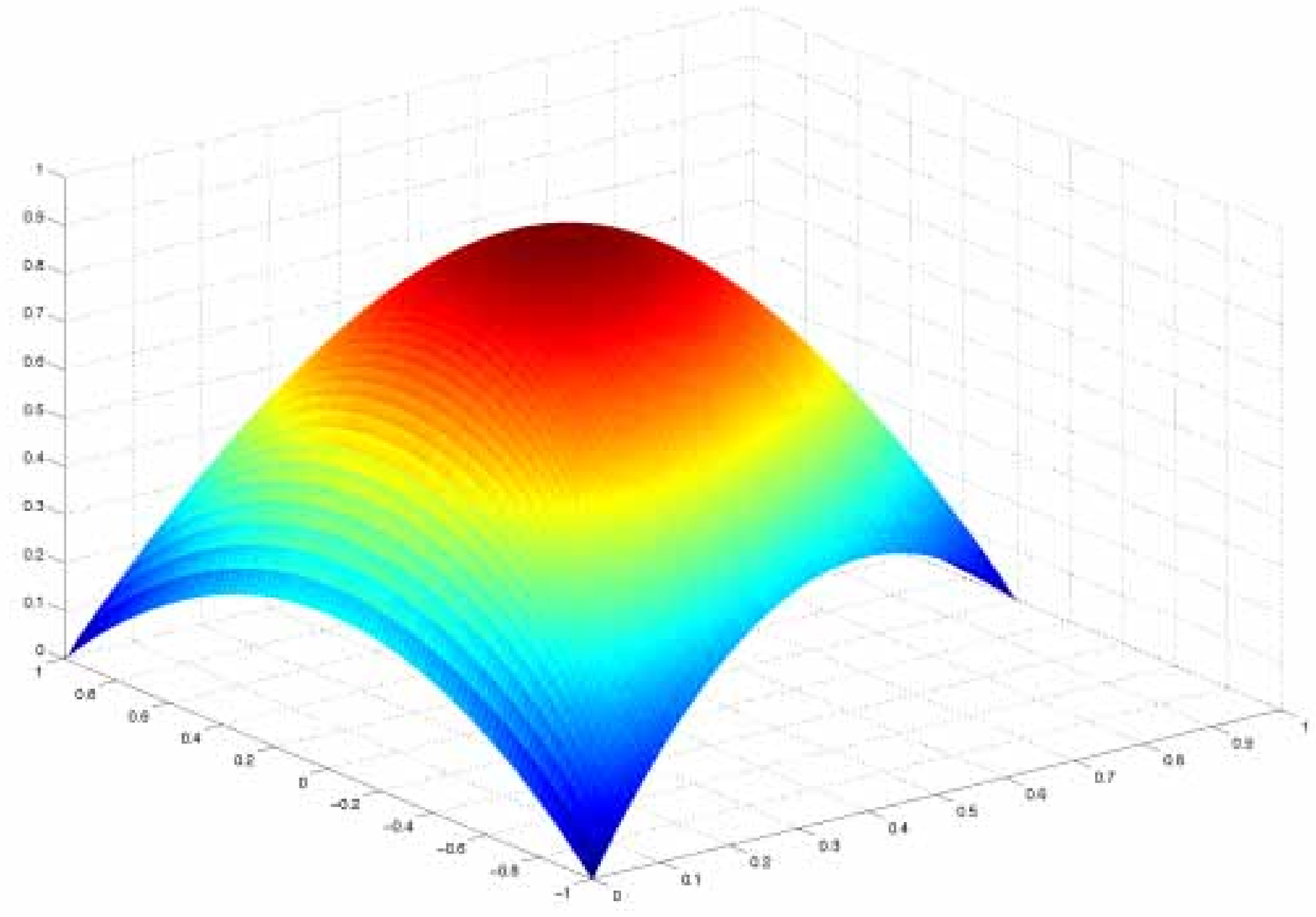} \\
\includegraphics[width=0.5\linewidth]{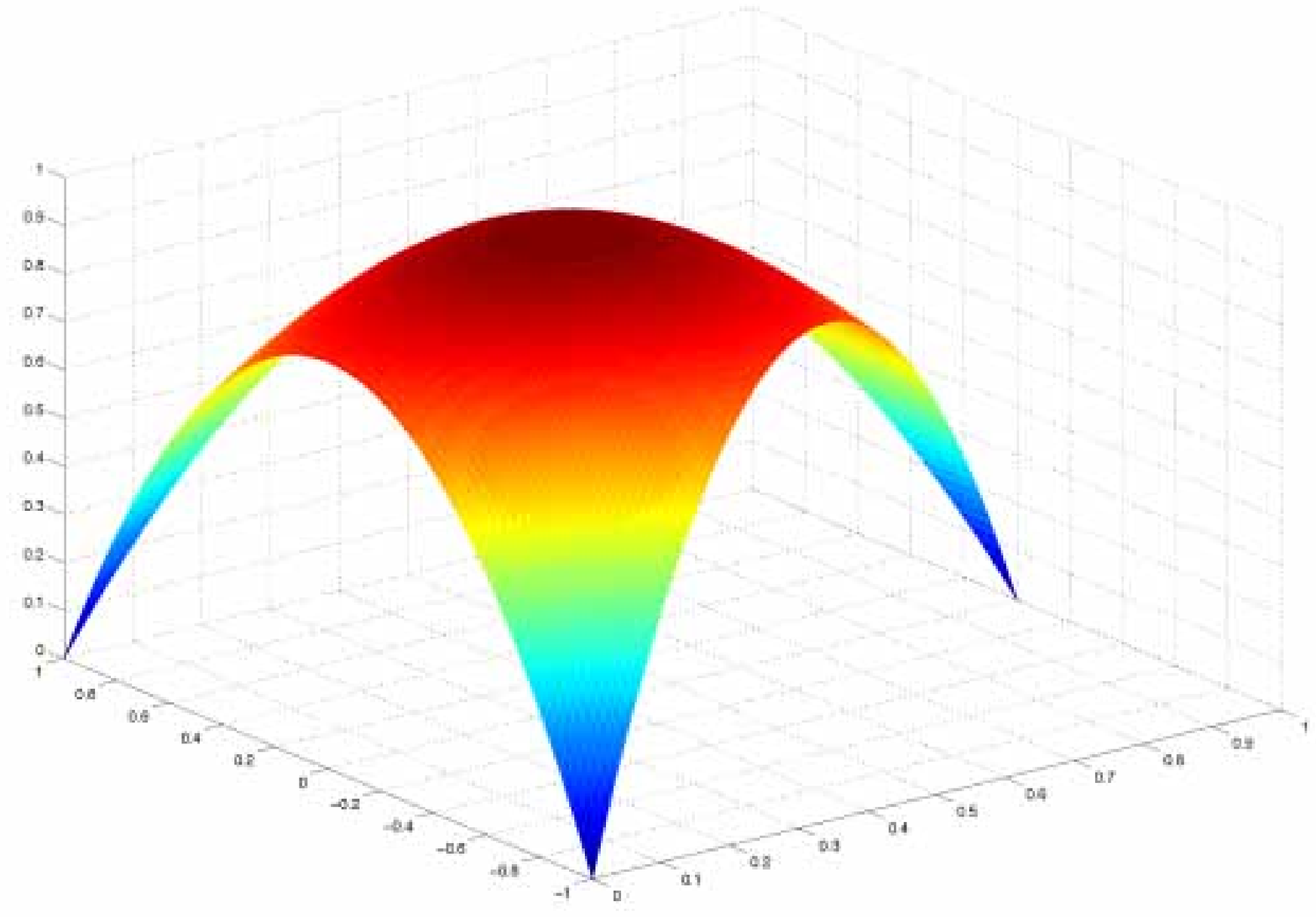} &
\includegraphics[width=0.5\linewidth]{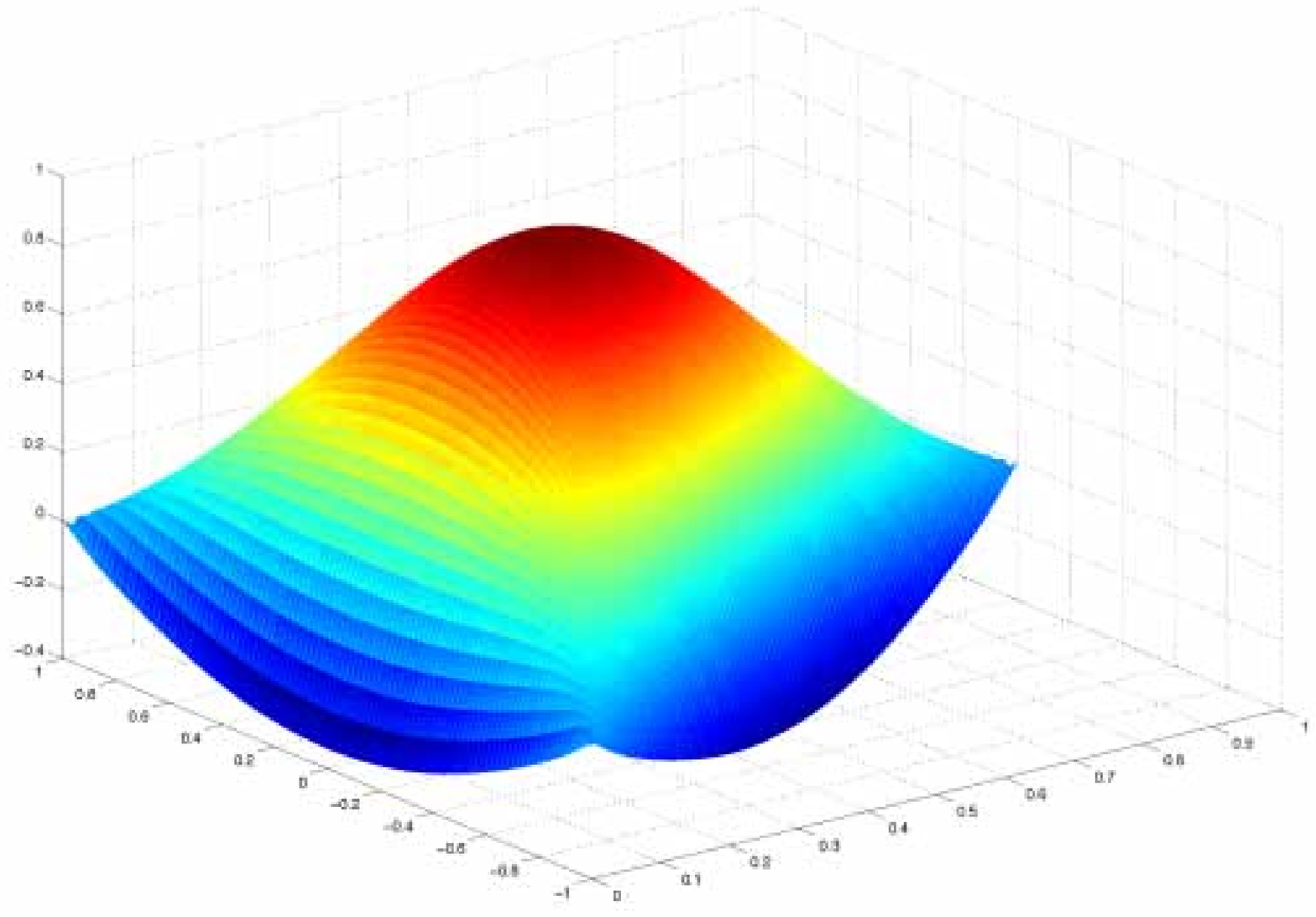}
\end{tabular}
\caption[The shape function of models in the equilateral class.]{\small The shape function of models in the equilateral class. Clockwise from top left we have the equilateral, DBI, single and ghost models. All four of these models have the majority of their signal concentrated in the equilateral limit corresponding to the centre of the triangle. Despite significant variations in the flattened limit, particularly around 
the edges of the triangle, all are strongly correlated by 96\% or greater to the equilateral model}
\label{fig:equipics}
\end{figure}

\subsection{Equilateral triangles -- centre-weighted models}

We begin this brief survey with the shape functions which are most easily characterised, those bispectra dominated by contributions from nearly equilateral triangle configurations, $k_1\approx k_2\approx k_3$. While these might be well-behaved shapes, they are not necessarily the best-motivated physically. Equilateral non-Gaussianity requires the amplification of nonlinear effects around the time modes exit the horizon, which is not possible in a slow-roll context for vanilla single field inflation. Instead, the kinetic terms in the effective action must be modified as in the Dirac-Born-Infeld (DBI) model \cite{0404084} or by explicitly adding higher derivative terms, such as in K-inflation (see, for example, ref.~\cite{0605045}). The resulting corrections modify the sound speed $c_s$, acting to slow the scalar field motion and, when the field theory is coupled to gravity, inflation is able to take place in steep potentials. For DBI inflation, this leads to non-Gaussianity being produced with a shape function of the form \cite{0306122, 0404084}
\begin{align} \label{eq:dbi}
S^{DBI}(k_1,k_2,k_3) \propto \frac{1}{K_{111}K^2} \(K_5 + 2 K_{14} - 3 K_{23} + 2 K_{113} - 8 K_{122}\)\,,
\end{align}
where we have used the compact notation of equation (\ref{eq:compactnotn}). Another example of a model with non-standard kinetic terms is ghost inflation \cite{0312100}. Here, a derivatively-coupled ghost scalar field $\o$ is responsible for driving inflation. When $\o<0$ the potential can be thought of as flat, but $\dot{\o} \neq 0$ and so the field continuously evolves towards $\o=0$, where inflation ends. In this model the dominant effect for the perturbations comes from the trilinear term in the Lagrangian which naturally leads to a non-zero bispectrum. The shape function for this model is of the form,
\begin{align} \label{eq:ghost}
S^{ghost}(k_1,k_2,k_3) \propto 2 \frac{1}{K_{111}} \mbox{Re}\[ \int^0_{-\infty} \frac{d\n}{\n} F^*(\n) F^*\(\frac{k_2}{k_1}\n\) F^{\pr*}\(\frac{k_3}{k_1}\n\)k_3 \tilde{k}_{32} \]\,,
\end{align}
plus two permutations, where,
\begin{align}
F(\n) = \sqrt{\frac{\pi}{8}}(-\n)^{\frac{3}{2}}H^{(1)}_{\frac{3}{4}}\(\frac{\n^2}{2}\)\,.
\end{align}

General non-Gaussian shapes arising from modifications to single field inflation have been extensively reviewed in ref.~\cite{0605045}. Using a Lagrangian that was an arbitrary function of the field and its first derivative, they were able to identify six distinct shapes describing the possible non-Gaussian contributions. Half of these had negligible amplitude being of the order of slow-roll parameters (two already given in (\ref{eq:Maldshape})). Of the remaining three shapes \cite{0605045} (see also
\cite{0503692}), one was believed to be subdominant, the second recovered the DBI shape (\ref{eq:dbi}), leaving a third distinct single field shape of the form,
\begin{align}\label{eq:single}
S^{single}(k_1,k_2,k_3) \propto \frac{K_{111}}{K^3}\,.
\end{align}
The sub-dominant term is a complex combination of special functions (somewhat like ghost inflation (\ref{eq:ghost})) with indeterminate parameter values; we will neglect it in the subsequent discussion. Finally, we recall the original equilateral shape (\ref{eq:equi}), noting that it was introduced not because of a fundamental physical motivation, but as a separable approximation to the DBI shape (\ref{eq:dbi}) \cite{0405356}.

Using the shape correlator (\ref{eq:shapecor}) and the shape decomposition (\ref{eq:shapedec}) introduced above, we can make a preliminary comparison of the four equilateral shapes---DBI, ghost, single and equilateral---which are illustrated on a $k=\hbox{const.}$ slice in figure~\ref{fig:equipics}. The results of the shape correlators are given in the table~\ref{tb:shapecorrelator}. Despite the apparent visual differences between these shapes, particularly near the edges of the triangular domain, there is 
at least a 96\% or greater correlation of each to the equilateral shape (\ref{eq:equi}). These particular centre-weighted shapes must then be regarded as a single class which would-be extremely difficult to distinguish observationally (see later for the CMB bispectrum discussion). 

The underlying similarity of the shapes is evident from the magnitudes of the eigenvalues illustrated in figure~\ref{fig:eigen_all}. Each shape has a dominant constant term $c_{00}$ and can be very well-approximated by a linear polynomial $ S(x,y) \approx c_{00} + c_{10}R_1(x) + c_{01}P_1(y)$ on the relevant subdomain (within a 96\% correlation as previously (\ref{eq:equillinear}) for the equilateral shape). The greatest difference occurs between the single and ghost shapes, with the former completely dominated by the $c_{00}$ and the latter having significant $c_{01}$ and $c_{11}$ eigenvalues. Independent equilateral shapes are possible, in principle, but they must go further than the ghost shape in suppressing the constant term $c_{00}$ in favour of eigenmodes with greater internal structure ($n,m\ge1$).

\begin{figure}[t]
\centering
\begin{tabular}{@{}c@{}c@{}}
\includegraphics[width=0.5\linewidth]{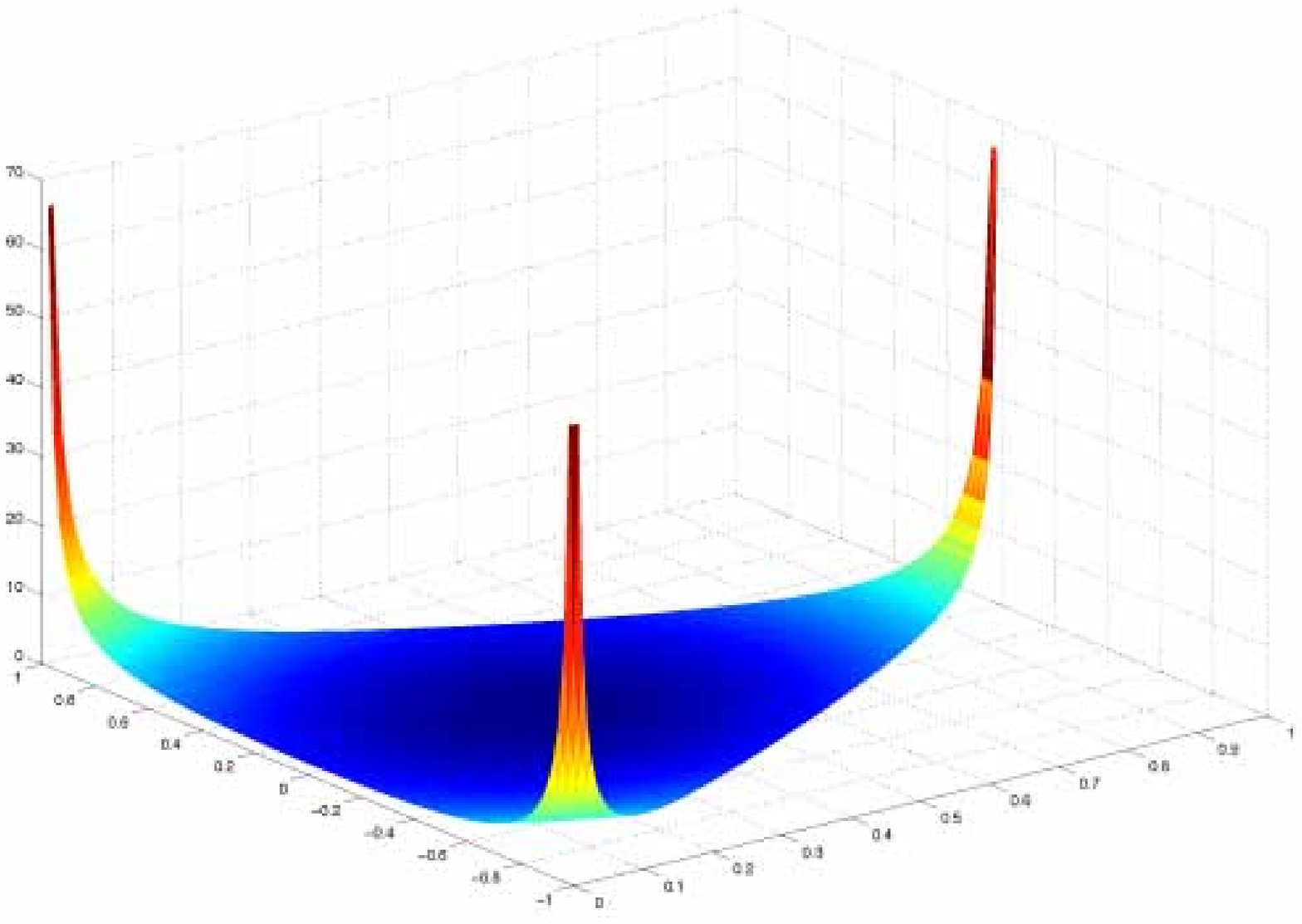} &
\includegraphics[width=0.5\linewidth]{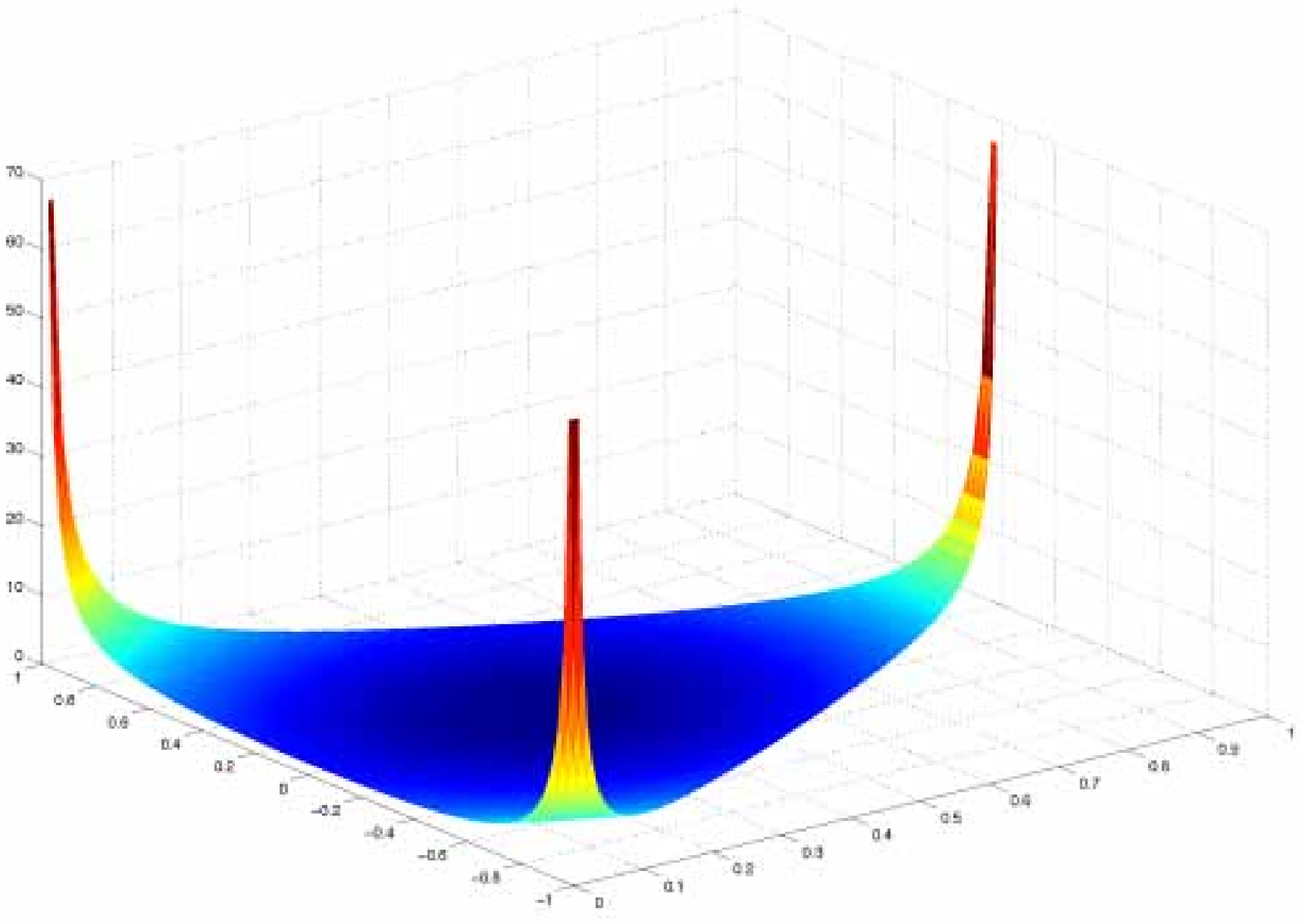}
\end{tabular}
\caption[The shape function of models in the local class.]{\small The shape function of models in the local class. On the left is the usual local model, while on the right is the non-local model, which is virtually identical. These two models are very highly correlated and so fall into the same class}
\label{fig:localslice}
\end{figure}

\subsection{Squeezed triangles -- corner-weighted models}

The local shape covers a wide range of models where the non-Gaussianity is produced by local interactions. These models have their peak signal in ``squeezed" states where one $k_i$ is much smaller than the other two, this is because non-Gaussianity is typically produced on superhorizon scales. The simplest case is that of single-field slow-roll inflation (\ref{eq:Maldshape}) \cite{0406398}, which as we have seen is dominated by the local shape. The non-linearities produced are tiny and $f^{local}_{NL}$ is constrained to be of order slow-roll parameters \cite{0210603,0406398,811001,820101}. The production of non-Gaussianity during multiple field inflation \cite{0207295,0209330,0504045,0504508,0506056,0511041} shows much greater promise (see, for example, recent work in refs.~\cite{08071101, 08070180} and references therein). Here non-Gaussianity is created by the inflaton when it follows a curved trajectory in phase space, during which isocurvature perturbations are converted into adiabatic perturbations \cite{0506704, 0603799}. The magnitude of the non-Gaussianity generated is normally around $f^{local}_{NL} \px O(1)$, which is at the limit for Planck detection. Significant $f^{local}_{NL}$ can be produced in curvaton models \cite{0511736,0208055,0309033} where the adiabatic density perturbation is generated after inflation by an initially isocurvature perturbation in a light scalar field, different from the inflaton. The non-Gaussianity generated in this scenario can be as large as, $f^{local}_{NL} \px O(100)$. Large $f^{local}_{NL}$ can be generated at the end of inflation from massless preheating or other reheating mechanisms \cite{0411394,0501076,08054795,0601481,0611750}. After slow-roll inflation ends, the inflaton oscillates about its minimum and decays. Preheating occurs when a light field oscillates in resonance with it, taking energy from the inflaton, so its amplitude grows. The amplitude of the resonant field eventually becomes so large that its dynamics become non-linear and this non-linearity is transferred to the density perturbations. It is claimed this process can generate enormous non-Gaussianity, $f^{local}_{NL} \px O(1000)$, which is already tightly constrained by observation. 

The local shape is strongly motivated because it appears in models that use standard kinetic terms in the action, smooth potentials without exotic couplings and which assume the standard Bunch-Davies vacuum. We note, however, that it also occurs in other contexts. Significant local non-Gaussianity can appear in models based on non-local field theory, such as $p$-adic inflation \cite{08023218}. In these models slow roll inflation is again able to occur in very steep potentials. Like single field slow-roll inflation, the predicted non-local shape function is a combination of local (\ref{eq:local}) and equilateral-like (\ref{eq:equi}) shape functions (see also refs~\cite{08020588,0306006,9208001} for its origin).  However, the combination is even more heavily weighted than (\ref{eq:Maldshape}) towards the local shape (with the relative ratio given roughly by the number of e-foldings). Consequently, the non-local shape is (paradoxically) completely indistinguishable from the local shape (and is subsumed in this class henceforth). A comparison of the two shapes can be seen in figure (\ref{fig:localslice}). The ekpyrotic model can also generate significant $f^{local}_{NL}$ \cite{0702165,07084321,07105172, 07123779,08041293}. Here the density perturbations are generated by a scalar field rolling in a negative exponential potential, so non-linear interactions are important, and large local non-Gaussianity can be produced, $f^{local}_{NL} \px O(100)$.

\begin{figure}[t]
\centering
\begin{tabular}{@{}c@{}c@{}}
\includegraphics[width=0.5\linewidth]{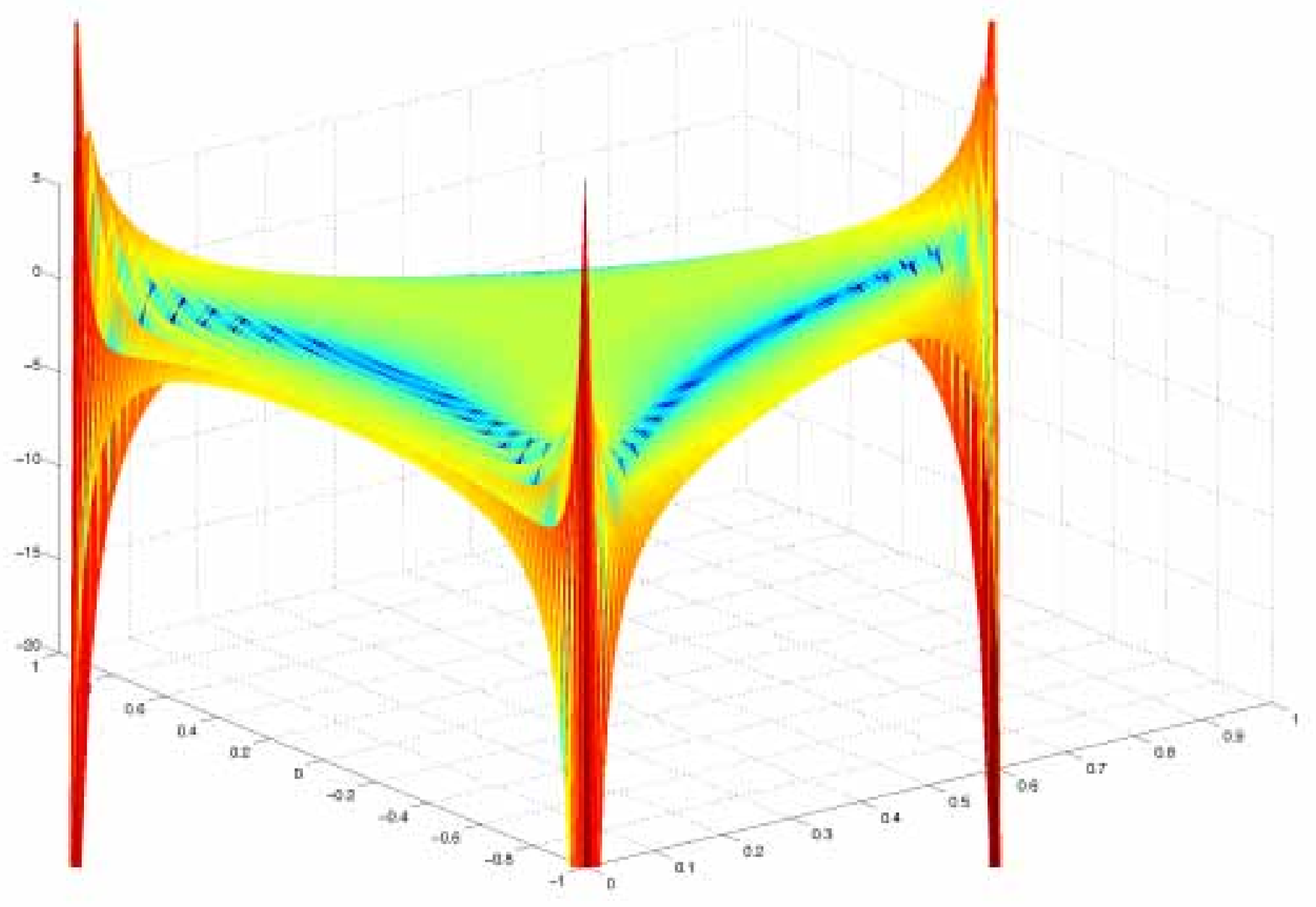} &
\includegraphics[width=0.5\linewidth]{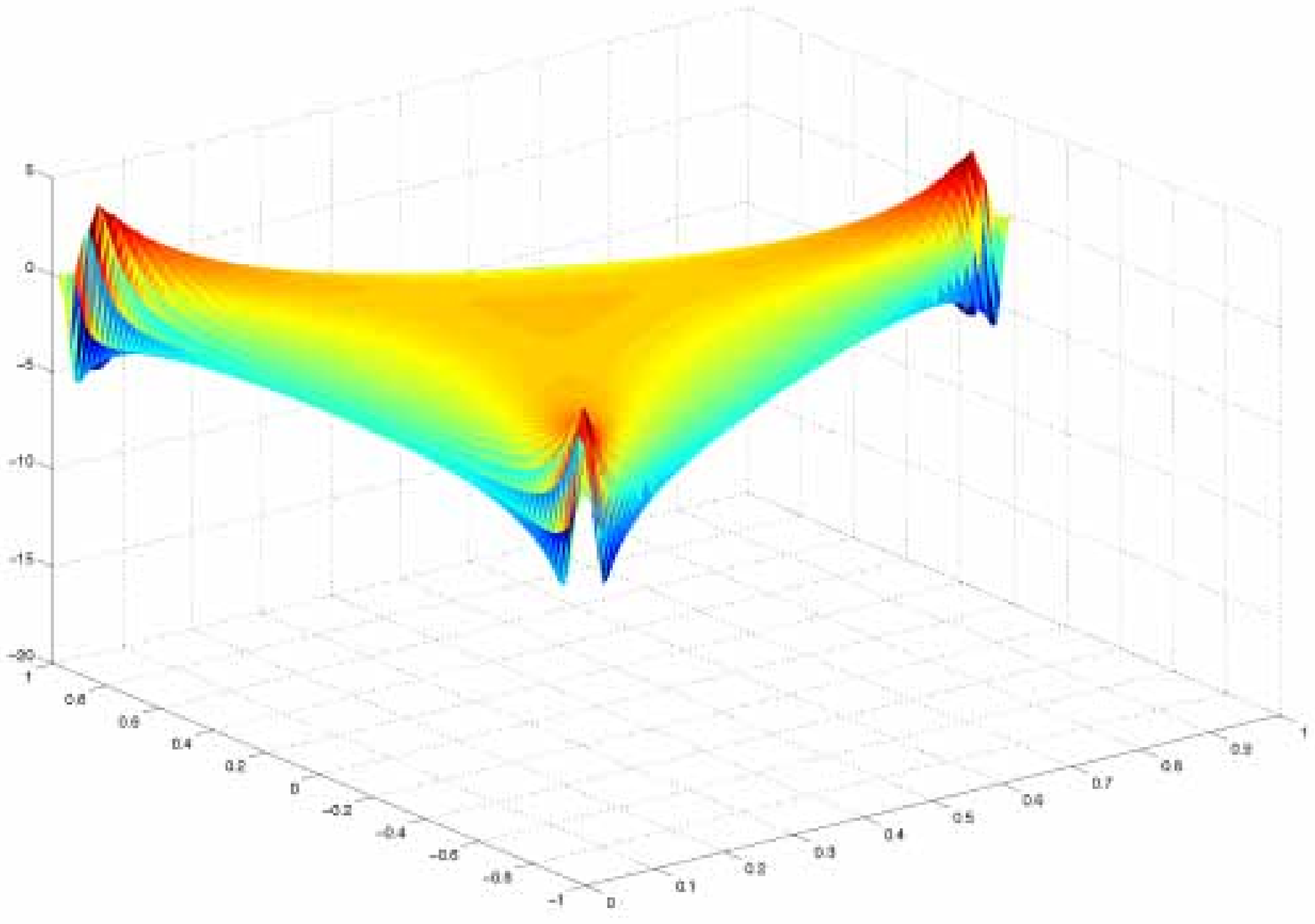}
\end{tabular}
\caption[The shape function of warm inflation before and after smoothing.]{\small The shape function of warm inflation before (left) and after smoothing (right).
Note the pathological behaviour of the squeezed states in the corners for the unsmoothed model. The derivation for this shape breaks down when $\frac{k1}{k2} < \sqrt{\frac{H}{\G}}$ as $k1 \rightarrow 0$, where $\G$ is the friction coefficent due to dissipative effects. The smoothed model presents a more reasonable profile, but results are cut-off dependent.}
\label{fig:warmprim}
\end{figure}

Finally, we note that warm inflation scenarios, i.e.\ models in which dissipative effects play a dynamical role, are also predicted to produce significant non-Gaussianity \cite{0205152, 9312033, 0701302}. Contributions are again dominated by squeezed configurations but with a different more complex shape,
\begin{align} \label{eq:warm}
S^{warm}(k_1,k_2,k_3) \propto \frac{1}{K_{333}}(K_{45} - K_{27} + 2K_{225})\,.
\end{align}
As we can see from figure~\ref{fig:coordtransf}, the squeezed limit contains an orthogonal sign change as the squeezed limit is approached $k_3\rightarrow 0$.

It is immediately apparent that we need to introduce a cut-off in order to normalize the squeezed shape functions for the correlator (\ref{eq:shapecor}). This logarithmic divergence does not afflict the CMB bispectrum $B_{l_1l_2l_3}$ because we do not consider contributions below the quadrupole $l=2$. Given the cut-off at large wavenumbers where $k_{max}$ is related through a flat sky approximation to the largest multipole $l_{max}$, we can similarly define $k_{min} \approx (2/l_{max})\, k_{max}$. There is only a weak dependence on the precise value of $k_{min}$. We note that a more serious concern is the apparent breakdown of the approximations used to calculate the warm inflation shape near the corners. In the absence of a specific prescription for this asymptotic regime, we have to explore the dependence of an edge cut-off and/or smoothing. We remove the divergence in the squeezed limit, $k_1\rightarrow0$, by truncating the shape function when $k_1/(k_2 + k_3) < 0.015$. We smooth the resulting discontinuity by applying a Gaussian window function on the cross sectional slices with a FWHM of $0.03/(k_1+k_2+k_3)$. The result of this applying this process to the shape function can be seen in figure (\ref{fig:warmprim}), with the model denoted warmS.

We note that the local and warm shape functions are only correlated at the 33\% level, despite the primary contribution coming from squeezed states in both cases. As discussed previously, this is because their dominant eigenvalues $c_{m0}$ (local) and $c_{m1}$ (warm) correspond to orthogonal eigenmodes. 

The local shape is modestly correlated at the 40-55\% level with the equilateral shapes (qualitatively in 
agreement with \cite{0405356}, though not quantitatively presumably because of the different weighting). In contrast, however, the local contribution from the constant term $c_{00}\approx 0.55$ is relatively small. Thus removing the $c_{00}$ term from the local estimator eliminates most of the equilateral correlations while leaving 70\% of the local signal (i.e. in the autocorrelator $\bar C_{\curl{S}_k}(S,S)$). Thus, we propose subtraction of the constant term as a significant test of the local model.

\begin{figure}[t]
\centering
\begin{tabular}{@{}c@{}c@{}}
\includegraphics[width=0.5\linewidth]{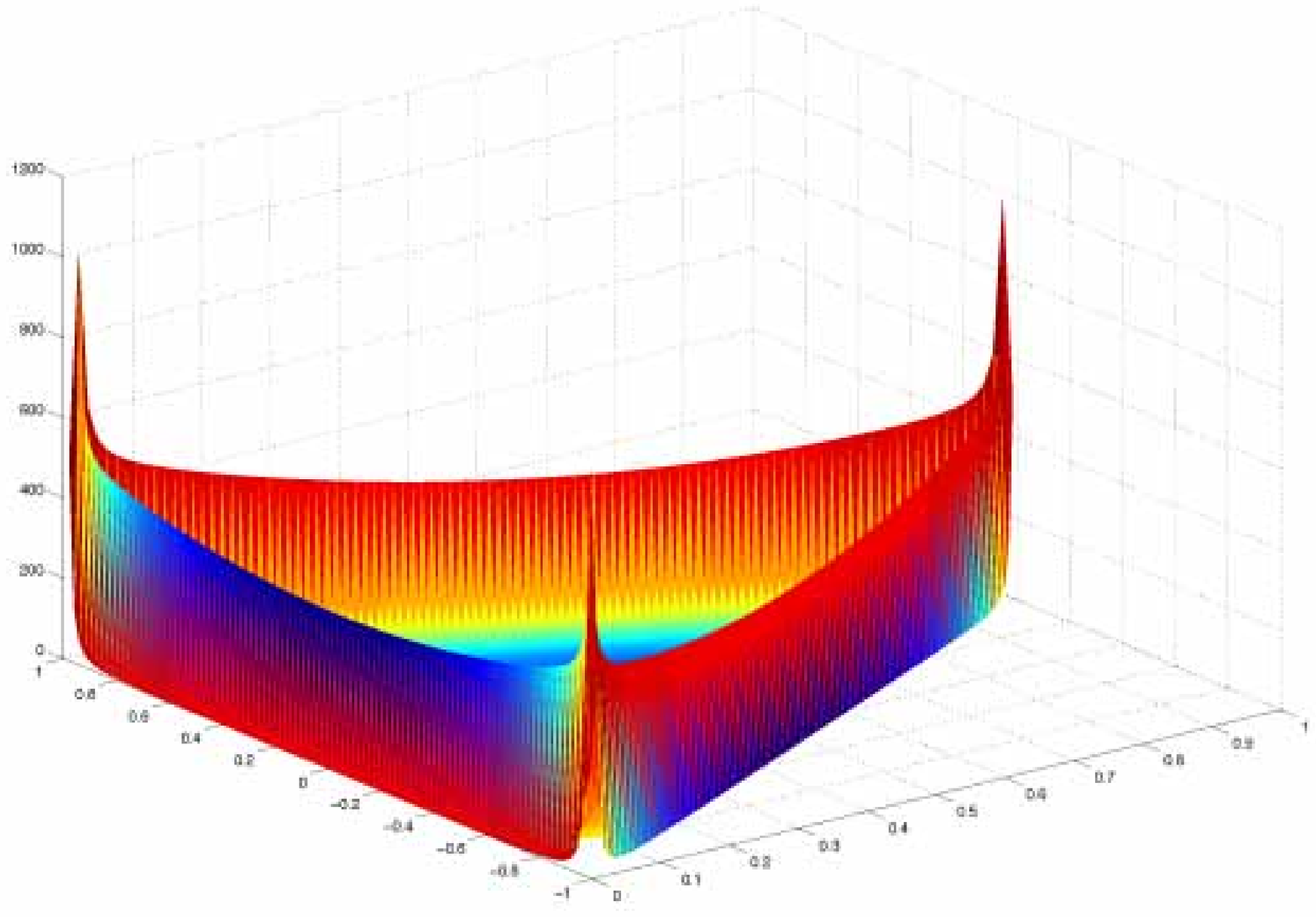} &
\includegraphics[width=0.5\linewidth]{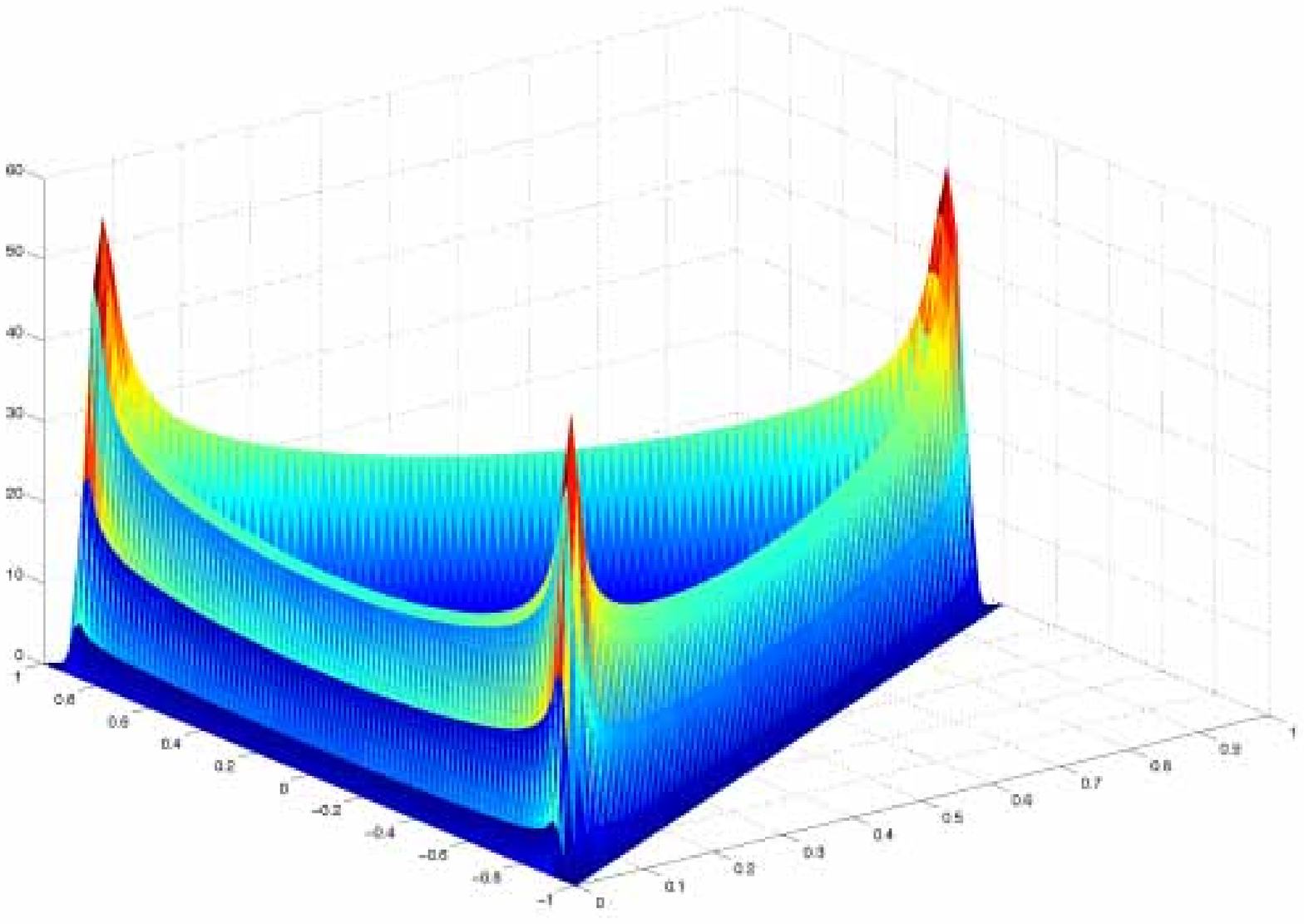}
\end{tabular}
\caption[The shape function of the flattened model before and after smoothing.]{\small The shape function of the flattened model before (left) and after smoothing (right). The approximation diverges in the flattened limit, whereas it should oscillate and decay to zero. Here, we set the boundary to zero then apply a Gaussian smoothing to obtain a more reasonable profile.}
\label{fig:flatprim}
\end{figure}

\subsection{Flattened triangles -- edge-weighted models}

It is possible to consider inflationary vacuum states which are more general than the Bunch-Davies vacuum, such as an excited Gaussian (and Hadamard) state \cite{07101302}. Observations of non-Gaussianity in this case might provide insight into trans-Planckian physics. The bispectrum contribution from early times is strongest for flattened triangles with, for example, $ k_3 \approx k_1+k_2$. In the small sound speed limit $c_s\ll 1$, the primordial bispectrum could be significant with a shape given by \cite{0605045}
\begin{align}\label{eq:flat}
S^{flat}(k_1,k_2,k_3) \propto \frac{1}{K_{111}} \(K_{12} - K_3\) + 4\frac{K_2}{\tilde{k}^2_1\tilde{k}^2_2\tilde{k}^2_3}\,.
\end{align}
Unfortunately, as this analytic approximation diverges on the entire boundary of the allowed region, any integrals over the bispectrum are unbounded. In principle, however, this divergence should be cut-off near the edges, though the nature of the asymptotic behaviour which replaces it is poorly understood. In order to obtain even a qualitative picture for the flat shape function we must truncate it in some way. We follow the same procedure as for the warm model, removing the section $\tilde{k}_i / K < 0.03$, then applying the same Gaussian filter to remove the discontinuity. We refer to this shape as $S^{flat S}$. Plots of the flattened model before and after smoothing can be seen in figure (\ref{fig:flatprim}).

Reflecting its distinctive properties, the flat shape is poorly correlated with most of the other shapes, with a particularly striking absence of any correlation with the orthogonal warm shape. Having a dual divergence $(xy)^{-1}$ means that the eigenvalues are spread more thinly and widely than the corner-weighted models with a smaller constant term. Nevertheless, the flat shape would be most susceptible to confusion with the local shape with which it has a 62\% correlation.

\subsection{Features -- scale-dependent models}

\begin{figure}[ht]
\centering
\begin{tabular}{@{}c@{}c@{}}
\includegraphics[width=0.5\linewidth]{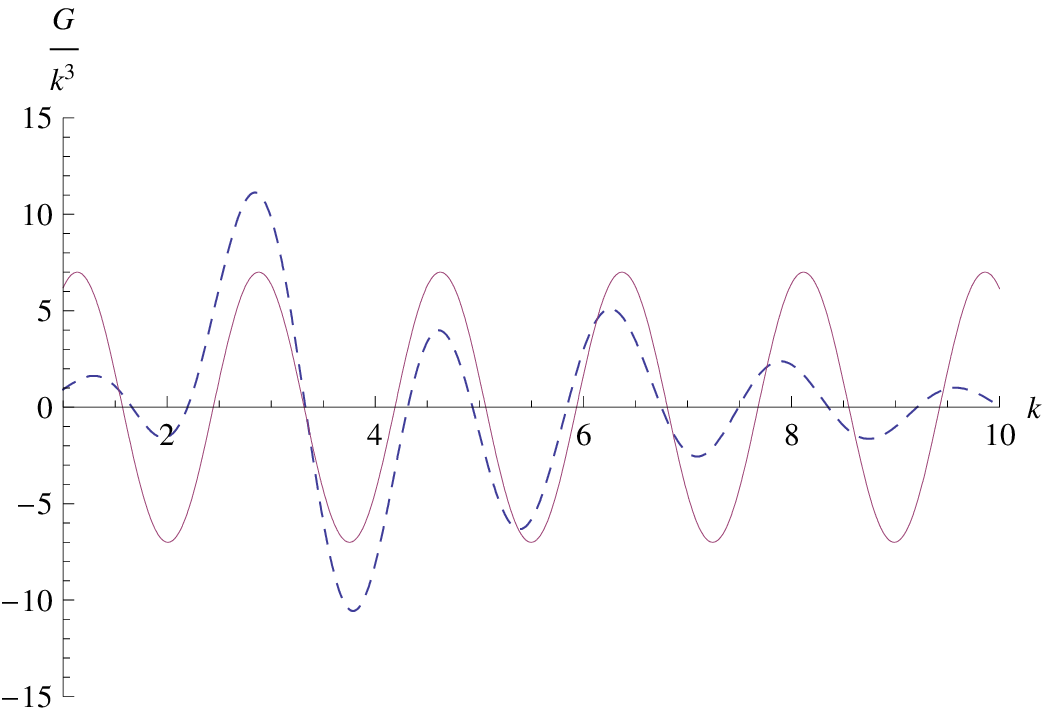} &
\includegraphics[width=0.5\linewidth]{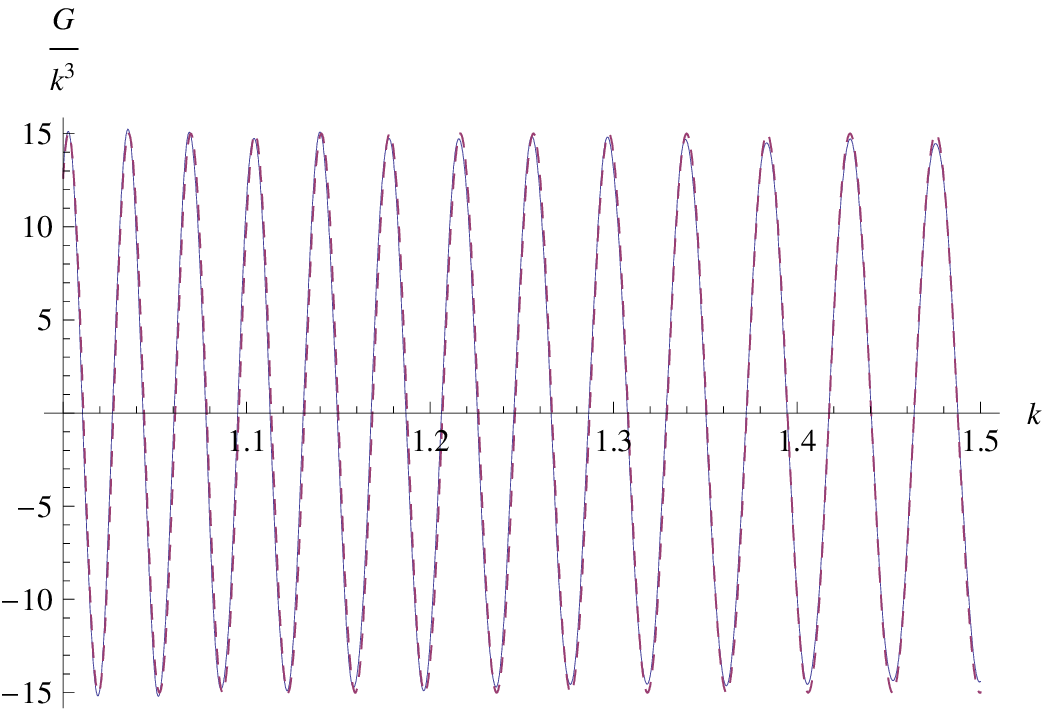}
\end{tabular}
\caption[The scaling of the shape functions of models with features in the potentials.]{\small The scaling of the shape functions of models with features in the potentials. These are plots of the scaling of the central value $S(k,k,k)$ for a model with a single feature on the left, and a potential with an oscillatory component on the right. The blue dashed line is the correct numerical result, the red solid line is the simple approximation quoted earlier (these plots approximate those in \cite{08013295}).}
\label{fig:featprim}
\end{figure}

Finally, there are models in which the inflation potential has a feature, providing a break from scale-invariance and introducing large scale power where it is deemed to be indicated by observation. This can take the form of a either a step \cite{0611645} or a small oscillation superimposed onto the potential \cite{08020491}. Analytic forms for both these three point functions have been produced in \cite{08013295}. However, these approximations are both somewhat simplistic and so are unsuitable for a detailed analysis, other than as preliminary check of their correlation to other models. The two approximations are of the form,
\begin{align}\label{eq:feature}
S^{feat}(k_1,k_2,k_3) &\propto \sin(\frac{K}{k^*} + P)\,, \\
S^{osci}(k_1,k_2,k_3) &\propto \sin\( C \ln(K) + P \)\,, \label{eq:osci}
\end{align}
where $k^*$ is the associated scale of the feature in question, $C$ is a constant and $P$ is a phase factor. The correspondence of these analytic approximations to the full shape function can be seen in figure (\ref{fig:featprim}) 

Results for the shape correlator for a particular feature model (with $k^* \approx l^*/\tau_0$ and $l^*=50$), are given in table (\ref{tb:shapecorrelator}). It can be seen to be essentially independent of all the other shapes. Obviously this is because all variations in feature model occur in the $K$-direction which is orthogonal to the $(\alpha, \,\beta)$-slice -- it only shares the constant term in common.

We conclude, from this brief survey of the literature, that we can identify the feature model in a fifth distinct category beyond the equilateral, local, warm and flat shapes. We shall now turn to the much more formidable task of calculating the CMB correlators directly in order to determine the accuracy of our shape correlator analysis.

\section{CMB bispectrum calculation methodology}

\subsection{Numerical approach}

It is not feasible to directly evaluate the bispectrum for a completely general model. However, provided the shape function obeys a mild separability ansatz then the reduced bispectrum integral can be re-written in a tractable form.  The method is based on the splitting of the shape function (\ref{eq:shapefn}) into scale and scale-free parts (\ref{eq:shapesplit2}),
$
S(k_1,k_2,k_3) = f(k) \bar{S}(\a,\b)\,,
$
as discussed in the previous section, that is, an ansatz which applies to all the models under discussion. 
By using this decomposition with the reparameterisation into rescaled wavenumbers $\hat k_1,\, \hat k_2,\, \hat k_3$ from (\ref{eq:hatparam}), 
we can rewrite the integral for the reduced bispectrum (\ref{eq:biint}) in a simple form
\begin{align} \label{eq:biint2}
\nn b_{l_1 l_2 l_3} = f_{NL} \(\frac{2}{\pi}\)^3 \int_{\curl{V}_k} & dk d\curl{S}_k\, k^2 f(k) \bar{S}(\hat{k}_1,\hat{k}_2,\hat{k}_3) \D_{l_1}(k\hat{k}_1) \D_{l_2}(k\hat{k}_2) \D_{l_3}(k\hat{k}_3)\\
\nn &\times I^G_{l_1 l_2 l_3}(k\hat{k}_1,k\hat{k}_2,k\hat{k}_3) \\
= f_{NL} \(\frac{2}{\pi}\)^3 \int_{\curl{S}_k} & d\curl{S}_k \bar{S}(\a,\b) I^T_{l_1 l_2 l_3}(\a,\b) I^G_{l_1 l_2 l_3}(\a,\b)\,,
\end{align}
where $\curl{S}_k$ is the cross-section spanned by $\a$ and $\b$ from (\ref{eq:parameters}) and $d\curl{S}_k = d\a d\b$. Here we have made the definitions,
\begin{align}
\label{eq:transferint} I^T_{l_1 l_2 l_3}(\hat{k}_1,\hat{k}_2,\hat{k}_3) &= \int^\infty_0 dx\, \frac{f(x)}{x} \D_{l_1}(x\hat{k}_1) \D_{l_2}(x\hat{k}_2) \D_{l_3}(x\hat{k}_3)\,, \\
\label{eq:geometricint} I^G_{l_1 l_2 l_3}(\hat{k}_1,\hat{k}_2,\hat{k}_3) &= \int^\infty_0 dx\, x^2 j_{l_1}(x\hat{k}_1) j_{l_2}(x\hat{k}_2) j_{l_3}(x\hat{k}_3)\,.
\end{align}
We refer to eqn (\ref{eq:transferint}) as the transfer integral and eqn (\ref{eq:geometricint}) as the geometric integral. With this decomposition we have reduced the number of dimensions in the integral from four to three and we have also trapped all the highly oscillatory behaviour into the two one-dimensional integrals, $I^T$ and $I^G$. Having achieved this, the remaining two-dimensional integral over $\curl{S}_k$ has very mild oscillatory behaviour and only requires a similar number of points as the one-dimensional integrals to evaluate accurately.

\begin{figure}[ht]
\centering
\includegraphics[width=0.75\linewidth]{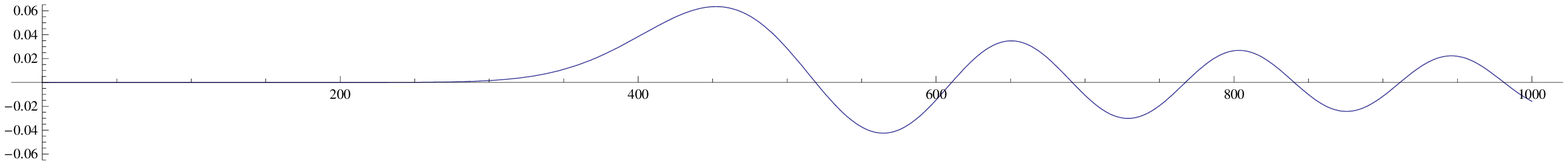}\\
\includegraphics[width=0.75\linewidth]{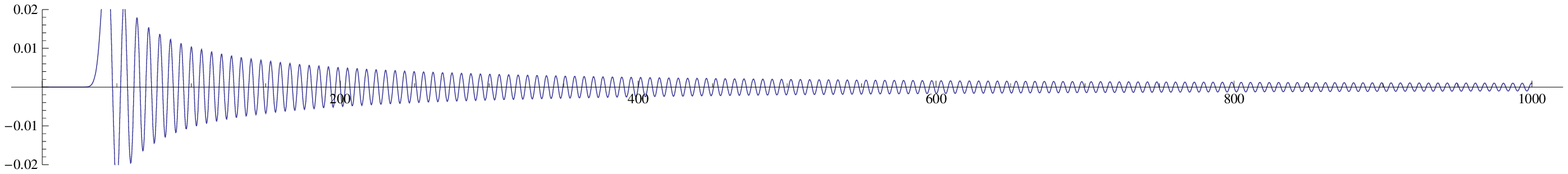}\\
\includegraphics[width=0.75\linewidth]{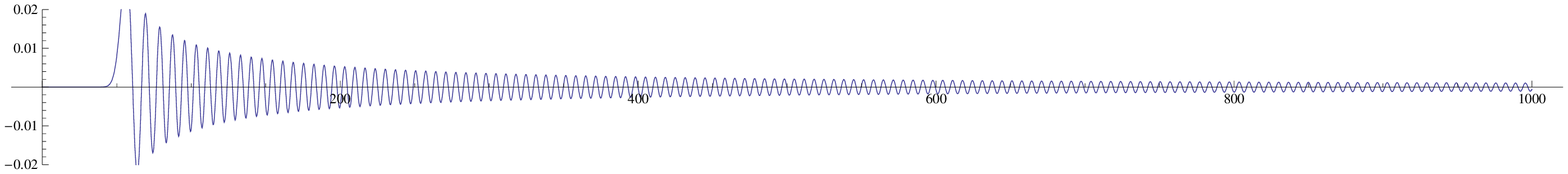}\\
\includegraphics[width=0.75\linewidth]{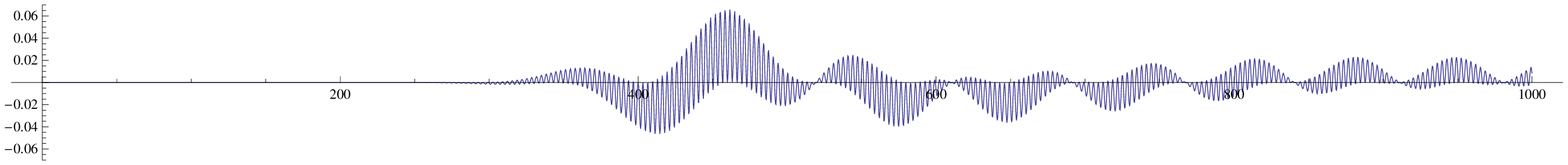}
\caption[A plot of the three spherical Bessel functions and the resulting integrand $I^G$.]{\small A plot of the three spherical Bessel functions and the resulting integrand $I^G$. The top three plots are of $j_l(a x)$ for different values of $l$ and $a$, with (20,0.05) (top),  (40,1.0) (second) and (50,0.95) (third). 
In the bottom diagram, the resulting integrand for $I^G$ is plotted. From this we can see that the peak sections of the two middle plots are ignored and it is their tails that are picked out in the integrand. Thus, for accurate evaluation we must calculate the transfer functions and Bessel functions over a much larger range than is required to evaluate the power spectrum.}
\label{fig:bessel}
\end{figure}

 We will now detail the numerical methods used to evaluate both the one-dimensional transfer and geometric integrals and the two-dimensional integral over the triangular domain $\curl{S}_k$.  We use a modified form of the CMB bispectrum code already presented in \cite{0612713}, so here we will focus
on substantial recent improvements.

To evaluate the bispectrum using this  formalism we must first be able to compute the two 1D integrals, (\ref{eq:transferint}) and (\ref{eq:geometricint}), for every combination of the three $\hat{k}_i$ possible in the triangular domain $\curl{S}_k$. The transfer integral converges quickly, as $1/k^4$, so while it is highly oscillatory it does not pose an enormous challenge. Also the transfer functions truncate at large values of $k$ due to photon diffusion and so the integral naturally terminates. The geometric integral only converges slowly, as $1/k$, and so constitutes the majority of time in calculating the integrand for each point in the triangular domain $\curl{S}_k$.

To evaluate these two one-dimensional integrals we first need to obtain the transfer functions and Bessel functions that make up their integrands. We cannot simply output them from the currently available CMB temperature anisotropy codes, like CAMB and CMBFast, as their ranges are insufficient for our purposes. This is due to the rescaling of the functions by the three $\hat{k}_i$, which range between [0,1]. Although, due to the constraint $\hat{k}_1+\hat{k}_2+\hat{k}_3=2$, only one can be small at a time. This has the effect of stretching the functions relative to each other. Both the transfer functions and Bessel functions have a similar form. They begin with a long section which is approximately zero before beginning oscillations which decay as $k^{-1}$. This means that they can pick out sections of the other functions that would have been unimportant for the calculation of the power spectrum. This effect can be clearly seen when we plot, in figure (\ref{fig:bessel}), the three individual functions $j_l(x\hat{k})$ and the integrand of the geometric integral $I^G$ for a point in the triangular domain $\curl{S}_k$ generated for the reduced bispectrum point $b_{20\,40\,50}$. It is clear that it is the tail of the second and third Bessel functions which is important for calculation, rather than the initial region where their individual signals are largest. Thus, to accurately calculate the two 1D integrals we need both the transfer functions, and Bessel functions, to cover a much larger range of $k$, with a much better resolution, than is required to evaluate the power spectrum.

Our approach is to output the source function as calculated by one of the CMB anisotropy codes, having first increased the range and resolution of $k$ and the resolution of $\t$. We then calculate the transfer functions ourselves. The Bessel functions are calculated using a similar method to that in CMBFast. Both are then stored in tables for later use. This fixes us to a particular cosmological model unless we wish to keep regenerating the tables. As as we are primarily concerned with selecting primordial models, and the bispectrum is too small to usefully constrain cosmological parameters, this is a minor concern at this stage.

At the beginning of a calculation, we read in the tables selecting the rows that correspond to the relevant $l$'s and interpolate them into a cubic spline. For each point in the triangular domain $\curl{S}_k$ we then take the three $\hat{k}_i$ and calculate the three function values corresponding to the rescaled points, $\hat{k}_i x$. The three rescaled functions can then be multiplied together to form the integrand and we then use a cubic spline to evaluate the integral. The use of splines significantly reduces the resolution needed for accurate evaluation as compared to using simple linear interpolation (used 
previously \cite{0612713}), usually by an order of magnitude. Developing faster integration methods, like the search for an analytic solution to the geometric integral, remains an interesting avenue for future work with scope for significant efficiency gains. With this method there are only two main parameters that control convergence, the resolution and the range. These have undergone extensive experimentation and minimal values for sub 1\% accuracy have been found. The calculation for the 2D integral was completed using the same adaptive method used in \cite{0612713}.

This approach allows us to accurately calculate the bispectrum for a broad range of primordial non-Gaussian models. If we wish to determine the bispectrum at the resolution of Planck, $l_{max} \px 2000$, then the possible allowed $l$ configurations require over 600 million integrations. Fortunately, there are several techniques we can use to make this problem tractable.
These calculations naturally coarse-grain the computational work either through the sampling of 1D integrals on the 2D triangular grid or, at a higher level, simply by evaluating the bispectrum at different multipole values. Secondly, the problem is well-suited to parallelisation on a large supercomputer or cluster and this has been achieved with the present code using a message passing interface (MPI) implementation which significantly reduces calculation time-scales. However, there are two further methods we use to dramatically speed up calculation which are detailed in the following subsections.

\subsection{Flat sky approximation}
If we calculate the reduced bispectrum when all three $l_i$ are large then we are considering very small angles in the sky. If the angles are small then the curvature of the surface of last scattering is small and the so the sky can be approximated as flat. This allows us to greatly simplify the integral for the reduced bispectrum. We use the flat sky methodology which first appeared in \cite{0405356} beginning by expanding the temperature perturbation into plane waves,
\begin{align}
\H (\bx,\bn) = \int \frac{d^2l}{(2\pi)^2} a(\bl) e^{-i\bl \cdot \bn} \imp a(\bl) = \int d^2n \H(\bx,\bn) e^{i\bl\cdot\bn}\,.
\end{align}
and so,
\begin{align} a(\bl)
&= \int \frac{d^3 k}{2\pi} \O(k) \int^{\t_0}_0 d\t S(k,\t) e^{-ik^z(\t_0-\t)} \d^2\(\bk^\pl(\t_0-\t) - \bl\)\,,
\end{align}
where we have split $\bk$ into the part parallel to the tangent plane, $\bk^\pl$, and the part perpendicular, $k^z$. We now form the three point correlator for the flat sky $a(\bl)$'s,
\begin{align} \<a(\bl_1)a(\bl_2)a(\bl_3)\>
\nn =\& \int d\t_1 d\t_2 d\t_3 d^3 k_1 d^3 k_2 d^3 k_3 \d\(\sum k^z_i\) \d^2(\sum \bk^\pl_i) B_\O(k_1,k_2,k_3) \\
\nn & S(k_1,\t_1) S(k_2,\t_2) S(k_3,\t_3) e^{-ik^z_1(\t_0-\t_1)} e^{-ik^z_2(\t_0-\t_2)} e^{-ik^z_3(\t_0-\t_3)} \\
& \d^2\(\bk_1^\pl(\t_0-\t_1) - \bl_1\) \d^2\(\bk_2^\pl(\t_0-\t_2) - \bl_2\) \d^2\(\bk_3^\pl(\t_0-\t_3) - \bl_3\)\,.
\end{align}
To integrate out the three delta functions for $\bk^\pl$ we must assume that the variation in $B_\O(k_1,k_2,k_3)$ is small in the $\bk^\pl$ direction across the width of last scattering. This allows us to use an average value for $\bk^\pl$ of $\bk^\pl = \bl / (\t_0 - \t_R)$. After substitution this gives,
\begin{align} \<a(\bl_1)a(\bl_2)a(\bl_3)\>
\nn =\& (\t_0 - \t_R)^2 \d^2\(\sum \bl_i\)\int d\t_1 d\t_2 d\t_3 d k^z_1 d k^z_2 d k^z_3 \d\(\sum k^z_i\) B_\O(k'_1,k'_2,k'_3)\\
\& S(k'_1,\t_1) S(k'_2,\t_2) S(k'_3,\t_3) e^{-ik^z_1(\t_0-\t_1)} e^{-ik^z_2(\t_0-\t_2)} e^{-ik^z_3(\t_0-\t_3)}\,.
\end{align}
where $k'$ is k evaluated with $\bk^\pl = \bl / (\t_0 - \t_R)$. If we define a new transfer function,
\begin{align}
\D(l,k^z) = \int \frac{d\t}{(\t_0 - \t)^2} S(\sqrt{(k^z)^2 + l^2/(\t_0 - \t)^2},\t)e^{ik^z \t}\,,
\end{align}
and use, $\<a(\bl_1)a(\bl_2)a(\bl_3)\> \px (2\pi)^2 \d^2\left(\sum \bl_i\right) b^{flat}_{l_1 l_2 l_3}$, then we find that, in the flat sky limit, the expression for the reduced bispectrum is,
\begin{align}\label{eq:flatsky}
b^{flat}_{l_1 l_2 l_3} = \frac{(\t_0 - \t_R)^2}{(2\pi)^2}\int^{\infty}_{-\infty} d k^z_1 d k^z_2 d k^z_3 \d\(\sum k^z_i\) B_\O(k'_1,k'_2,k'_3)\D(l_1,k^z_1) \D(l_2,k^z_2)\D(l_3,k^z_3)\,.
\end{align}

\begin{figure}[ht]
\includegraphics[width=0.6\linewidth]{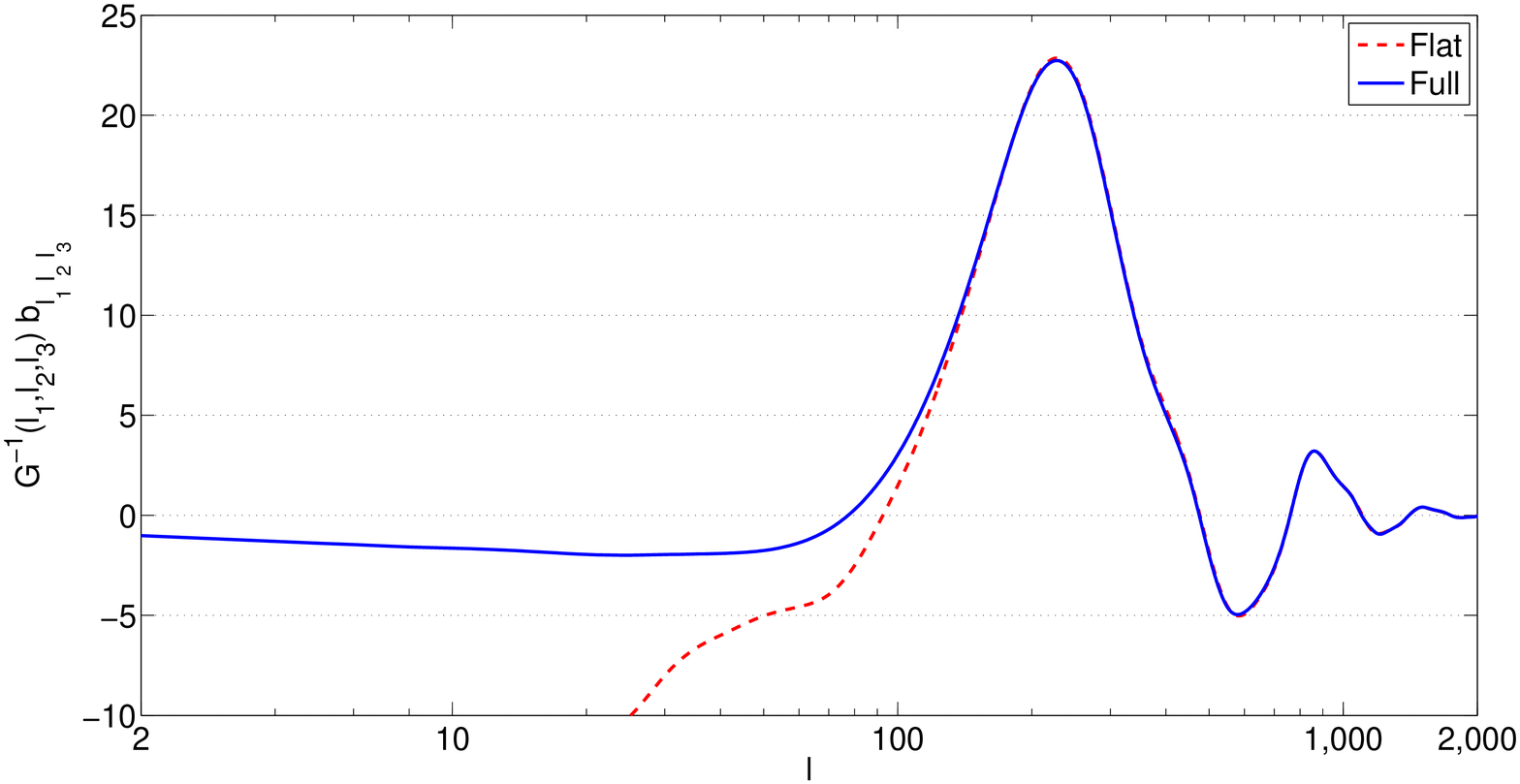}
\centering
\begin{tabular}{@{}c@{}c@{}}
\includegraphics[width=0.5\linewidth]{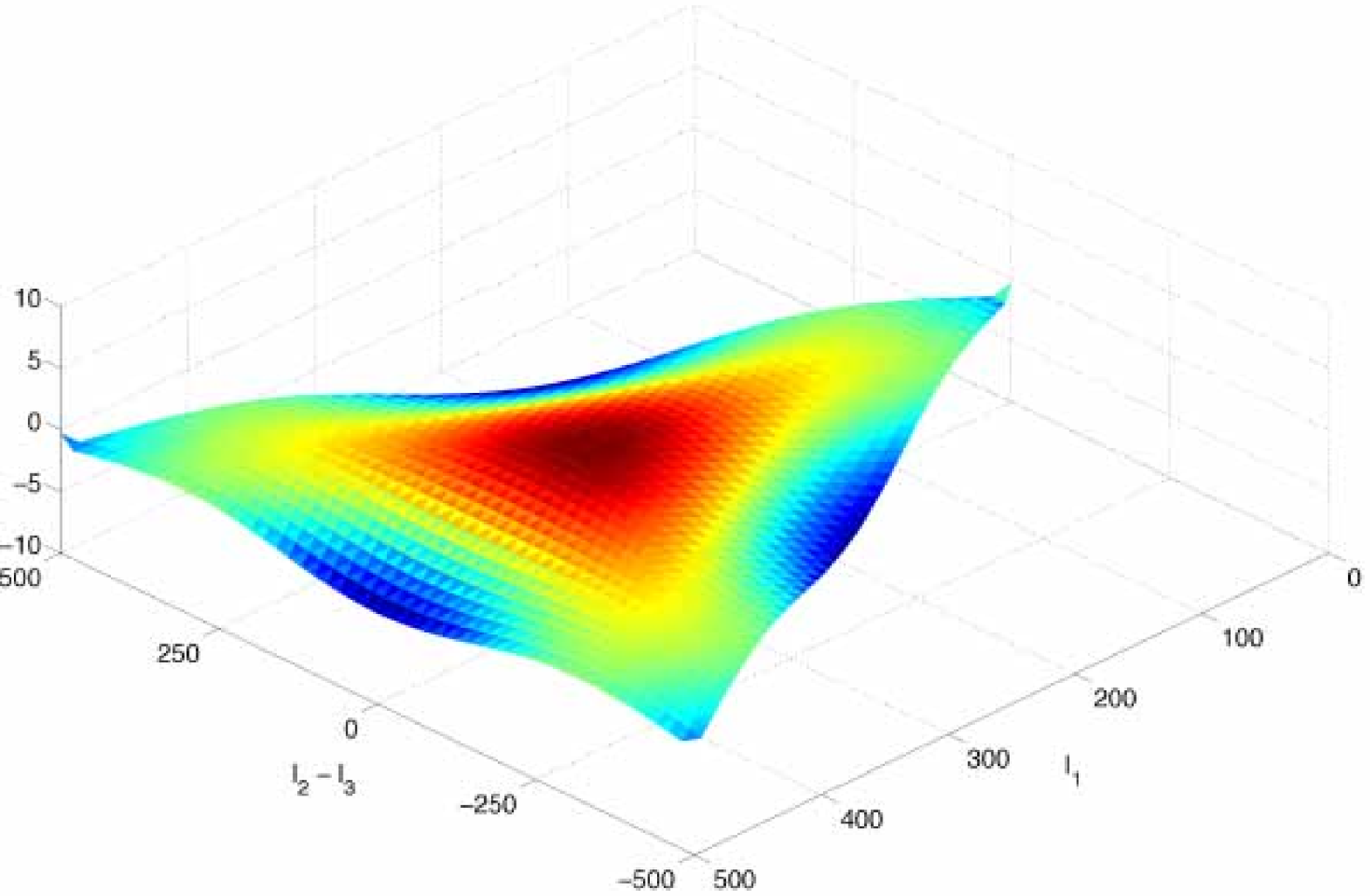} &
\includegraphics[width=0.5\linewidth]{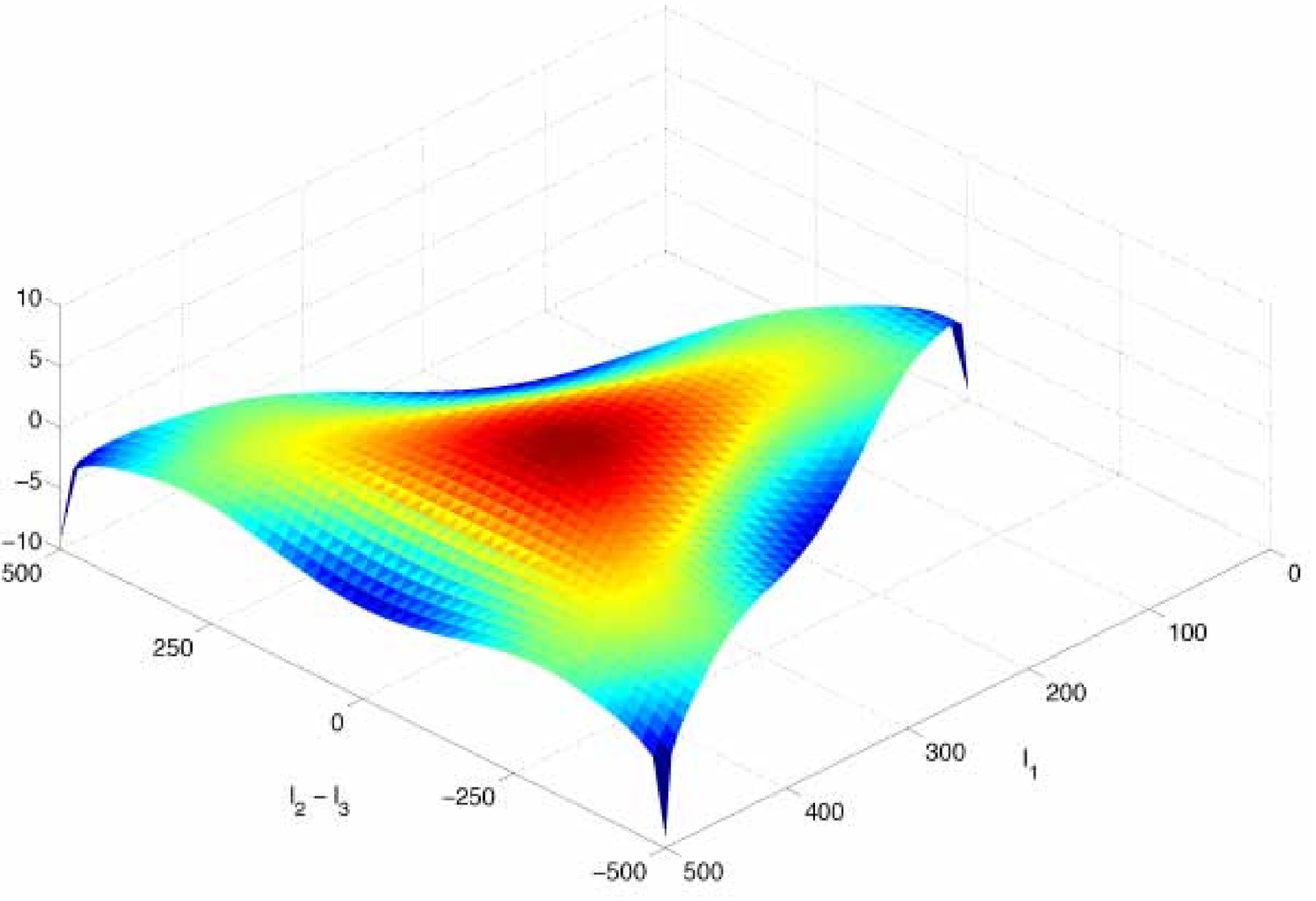} \\
\end{tabular}
\caption[A comparison of the full calculation of the reduced bispectrum with the flat sky approximation.]{A comparison of the full calculation of the reduced bispectrum with the flat sky approximation. Top is a comparison of $b_{lll}$ and we find excellent agreement when $l\ge 150$. The bottom two plots show a cross section through $l_1+l_2+l_3=1000$ with the full case on the left and the flat sky approximation on the right. Again we see that the two methods agree except in the corners when one of the $l_i<150$. The close agreement between the two independent methods establishes the accuracy of the two codes}
\label{fig:flatcomparison}
\end{figure}

\noindent For small angles $b^{flat}_{l_1 l_2 l_3} \px b_{l_1 l_2 l_3}$ and so we can use (\ref{eq:flatsky}) to calculate the bispectrum. We evaluate the two integrals using cubic splines.

We have calculated the reduced bispectrum using the full method detailed in the previous section, and again in the flat sky case,  comparing them in figure (\ref{fig:flatcomparison}). We find that the flat sky approximation becomes valid when all three $l_i \ge 150$, producing less than 1\% error. Also, as the two methods are completely independent of each other, this provides a powerful cross check of the accuracy.  The flat sky approximation allows us to calculate the bispectrum, when all three $l_i \ge 150$, more than 300 times faster. This is a dramatic improvement only leaving  small subregions near 
the corners and edges which require the full calculation. Nevertheless, even with this much faster method, the reduced bispectrum at Planck resolution represents a formidable challenge, unless
we significantly reduce the number of points at which it needs to be evaluated.

\begin{figure}[ht]
\centering
\begin{tabular}{@{}c@{}c@{}}
\includegraphics[width=0.425\linewidth]{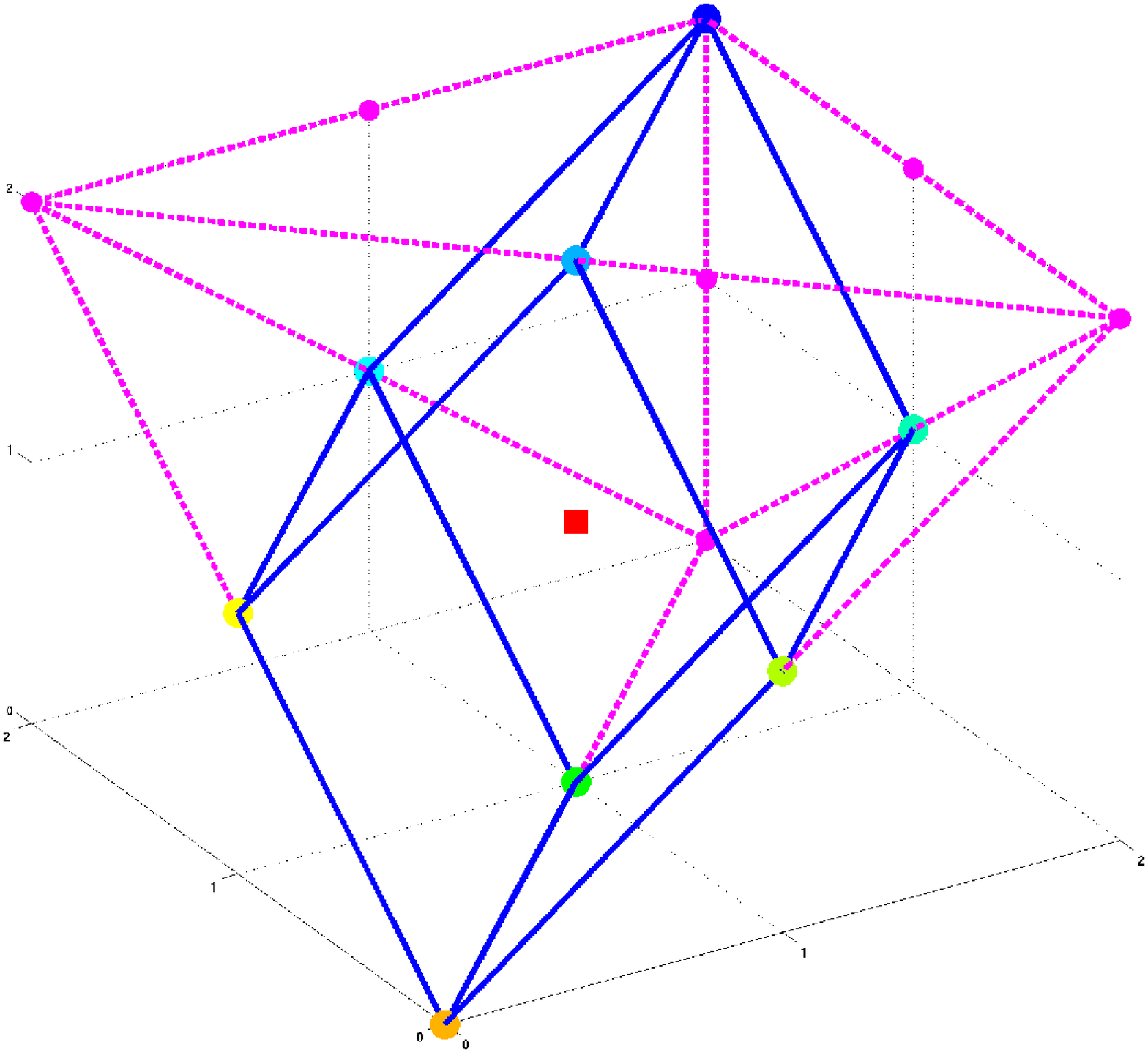} &
\hskip0.2in\includegraphics[width=0.4255\linewidth]{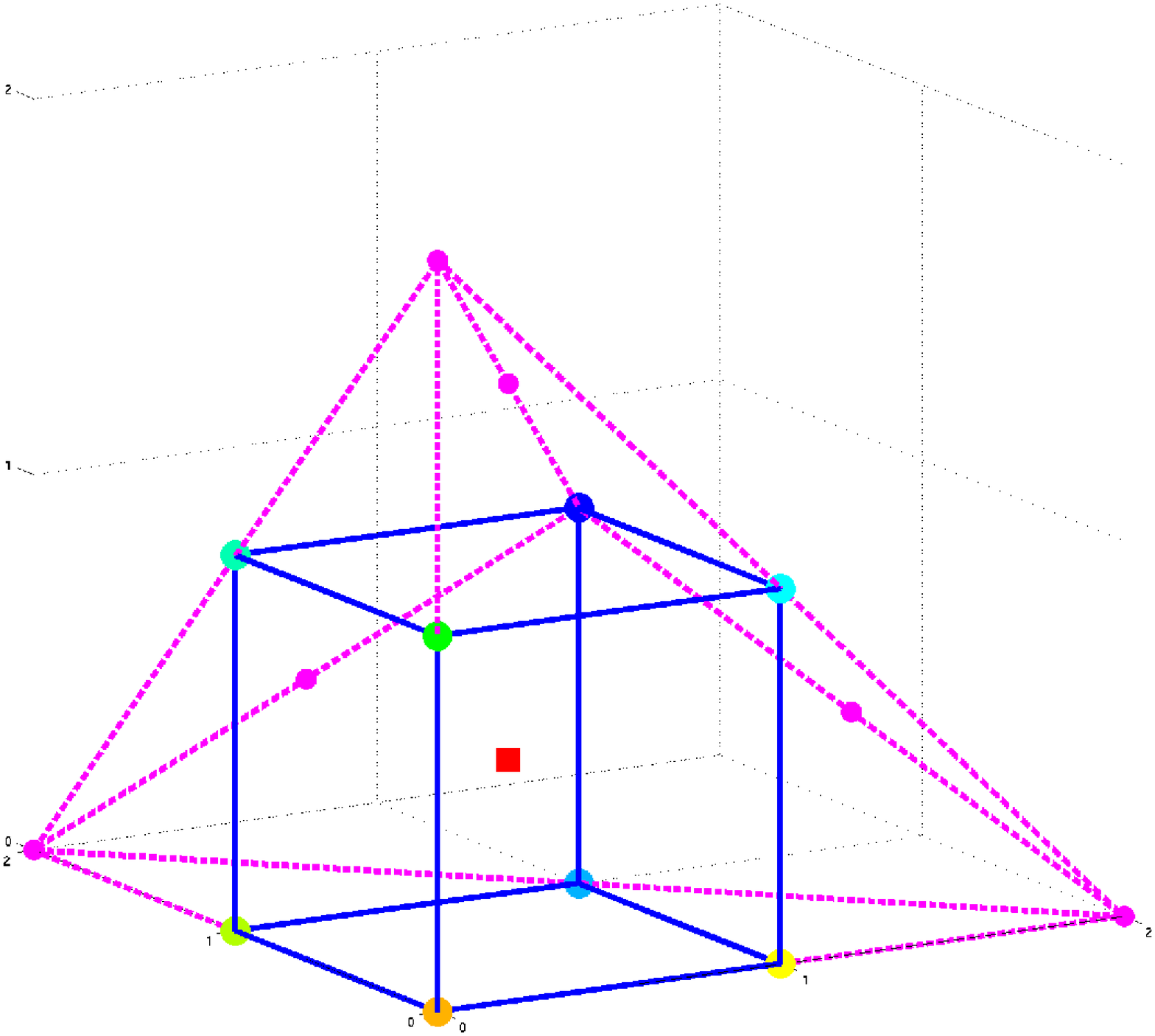} \\
\end{tabular}
\caption[Transformation to map the tetrahedron to a cube for interpolation]{Transformation to map the tetrahedron to a cube for interpolation. On the left we have the tetrahedron and on the right the tetrahedron after the mapping is applied. The blue lines (solid) define the cubic cell over which we interpolate. The red point (square) is the value that falls inside the cube after the transformation and is not used for interpolation. Pink points and lines (dashed) are the parts of the tetrahedron outside the cell being interpolated.}
\label{fig:transform}
\end{figure}

\subsection{Cubic interpolation}\label{se:cubic}

From the plots comparing the full calculation with the flat sky approximation in figure (\ref{fig:flatcomparison}), we note that $G^{-1}(l_1,l_2,l_3) b_{l_1 l_2 l_3}$ is actually very smooth. This is expected as for models with smooth shape functions all the structure present in the reduced bispectrum must be due to the acoustic peaks in the transfer functions. As a result the reduced bispectrum will only contain features that oscillate with periods of $l\px 200$. This indicates that we only need calculate the reduced bispectrum on a sparse grid and the remaining points could be generated via interpolation. One major problem is in selecting the grid to interpolate over. With the triangle condition on the three $l_i$ limiting us to a tetrahedron we cannot use the usual schemes as when they straddle the boundary they give incorrect results. The geometric integral returns zero for $l$ combinations that violate the triangle inequality creating a discontinuity which leads to poor convergence. 
We can circumvent this problem by rotating and stretching the allowed region so it forms a rectangular grid via the transform,
\begin{align}
l'_1 \= \frac{1}{2}\(l_2+l_3-l_1\)\,,\\
l'_2 \= \frac{1}{2}\(l_3+l_1-l_2\)\,,\\
l'_3 \= \frac{1}{2}\(l_1+l_2-l_3\)\,.
\end{align}
We can then use cubic interpolation to calculate the remaining points before transforming back to obtain the bispectrum for all combinations of $l_i$. There are some minor issues with this approach. For
example, not all points fall onto the grid when we rotate. If we are using a grid with steps in $l$ of 10 we would find that the first cell would be constructed from: $b_{2\,2\,2}$, 3 permutations of $b_{2\,10\,10}$, 3 permutations of $b_{10\,10\,20}$, and $b_{20\,20\,20}$. This leaves $b_{10\,10\,10}$ sitting in the centre of the cell and it is ignored in the subsequent interpolation, see figure (\ref{fig:transform}). As a result we must calculate the reduced bispectrum on twice the density of points that we require, but they are
not entirely redundant as we use them dynamically as a cross-check of the accuracy of the interpolation.
We encounter a similar issue when transforming back with the inverse mapping,
$l_1 = l'_2+l'_3\,,$ $l_2 = l'_3+l'_1\,,$ $l_3 = l'_1+l'_2\,.$

Interpolation reduces the number of points required to calculate the reduced bispectrum by several orders of magnitude and so together with the flat sky approximation and the parallelisation of the code, the problem becomes tractable for all the models reviewed in section~3.

\begin{figure}[hpt]
\includegraphics[width=0.7\linewidth]{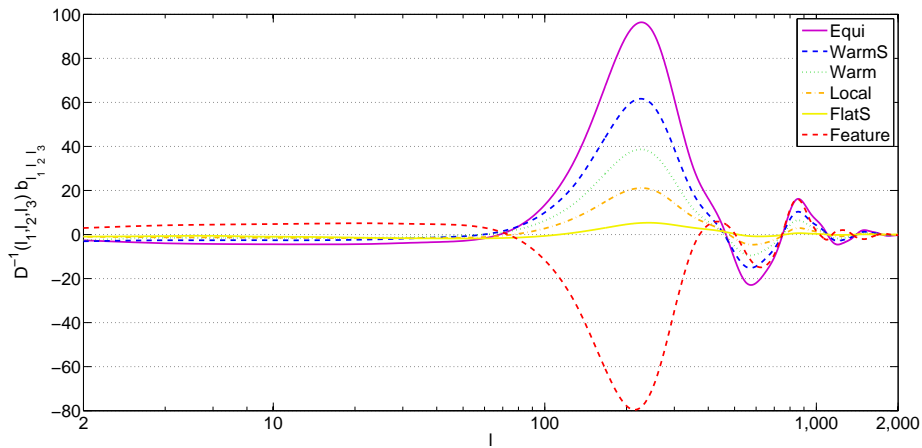}
\caption[The equal l bispectrum for all classes of models]{\small The equal l bispectrum for all classes of models. They are, from top to bottom at $l=220$: equilateral, smoothed warm, warm, local, smoothed flat, and feature. As all the models scale as $k^{-6}$ they all produce similar $b_{lll}$, with the exception of the feature model whose oscillations can be clearly seen.}
\label{fig:bmpiall}
\end{figure}

\begin{figure}[hpb]
\centering
\begin{tabular}{@{}c@{}c@{}}
\includegraphics[width=0.5\linewidth]{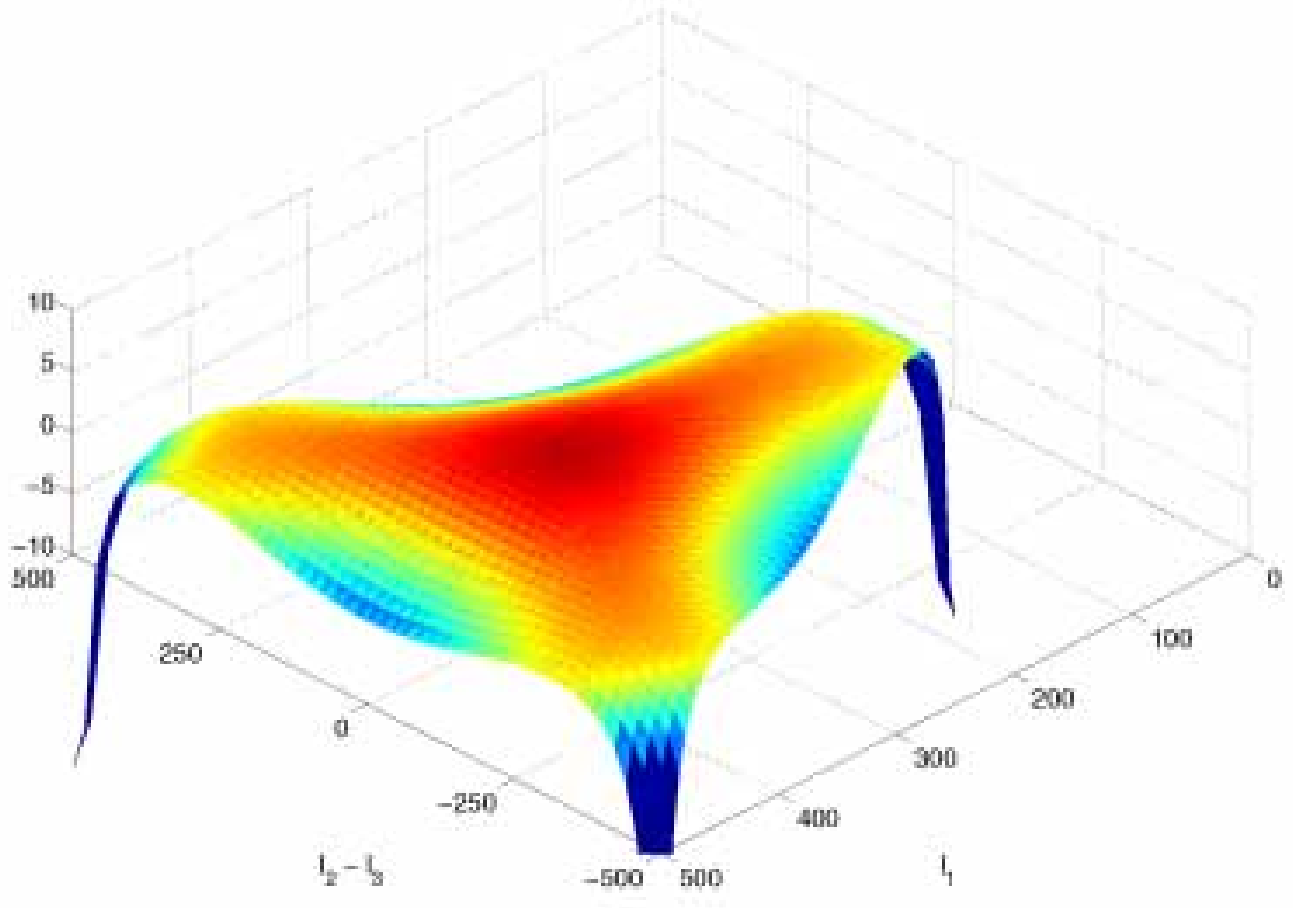} &
\includegraphics[width=0.5\linewidth]{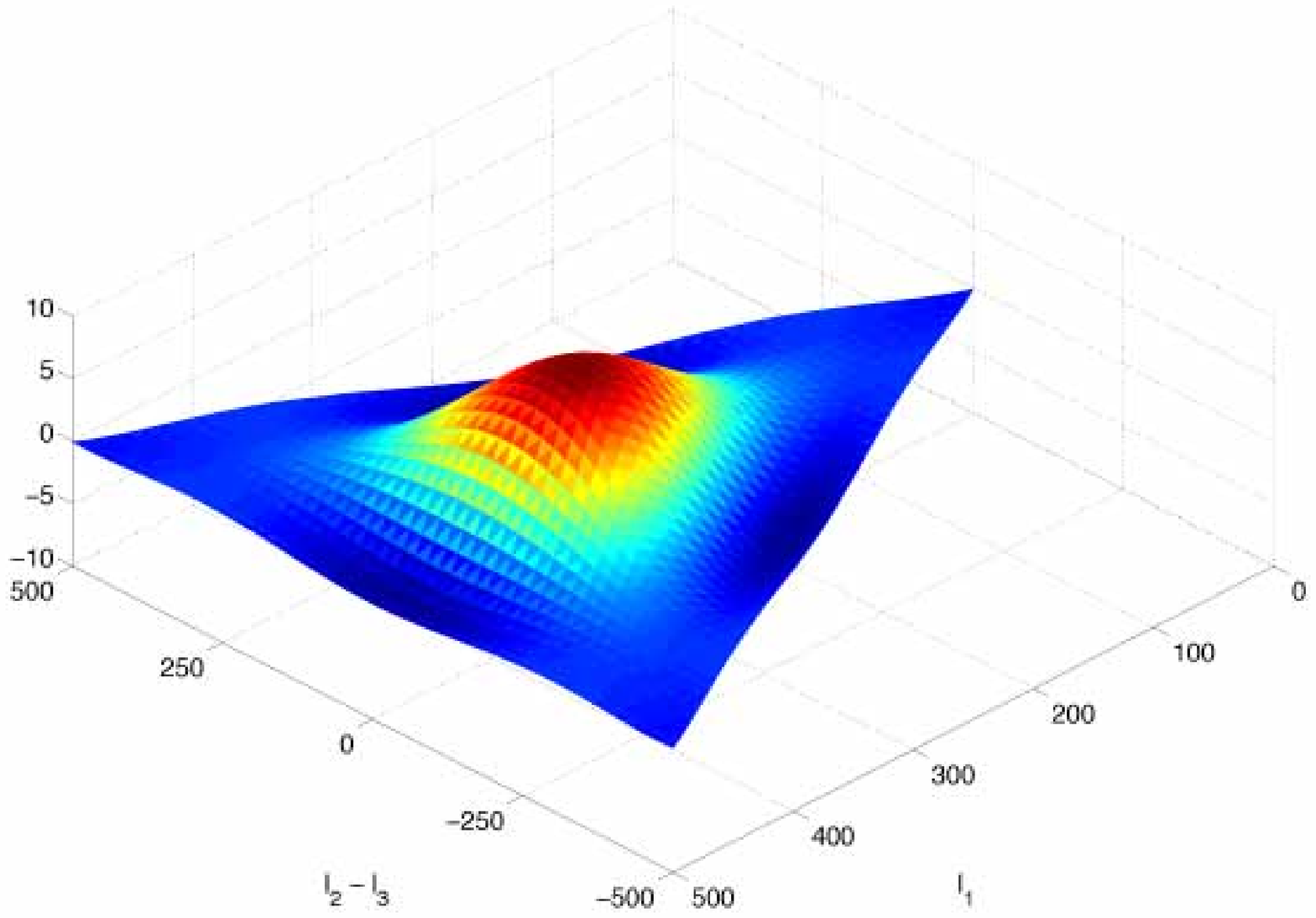} \\
\includegraphics[width=0.5\linewidth]{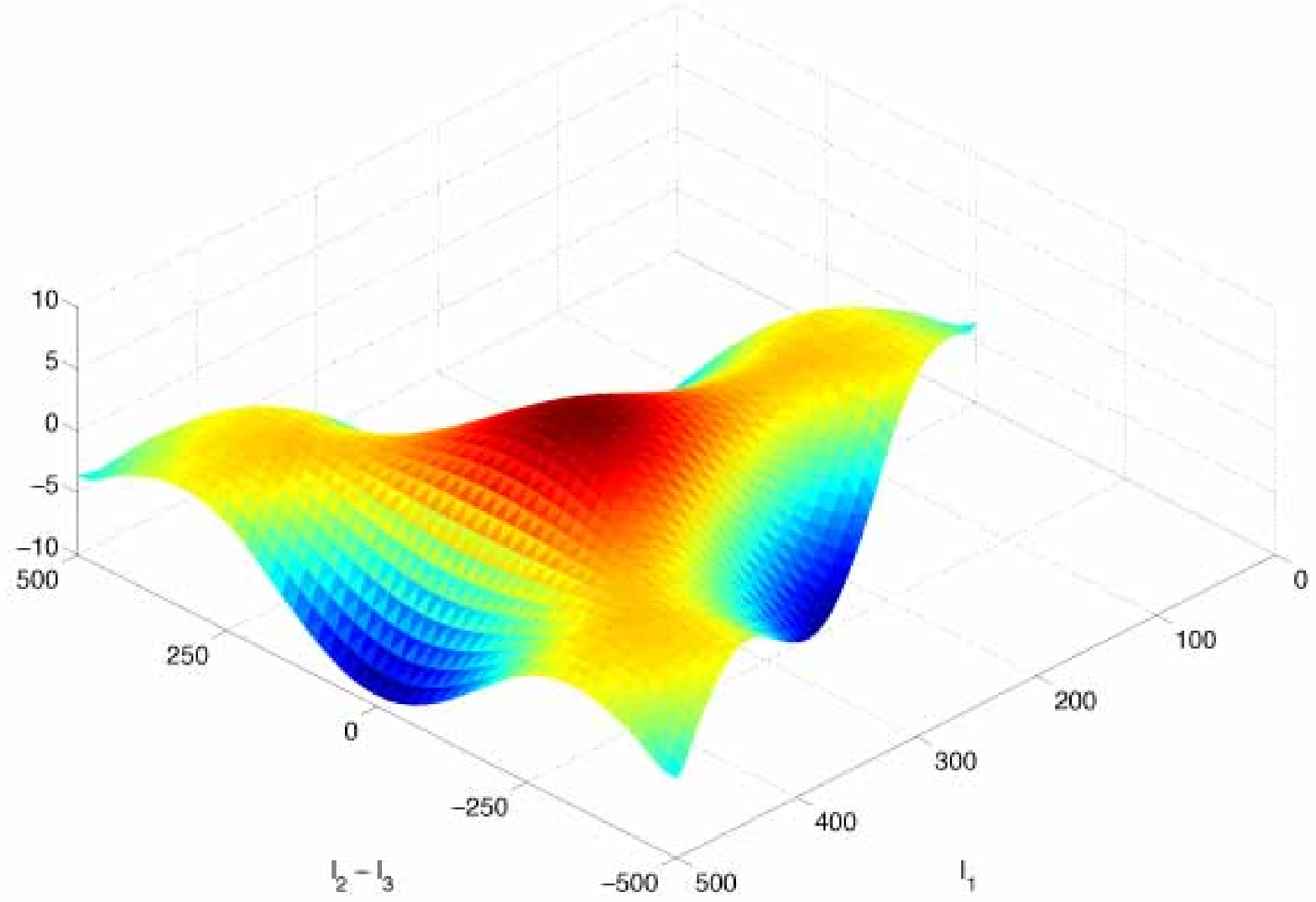}&
\includegraphics[width=0.5\linewidth]{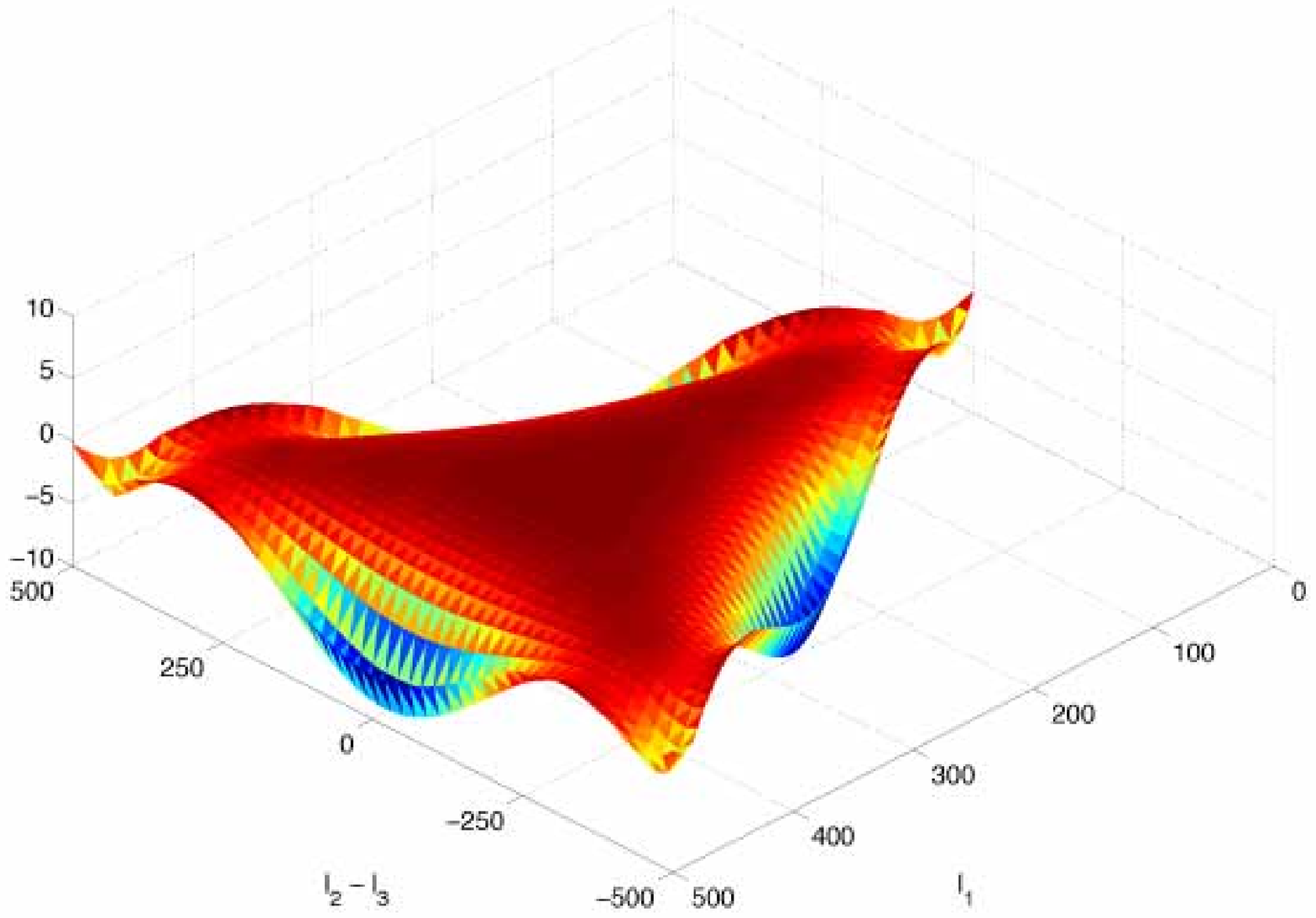}\\
\end{tabular}
\caption[The reduced bispectrum for different classes of models plotted on slices where $l_1+l_2+l_3=1000$]{\small The reduced bispectrum for different classes of models plotted on slices where $l_1+l_2+l_3=1000$. Clockwise from top left, Local, Equilateral, FlatS, and Feature. The effect of the primordial shape function can be clearly seen in the resulting bispectra.}
\label{fig:bmpislices}
\end{figure}

\begin{figure}[t]
\centering
\begin{tabular}{@{}c@{}c@{}}
\includegraphics[width=0.5\linewidth]{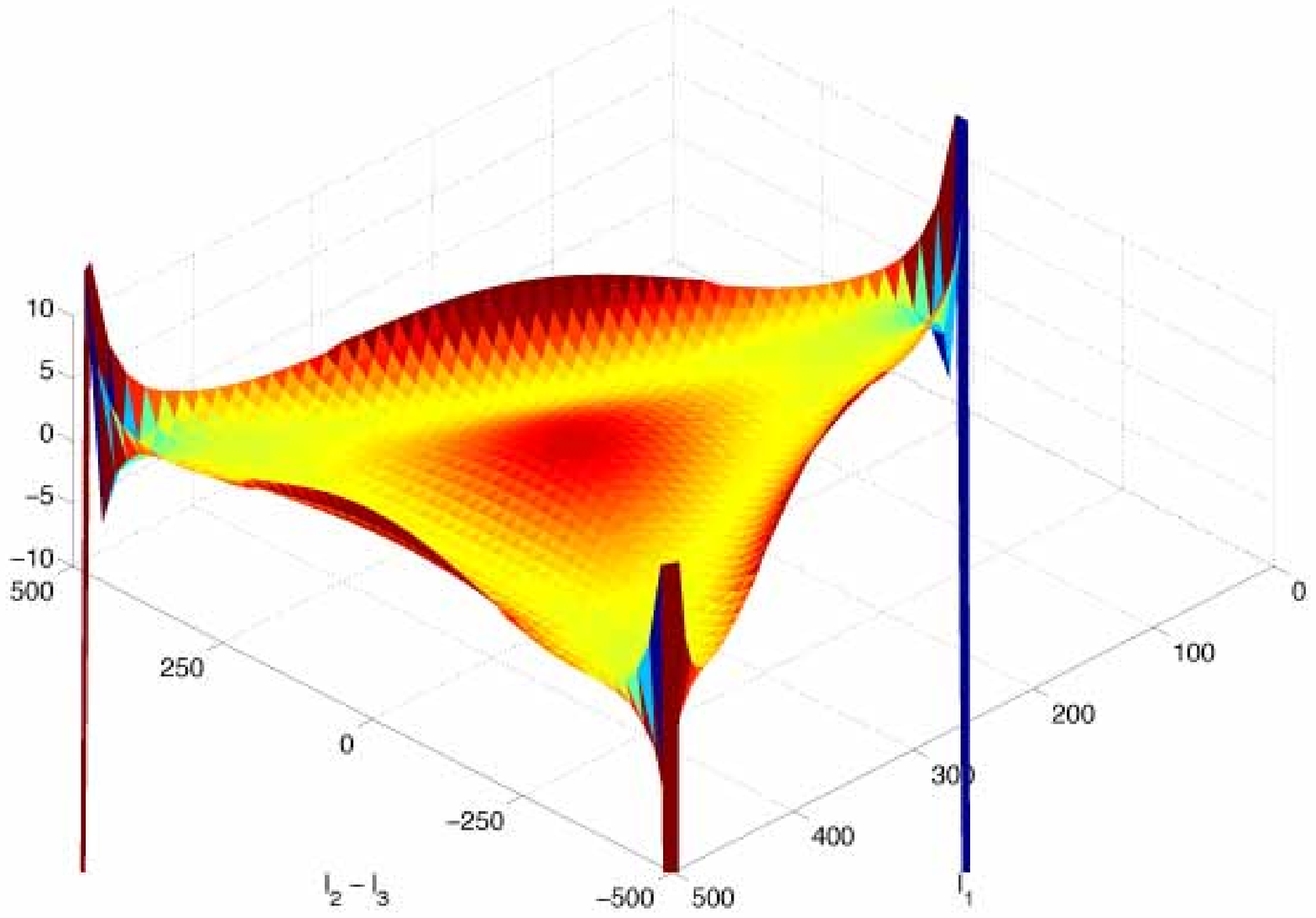} &
\includegraphics[width=0.5\linewidth]{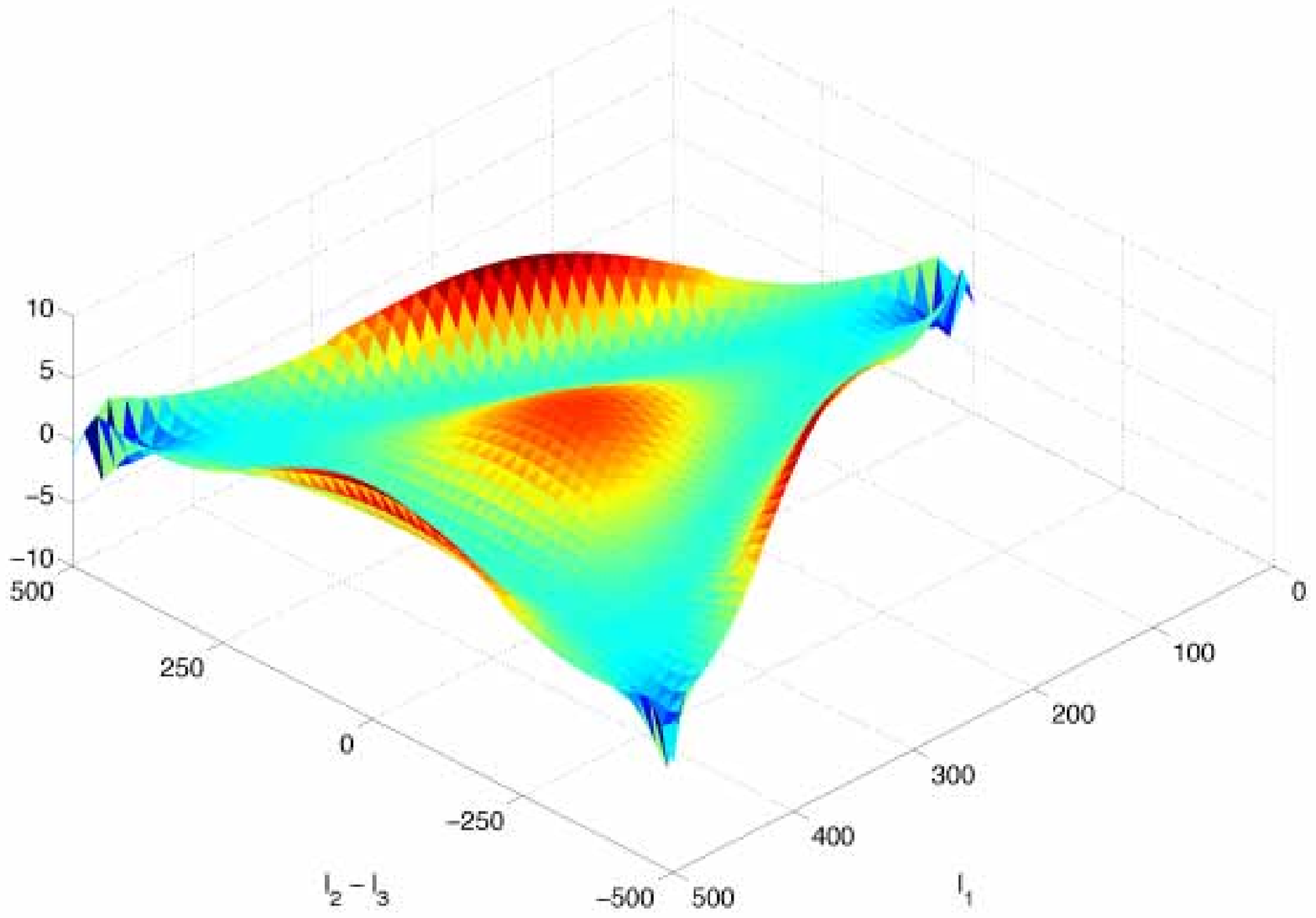}
\end{tabular}
\caption[The reduced bispectrum for the Warm (left) and WarmS models plotted on slices where $l_1+l_2+l_3=1000$]{\small The reduced bispectrum for warm (left) and warmS (right) models plotted on slices where $l_1+l_2+l_3=1000$. }
\label{fig:bmpiwarmslices}
\end{figure}

\section{CMB bispectrum model comparison}\label{se:results}

We have discussed the smooth nature of the CMB bispectrum and its general properties in a previous publication \cite{0612713}, so our purpose here is to investigate the distinguishing signatures of the models we outlined in section~3. With the method developed in the previous sections we can evaluate the entire reduced bispectrum for any primordial model, calculating up to $l=2000$ in approximately 100 processor hours; this has been achieved to at least 1\% accuracy for all the models investigated. In figure~\ref{fig:bmpiall}, we plot the central value $b_{lll}$ for all five primordial classes of shapes -- equilateral (\ref{eq:equi}), local (\ref{eq:local}), warm (\ref{eq:warm}), flat (\ref{eq:flat}) and feature (\ref{eq:feature}). The first four of these models are scale-invariant, so the $b_{lll}$ all take broadly the same profile but with different normalisations. We note that this figure demonstrates the oscillatory properties of the transfer functions which, as for the CMB power spectrum, create a series of acoustic peaks around $l = 200, 500, 800, ...$. There is a stark contrast, however, with the feature model which has a non-trvial scaling. Initially, it is anticorrelated with the other shapes, so that the primary peak has opposite sign, however, for increasing $l$ the phase of the oscillations becomes positively correlated by the second and third peaks (this, of course, reflects the particular choice of $k^*$ in (\ref{eq:feature})).

\begin{figure}[ht]
\centering
\begin{tabular}{@{}c@{}c@{}}
\includegraphics[width=0.5\linewidth]{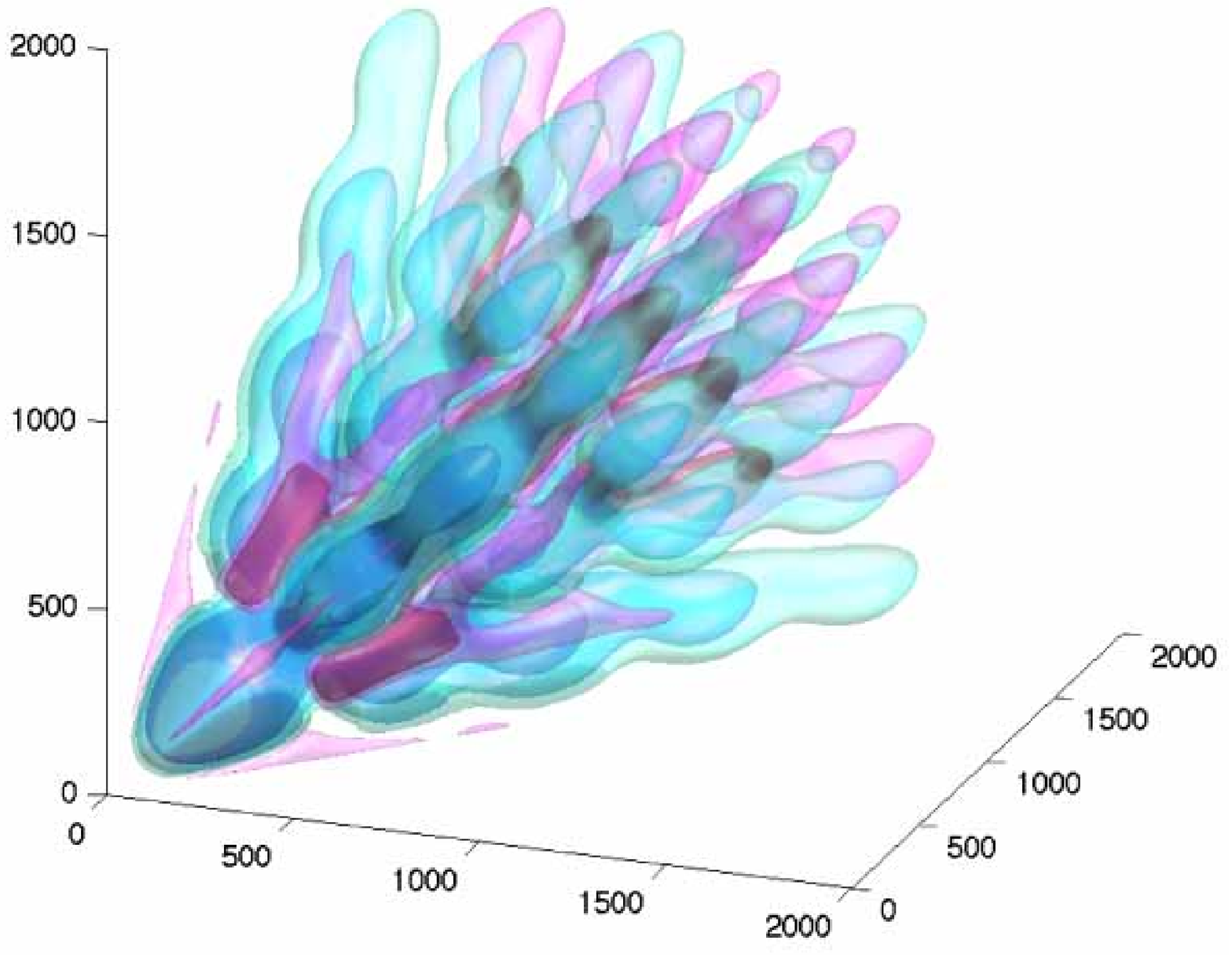} &
\includegraphics[width=0.5\linewidth]{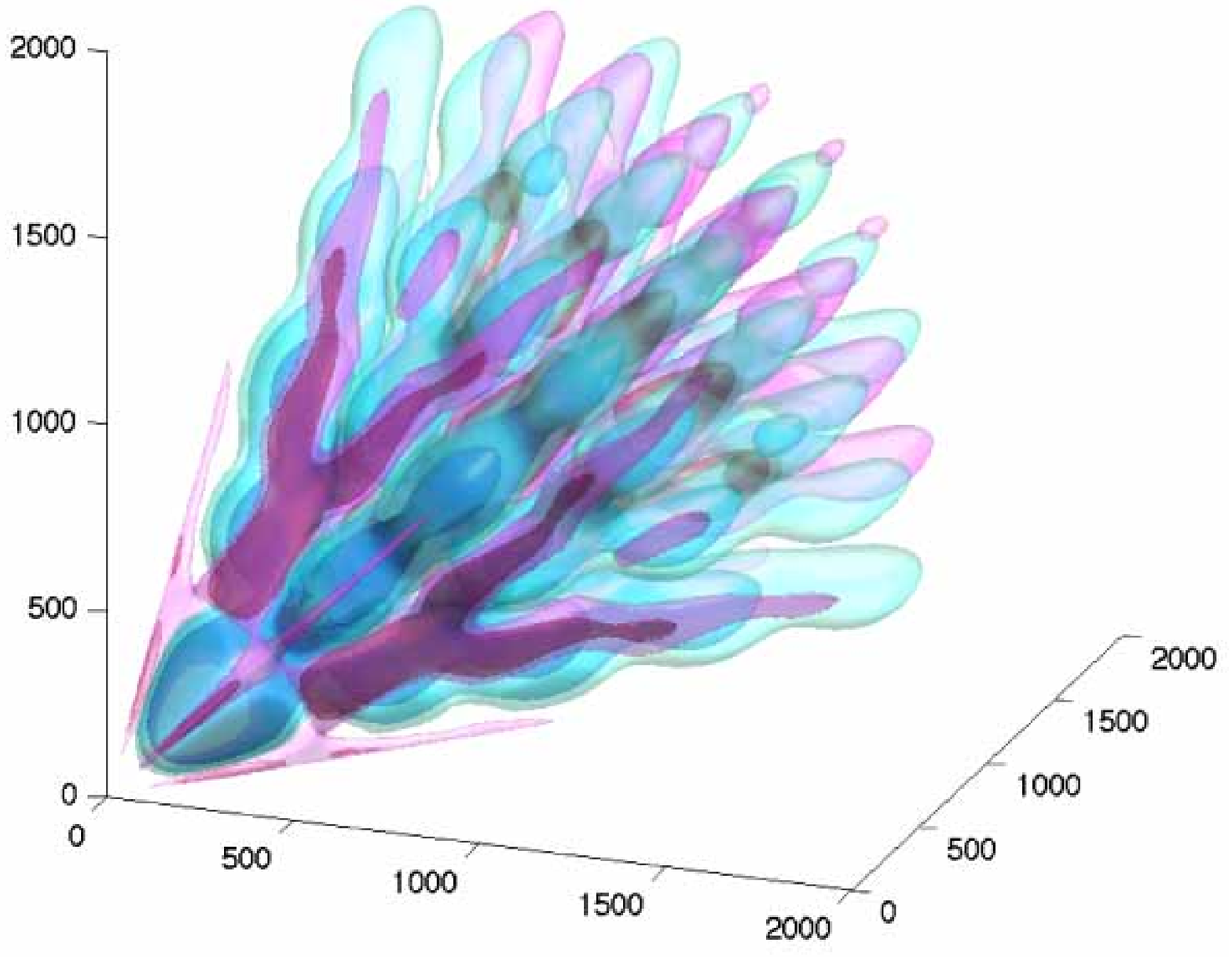} \\
\includegraphics[width=0.5\linewidth]{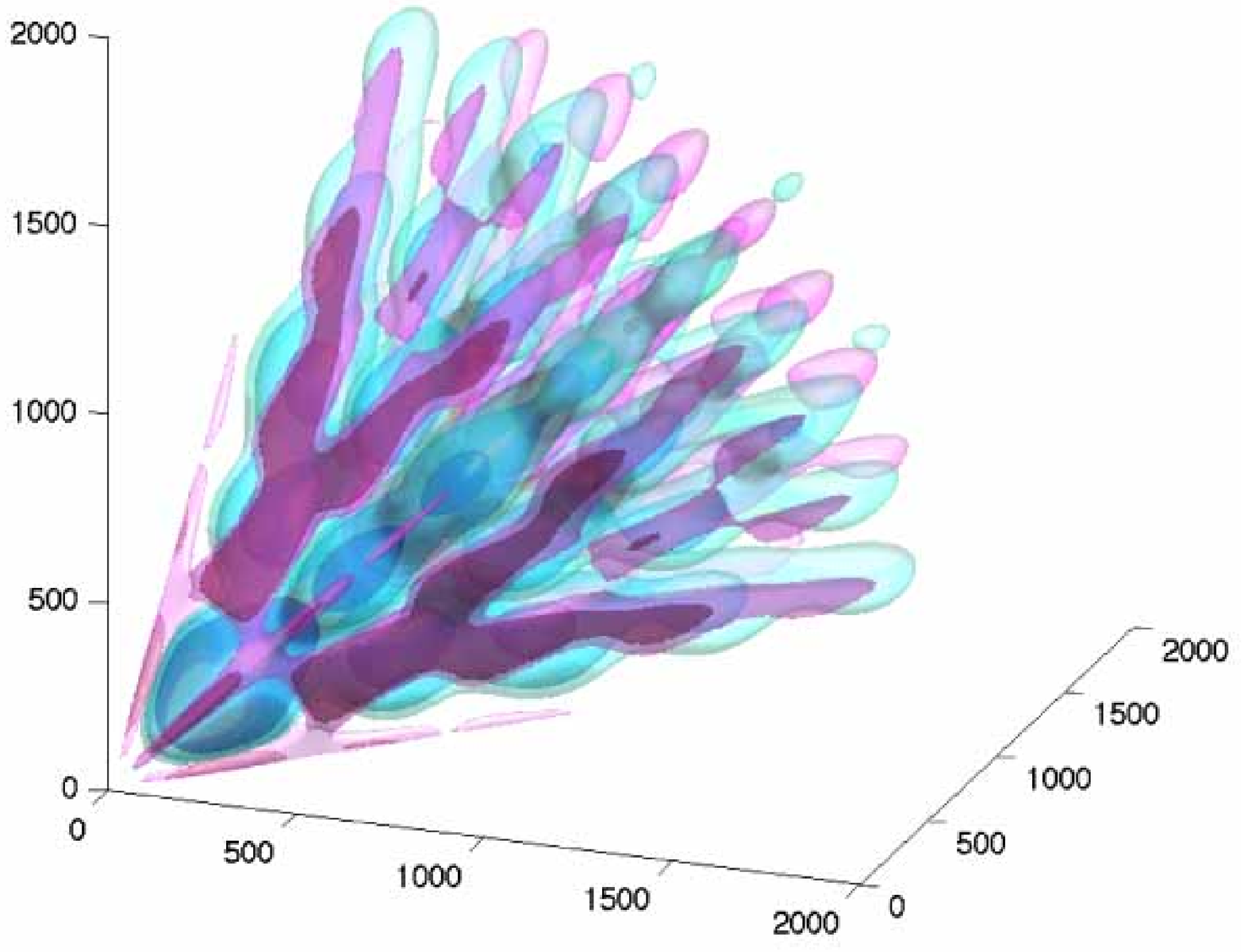} &
\includegraphics[width=0.5\linewidth]{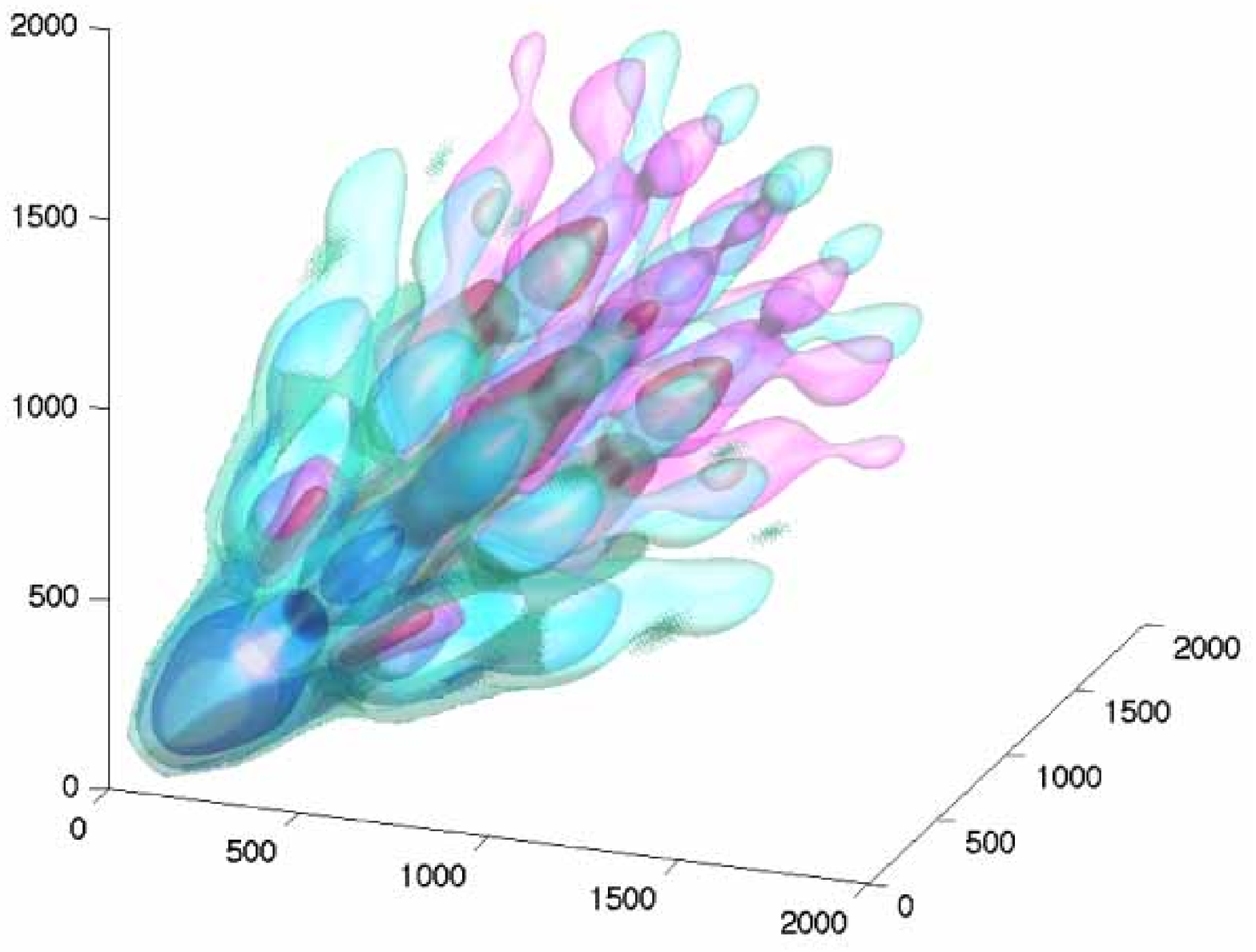}
\end{tabular}
\caption[3D Plots of the reduced bispectrum for models in the equilateral class.]{\small 3D Plots of the reduced bispectrum for models in the equilateral class. Here, and in the following, we have plotted surfaces of equal bispectra after division by the constant analytic solution. Clockwise from top left we have equilateral, DBI, Ghost and Single.}
\label{fig:equi3d}
\end{figure}

\begin{figure}[ht]
\centering
\begin{tabular}{@{}c@{}c@{}}
\includegraphics[width=0.5\linewidth]{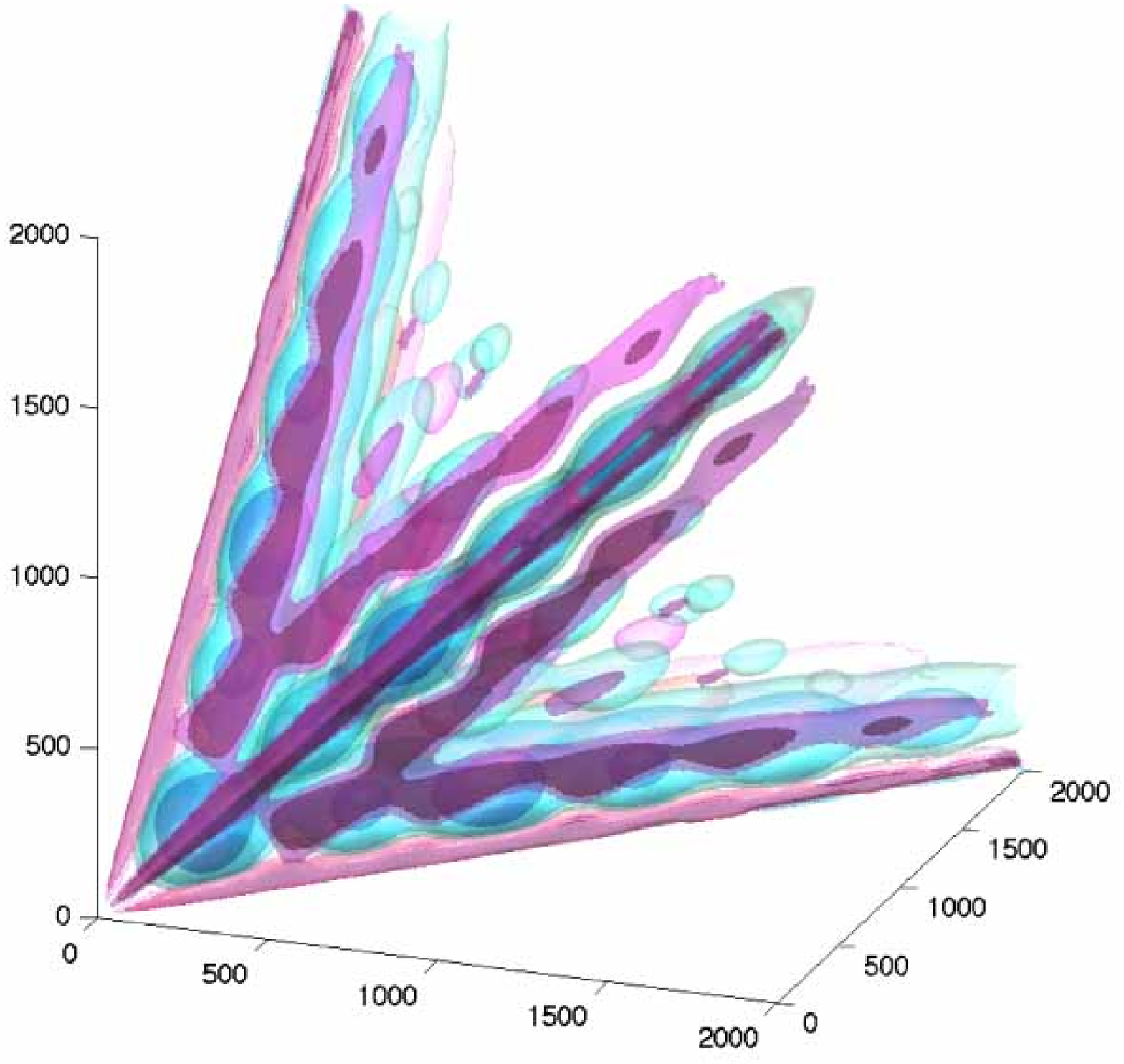} &
\includegraphics[width=0.5\linewidth]{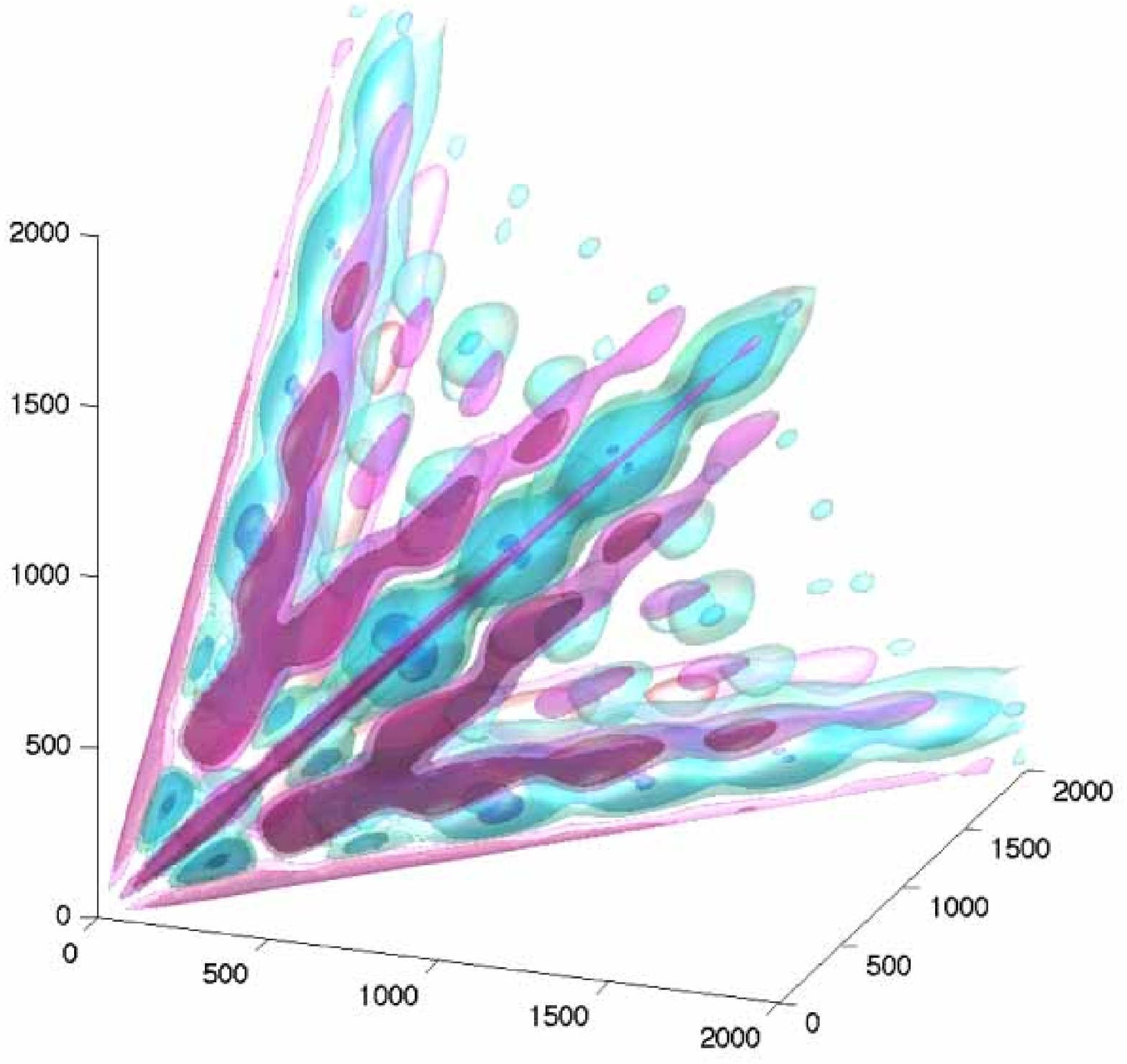}
\end{tabular}
\caption[3D Plots of the reduced bispectrum for the local and flattened models.]{\small 3D Plots of the reduced bispectrum for the local model (left) and the flattened model after smoothing (right).}
\label{fig:local3d}
\end{figure}

\begin{figure}[ht]
\centering
\begin{tabular}{@{}c@{}c@{}}
\includegraphics[width=0.5\linewidth]{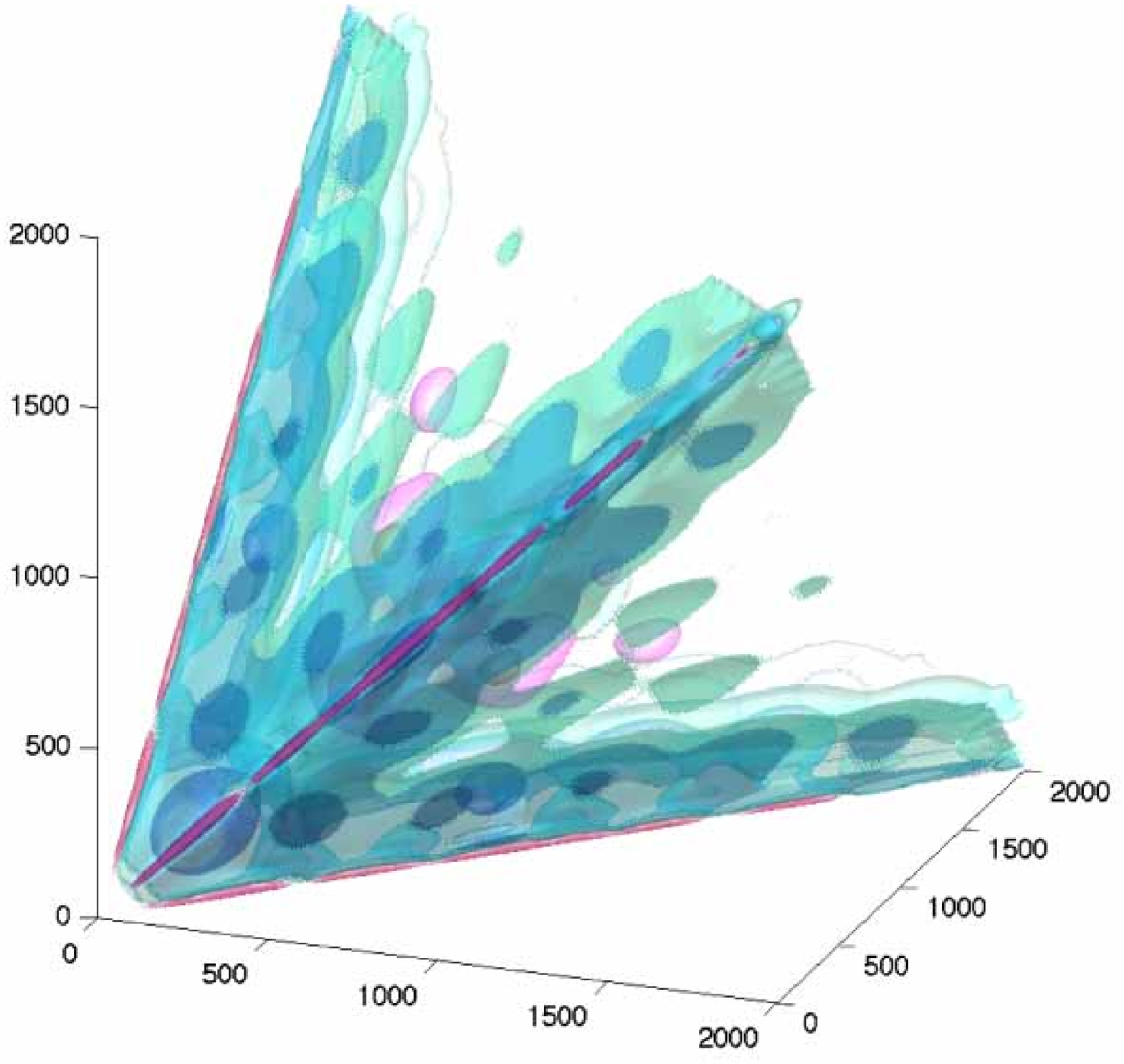} &
\includegraphics[width=0.5\linewidth]{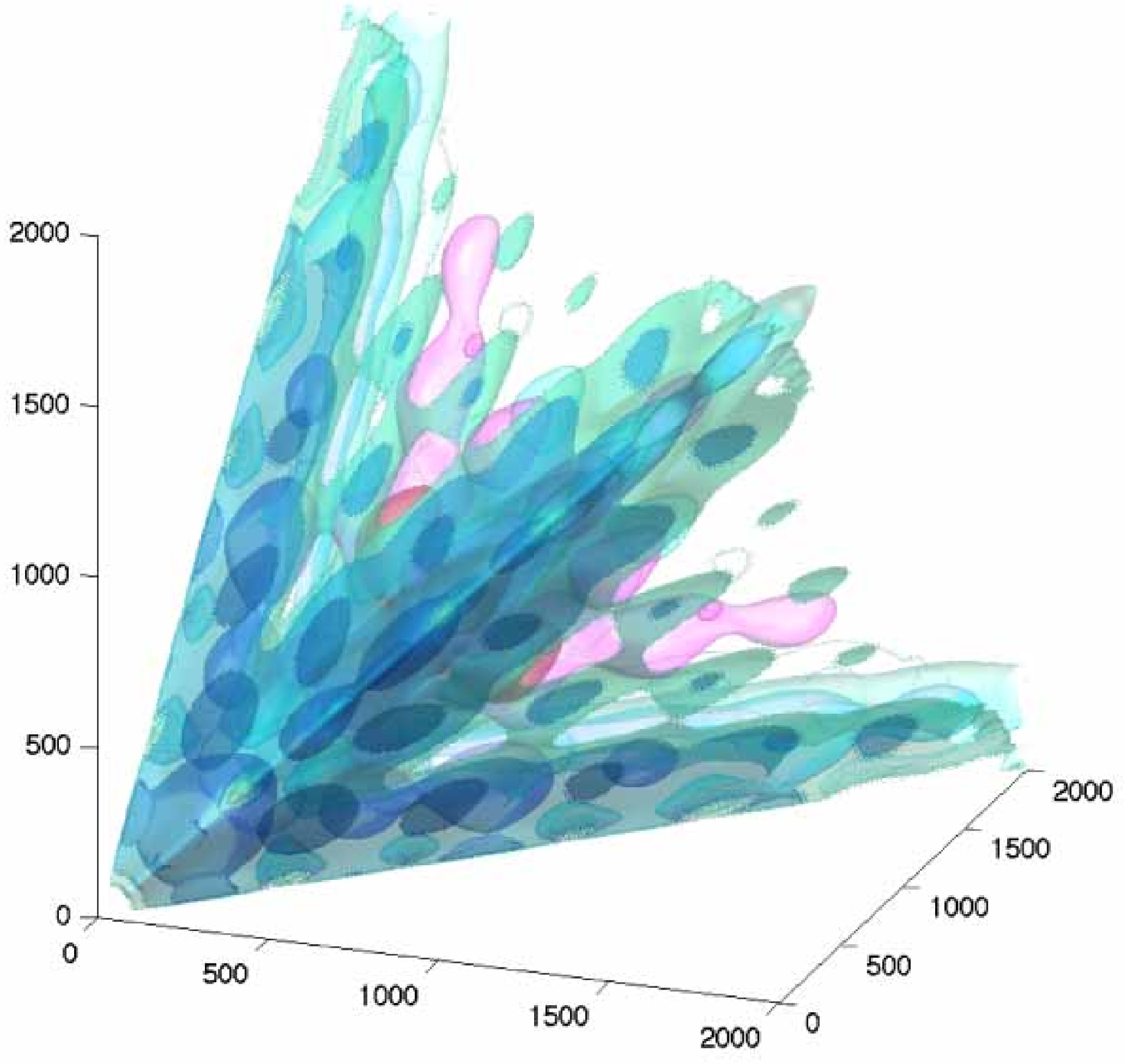}
\end{tabular}
\caption[3D Plots of the reduced bispectrum for warm inflation before and after smoothing.]{\small 3D Plots of the reduced bispectrum for warm inflation before smoothing, on the left, and after smoothing, on the right.}
\label{fig:warm3d}
\end{figure}

Of course, to observe the key differences between the models we must study the bispectrum in the plane orthogonal to the $(l,l,l)$-direction, that is, the directions reflecting changes in the primordial shape functions. To this end, in figures~\ref{fig:bmpislices} and \ref{fig:bmpiwarmslices} we plot cross-sectional slices through the reduced bispectrum; we choose triangles satisfying $l_1+l_2+l_3=1000$ just beyond the primary peak so as to reduce the effect of the transfer functions relative to the primordial shape. For these slices and in the subsequent figures \ref{fig:equi3d}--\ref{fig:warm3d} for the full three-dimensional bispectrum, we divide $b_{lll}$ by $D(l_1,l_2,l_3)$, the large-angle CMB bispectrum solution (\ref{eq:dfactor}) for a primordial shape which is \it{constant}. This is analogous to multiplying the power spectrum $C_l$'s by $l(l+1)$, because it serves to flatten the bispectrum except for the effect of the non-constant primordial shape (and the oscillating transfer functions). (We note that in our previous paper~\cite{0612713}, for plotting purposes we divided all the CMB bispectra by the large-angle local solution $G(l_1,l_2,l_3)$ given in (\ref{eq:gfactor}), but this is not as useful for comparison purposes.)

The bispectrum slices for the different models shown in figure~(\ref{fig:bmpislices}) directly mimic the primordial shapes from which they originated. The equilateral model has the majority of its signal under a prominent central peak (i.e.\ equilateral triangles), whereas the local model has a nearly flat interior with most signal at sharp peaks in the corners (i.e.\ for squeezed triangles). As expected, the (smoothed) flat model is strongly peaked along the edges for flattened triangle configurations. The last model illustrated in figure~(\ref{fig:bmpislices}) is the feature model whose primordial shape function is constant across $k={\rm const.}$ slices. On $l ={\rm const.}$ slices, therefore, the feature model should behave like the constant model, showing only the effect of the transfer functions in its structure. Finally, there is a slice through the warm model shown in figure~(\ref{fig:bmpiwarmslices}) in both original and smoothed versions. These have signal when the $l_i$ are in several configurations, squeezed, flattened and equilateral. Note the strong effect that smoothing has on suppressing the dominant contribution from squeezed states. A better understanding of the asymptotic behaviour in squeezed and flattened limits is clearly necessary if we are to make robust quantitative predictions for both the warm and flat models.

To understand the differences between the four models in the same equilateral class, it is easiest to look at the 3D plots of the reduced bispectrum shown in figure~\ref{fig:equi3d}. For all models we see the same peak structure with the maximums at positions where all the three $l_i$ correspond to peaks in the power spectrum. The largest peak is then when all three $l_i = 220$, corresponding to the large blue region near the origin. The four models are most strongly differentiated by their behaviour in the flattened limit, as can be seen in figure (\ref{fig:equi3d}). As a result they can be separated by their behaviour around the peak positioned where $l_1=l_2=250,\,l_3=500$, (the magenta part just above the $l_i = 220$ peak). In the equilateral case it is a modest feature but in DBI and single it is larger, connecting up to create a forked range of structures near the faces of the tetrahedron. For ghost inflation, which becomes negative as we approach the flattened limit, the peak is almost non-existent. 

Despite the visual differences between equilateral models, we note that the CMB correlator (\ref{eq:cmbcor}) accurately confirms the strong correlations forecast by our simple shape correlator (\ref{eq:shapecor}), see table~(\ref{tb:correlatorcomparison}). As predicted, all three shapes -- DBI, ghost and single -- correlate closely with the phenomenological equilateral shape to within 96\% (this agrees with \cite{0612571} for the 
DBI model, the only direct calculation elsewhere of a non-separable case).
Again the largest difference is between the ghost and single models with an 89\% correlation, which was slightly under-estimated by the shape correlator at 85\%, presumably because of subtle changes in weighting arising from the transfer functions, as well as smoothing effects. We confirm that these four equilateral shapes will be very difficult to differentiate without a bispectrum detection of high significance.

The local model is shown in figure~(\ref{fig:local3d}), demonstrating a marked constrast with the 3D equilateral bispectrum. The dominance of signal in the squeezed limit creates strong parallel ridges which connect up and emanate along the corner edges of the tetrahedron. The 51\% CMB correlation between the local and equilateral models is reasonably well predicted by the shape correlator at 41\%, though again underestimated. The full results for all the CMB cross-correlators can be seen in Table~\ref{tb:correlatorcomparison}.     (We note that the estimated 2D cosines between the local model and the
DBI and ghost models are in qualitative agreement with ref.~\cite{0405356}, see also \cite{0612571}.)

It remains to briefly discuss the bispectrum and correlations of the last three models. The two 3D plots for warm inflation, see figure (\ref{fig:warm3d}), seem superficially similar but the effect of smoothing is to move signal from the squeezed states into the centre. This apparently minor shift is enough to reduce the fisher correlation between the warm and warmS to 48\%, providing the most significant outlier for the shape correlator with an 80\% correlation. It demonstrates the sensitivity of the analysis to the exact cut-off used for the primordial bispectrum in the squeezed limit, but it may also be accentuated by an increased weight around the edges implicit in the CMB estimator (see earlier discussion in section~3. The smoothed flat model bispectrum shown in (\ref{fig:local3d}) has become visually more similar to the local bispectrum than might be expected from the primordial shapes, with the CMB correlator reflecting this change at 79\%. The flat and local shapes could be difficult to disentangle, though we note that this result depends on the imposition of an arbitrary cut-off to regularise the flat shape.

\begin{figure}[ht]
\centering
\includegraphics[width=0.5\linewidth]{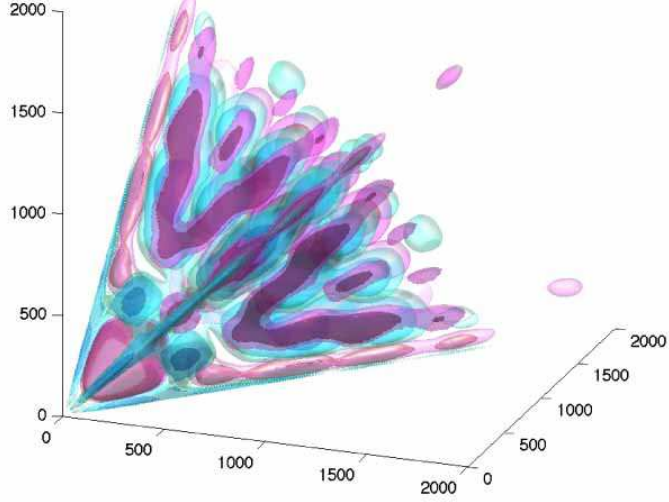}
\caption[3D Plot of the reduced bispectrum for the feature model.]{\small 3D Plot of the reduced bispectrum for the feature model.}
\label{fig:feat3d}
\end{figure}

Finally, the feature model clearly represents an entirely distinct type of bispectrum, which is evident from its very different behavior in the $(l,l,l)$-direction. The anticorrelation of the primary peak, relative to the other peaks, is clear in figure~(\ref{fig:feat3d}) from the nodal plane which cuts across the bispectrum. The poor correlation predicted with all the other shapes, is confirmed by the CMB cross-correlations which are all below 45\%.  (Here, the interplay with the nodal points introduced by the transfer functions makes these results strongly dependent on  $l_{max}$).  Preliminary analysis based on the approximate analytic shapes for an oscillatory model indicates a further independent shape which could be distinguished from the other classes given a reasonably significant bispectrum detection.

\begin{figure}[h]
\centering
\includegraphics[width=0.65\linewidth]{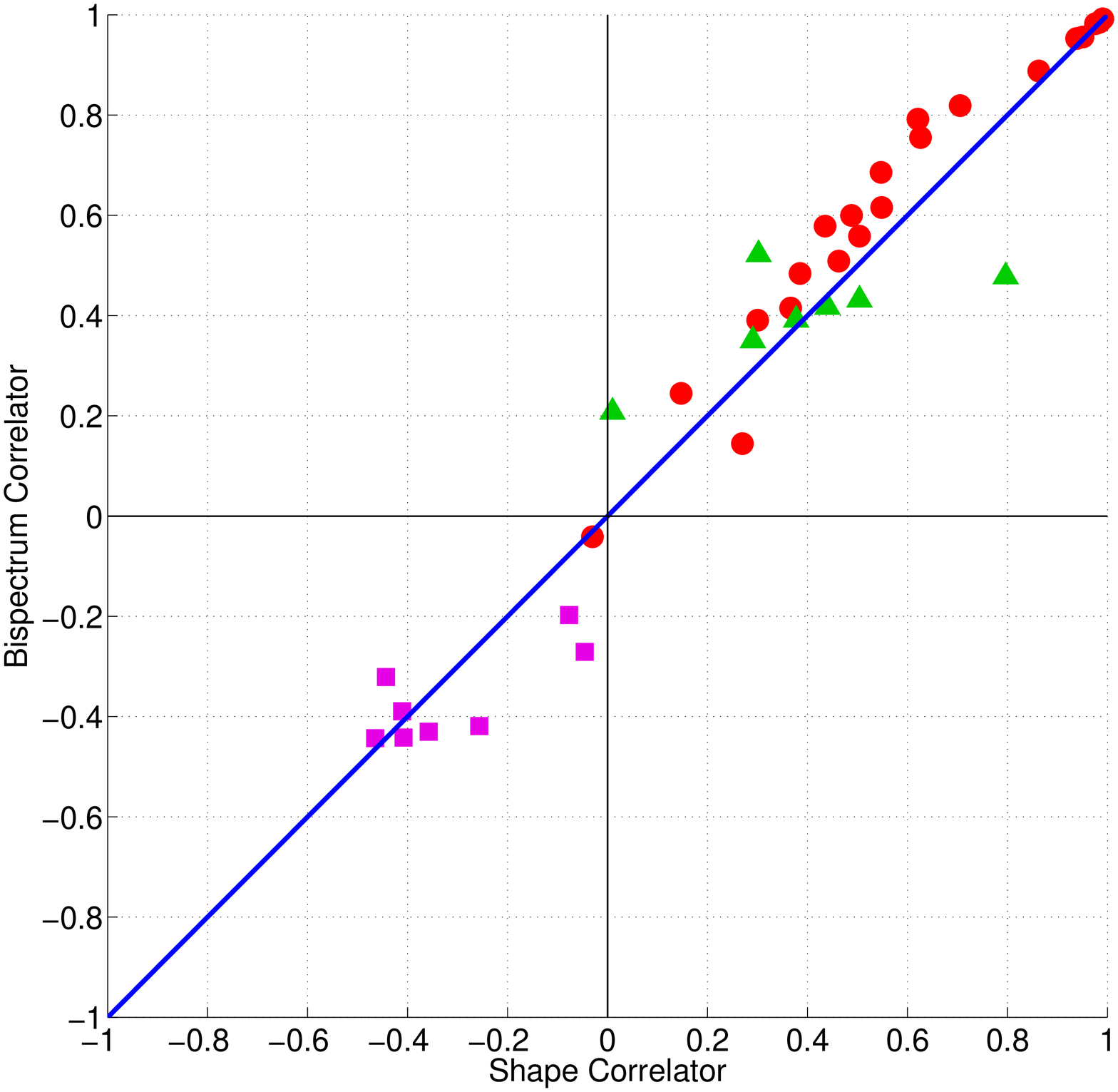}
\caption[Comparison or the bispectrum correlator and our primordial correlator.]{\small Comparison of the bispectrum correlator and primordial shape correlator. The blue line represents perfect agreement between the two, with the red data points (circles) showing correlations between: DBI, Equi, FlatS, Ghost, Local, Single, and WarmS.  The correlators are in good agreement in these cases.   The green data points (triangles) are correlations between these 7 models and the unsmoothed warm shape, while the purple data points (squares) are the correlations between the previous 8 models and the feature model. For the unsmoothed warm model, there is one significant outlier when the pathological behaviour in the squeezed limit for the warm case is being picked up in the shape correlator but not in the bispectrum (it has been smoothed by projection). Warm and WarmS models are poorly correlated indicating that the  squeezed limit must be understood fully before we can consider any constraints to be robust. The feature model is anti-correlated will all other models at about the 40\% level or below.}
\label{fig:correlatorcomparison}
\end{figure}

In figure~\ref{fig:correlatorcomparison}, we have plotted the full CMB bispectrum correlator (\ref{eq:cmbcor}) against the simple shape correlator (\ref{eq:shapecor}) for all the models investigated, demonstrating their remarkable concordance; highly correlated shapes agree accurately, while the shape correlator understimates the correlation of independent shapes (usually by about 5-15\%). Such a simple predictor of model correlations is important given the computational effort required to compare the CMB bispectra directly. This analysis confirms that there are indeed five distinct classes of models
among the cases reviewed: The equilateral class, the local class (which includes the non-local model), the warm model, the flat class and the feature model.

Here we note the importance of the weight, eqn.~(\ref{eq:weight}), in the shape correlator. For the feature model it is vital to achieve the correct correlation with any other choice producing extremely poor results. For the scale invariant models it is not as important to have the correct weight, as long as it weights each point on a slice equally. However, while tolerable results can still be achieved for the scale invariant models with weights $\w=1$ and $\w = 1/(k_1+k_2+k_3)^2$, we still see much better agreement between the two correlators when we use the correct weight, $\w=1/(k_1+k_2+k_3)$, due to the effect of the shape of the region of integration as discussed earlier.

\begin{table}[t]\label{table2}
\caption{Comaprison of bispectrum and shape correlators} 
\label{tb:correlatorcomparison}
\begin{tabular}[t]{|@{\hspace{1mm}}l@{\hspace{1mm}}|@{\hspace{1mm}}c@{\hspace{1mm}}|@{\hspace{1mm}}c@{\hspace{1mm}}|@{\hspace{1mm}}l@{\hspace{1mm}}|@{\hspace{1mm}}c@{\hspace{1mm}}|@{\hspace{1mm}}c@{\hspace{1mm}}|}
\hline
Model1- Model2 & $\bar{\curl{C}}(S,S^\pr)$ & $\curl{C}(B,B^\pr)$ & Model1- Model2 & $\bar{\curl{C}}(S,S^\pr)$ & $\curl{C}(B,B^\pr)$\\
\hline
DBI - Equi & 0.99 & 0.99 & Warm - DBI & 0.38 & 0.39 \\
DBI - Flat S & 0.39 & 0.48 & Warm - Equi & 0.44 & 0.42 \\
DBI - Ghost & 0.94 & 0.95 & Warm - Flat S & 0.01 & 0.21 \\
DBI - Local & 0.50 & 0.56 & Warm - Ghost & 0.50 & 0.43 \\
DBI - Single & 0.98 & 0.99 & Warm - Local & 0.30 & 0.52 \\
DBI - Warm S & 0.55 & 0.69 & Warm - Single & 0.29 & 0.35 \\
Equi - Flat S & 0.30 & 0.39 & Warm - Warm S & 0.80 & 0.48 \\
Equi - Ghost & 0.98 & 0.98 & & & \\
Equi - Local & 0.46 & 0.51 & Feature - DBI & -0.41 & -0.44 \\
Equi - Single & 0.95 & 0.96 & Feature - Equi & -0.36 & -0.43 \\
Equi - Warm S & 0.63 & 0.76 & Feature - Flat S & -0.44 & -0.32 \\
Flat S - Ghost & 0.15 & 0.24 & Feature - Ghost & -0.26 & -0.42 \\
Flat S - Local & 0.62 & 0.79 & Feature - Local & -0.41 & -0.39 \\
Flat S - Single & 0.49 & 0.60 & Feature - Single & -0.46 & -0.44 \\
Flat S - Warm S & -0.03 & -0.04 & Feature - Warm & -0.05 & -0.27 \\
Ghost - Local & 0.37 & 0.42 & Feature - Warm S & -0.08 & -0.20 \\
Ghost - Single & 0.86 & 0.89 & & & \\
Ghost - Warm S & 0.71 & 0.82 & & & \\
Local - Single & 0.55 & 0.62 & & & \\
Local - Warm S & 0.27 & 0.14 & & & \\
Single - Warm S & 0.44 & 0.58 & & & \\
\hline
\end{tabular}
\end{table}

\eject

\section{CMB bispectrum normalization}\label{se:norm}

The question now is how best to compare and contrast observational limits for such a wide variety of possible models. We need a framework for deciding a sensible definition of $f_{NL}$ for each class of models, normalised so that comparisons can be made. There have been many attempts at extending $f^{local}_{NL}$ to other models in the literature and the one most extensively used is to normalise the shape functions against a single point. This is done for the equilateral model and for the warm model, (although for the warm model, as they work with the curvature perturbation $\z$,  rather than $\O$, there is an additional factor of $3/5$ in their definition). Creating a new non-Gaussianity parameter for each model, $f^{equi}_{NL}$ and $f^{warm}_{NL}$, the models are normalised such that,
\begin{align}
S^{local}(k,k,k) = S^{equi}(k,k,k) = \frac{3}{5}S^{warm}(k,k,k)\,.
\end{align}
As all three shapes have the same scaling, this ratio is independent of $k$, and the method has the benefit of simplicity.  However, in the equilateral model we are normalising the maximum of one shape to the minimum of the other (local), so we find that a similar level of non-Gaussianity produces an $f^{equi}_{NL}$ that is over three times larger.  For the warm case, we have a $f^{warm}_{NL}$ that is almost four times larger, than $f^{local}_{NL}$. Consequently bounds on $f^{equi}_{NL}$ and $f^{warm}_{NL}$ seem far weaker than those for $f^{local}_{NL}$ which is entirely misleading, see eqns (\ref{eq:locallim}--\ref{eq:warmlim}).
The approach makes even less sense when the shapes have a running, in which case an arbitrary value for $k$ must be used for the normalisation, and there is no obvious extension to models with features or oscillatory behaviour.

One approach to normalisation would be to define $f_{NL}$ such that a given model with $f_{NL}=1$ produces the same level of bispectrum signal as the local case, also with $f^{local}_{NL}=1$, i.e.
\begin{align}
\curl{N}(B) = \curl{N}(B^{local})\,,
\end{align}
where $\curl{N}$ is defined by equation (\ref{eq:normalisation}) which we repeat here for convenience,
\begin{align}
\curl{N}(B) = \sqrt{\sum_{l_i} \frac{B^2_{l_1 l_2 l_3}}{C_{l_1}C_{l_2}C_{l_3}}}\,.
\end{align}
We could regard $\curl{N}$ as the total integrated bispectrum signal of a particular model. Note that here we calculate $\curl{N}$ for an ideal experiment without noise or beam effects covering the full $l$ range of the non-Gaussian signal; this would then give error bars which are approximately the same for all models. As it is defined at late times, this would also have the advantage that it could be used for the bispectrum from cosmic strings, or other general phenomena, e.g.~allowing an effective $f^{string}_{NL}$ to be defined and related to $G\u$, the string tension.

When we calculate the estimator from equation (\ref{eq:estimator}) for a particular experiment, including its particular noise and beam profiles, we will find any loss of sensitivity to the bispectrum for a particular model being reflected by an increase in the error bars. Thus the errors on models are now directly related to the sensitivity of the experiment to that model, rather than to an arbitrary choice for the definition of $f^{model}_{NL}$.

To avoid calculating the full bispectrum for each model it remains more straight forward to define the normalisation relative to the level of primordial signal instead. Given the success of the correlator based on the shape function (\ref{eq:shapecor}), we can use a normalisation based on the integrated non-Gaussianity in $k$-space. As $S(k_1,k_2,k_3)$ represents the primordial signal that we project forward to obtain the CMB bispectrum today, then using the same scaling as in the shape correlator should represent a more consistent measure of the actual non-Gaussianity measured in the CMB, that is,
\begin{align}\label{eq:normalise}
\curl{\bar{N}}^2(S) \equiv \int_{\curl{V}_k} {S^2(k_1,k_2,k_3)} w(k_1,k_2,k_3) d{\curl V}_k\,.
\end{align}
where $\curl {V}_k$ is the allowed region in $k$-space with the $k_i < k_{max}$ (see 
figure~\ref{fig:region}) and we take the weight function $w(k_1,k_2,k_3)=(k_1+k_2+k_3)^{-1}$, 
as before replicating the $l$-dependence of the CMB estimator (\ref{eq:estimatorscale}).  (A similar suggestion, but with a different weight, was made in ref.~\cite{0509029}.) We then propose to normalise the $f_{NL}$'s in each class of models relative to the local model, that is, such that
\begin{align}
\curl{\bar{N}}(S) = \curl{\bar{N}}(S^{local})\,.
\end{align}
We find then that the $2\s$ limits quoted above become,
\begin{align}\label{eq:obs}
-4 &< \bar f^{local}_{NL} < 80\,, \\  \label{eq:equiobs}
-34 &< \bar f^{equi}_{NL} < 57\,, \\
-107 &<  \bar f^{warm}_{NL} < 11\,,
\end{align}
where the standard deviation is now 21, 23, and 29 respectively and we use $\bar f_{NL}$ to denote that is it defined using the new approach to normalisation. We can see that the local and equilateral models are constrained at the same level with the warm constraint being weaker as the estimator used in \cite{07071647} is not optimal in the presence of inhomogeneous noise. This normalisation is applicable to all models regardless of their form, it is simple to calculate, and the $\bar f_{NL}$ for each class of models then represents a similar level of the measurable non-Gaussian signal.

Equally, we can use this standardised normalisation, together with the correlator results in table~\ref{tb:correlatorcomparison}, to naively forecast $ f_{NL}$ and its errors for alternative models which are not yet constrained.  Supposing our universe actually possessed significant local non-Gaussianity, then on
the basis of local estimator observations with $\bar f_{NL} = 38 \pm 21$ as in (\ref{eq:obs}),  the equilateral model should yield $ \bar f^{equil}_{NL} \approx 17 \pm 21$ (consistent with the observed result), while for the flat model  $ \bar f^{flat}_{NL} \approx 24 \pm 21$. Conversely, if our universe possessed flat non-Gaussianity, then given the local result (\ref{eq:obs}) we might obtain a marginally significant result  $ \bar f^{flat} \approx 61 \pm 21$.  We conclude that discovery potential remains, even with the present CMB data, for the independent shapes which have not yet been fully investigated, such as feature models\footnote{We note that the warm inflation results do not seem to match expectations from the shape correlator, suggesting an anticorrelation with the local model, rather than the 30\% correlation of table~\ref{tb:correlatorcomparison}.   However, we caution that the warm result has not been independently verified and it also depends sensitively on arbitrary cut-offs imposed on the shape function. We note that we have removed the factor of $3/5$ from the definition of $f^{warm}_{NL}$ in \ref{eq:obs}.}.

To reiterate the value of a standardised normalisation (\ref{eq:normalise}), we note an obvious failing of the previous method.  This comes from extending an observational result for one model to all the highly correlated models in the same class, such as equilateral to DBI, ghost and single. If we normalise the models using the centre point $S(k,k,k)$, as is done currently in the literature, then with the different $N(S)$ in each case the original equilateral limit $f^{equi}_{NL} = 51 \pm 101$ transfers to the following inconsistent limits
\begin{align}
f^{DBI}_{NL} &= 47 \pm 93 \,, \\
f^{sing}_{NL} &= 40 \pm 78 \,, \\
f^{ghost}_{NL} &= 59 \pm 116 \,.
\end{align}
In contrast, if we normalise using $N(S)$ in (\ref{eq:normalise}), then the equilateral limit (\ref{eq:equiobs}) on $\bar f_{NL}$ transfers across nearly identically in all these cases, because of their 95\% cross-correlation.

\section{Conclusions}

In this paper we have presented a comprehensive set of formalisms for comparing, evolving, and constraining primordial non-Gaussian models through the CMB bispectrum.

The primary goal was to directly calculate CMB bispectra from a general primordial model, 
enhancing methods previously outlined in ref.~\cite{0612713}.   This was achieved by 
exploiting common features of primordial bispectra to reduce the dimensionality of the 
transfer functions required to evaluate the CMB bispectra.   The new innovations reported
here include the use of the flat sky approximation when all $l_i\ge 150$, greatly reducing
computational times for most of the allowed region, and a cubic reparametrisation, 
significantly reducing the number of points required for accurate interpolation of the bispectrum. 
(We note that this CMB bispectrum code is being prepared for public distribution.)
These methods make feasible the repetitive calculation of highly accurate CMB bispectra at Planck 
resolution without specific assumptions about the separability of the underlying primordial bispectrum. 
 
Further, we have calculated the CMB bispectra for all the distinct primordial bispectrum shapes 
$S(k_1,k_2,k_3)$ currently presented in the literature, motivated by a range of inflationary and other
cosmological scenarios.   We have presented these, plotted relative to the large-angle CMB bispectrum
for the constant   primordial model ($S^ { const.}=1$) for which we presented an analytic solution
(\ref{eq:constbispectrum}).    The CMB bispectra from the different primordial models exhibit a 
close correspondence to their original shape modulated, of course, by the oscillatory transfer
functions. 

We were able to quantitatively determine the observational independence of 
the CMB bispectra by measuring their cross-correlations using the estimator (\ref{eq:estimator}).   
These results revealed five independent classes of shapes which it should be possible to 
distinguish from one another with a significant detection of non-Gaussianity in future experiments 
such as Planck.  
These were the equilateral (\ref{eq:equi}), local (\ref{eq:local}), warm (\ref{eq:warm}), and 
flat (\ref{eq:flat}) shapes, all described in section~3,
together with the feature model (\ref{eq:feature}) which is basically the constant shape (\ref{eq:constmodel}) plus broken scale-invariance.  Different models belonging within the same class will be difficult to 
distinguish, a fact best demonstrated by the 95\% correlation of the equilateral shape with 
DBI (\ref{eq:dbi}), ghost (\ref{eq:ghost}) and single (\ref{eq:single}) shapes.     
 
We have also presented a shape correlator (\ref{eq:shapecor}) which provides a fast and simple
method for determining the independence of different shapes.   In particular, for highly correlated 
models, it yielded results accurately reflecting those of the full CMB correlator (\ref{eq:cmbcor}), thus
avoiding unnecessary calculation.   The shape correlator also reliably identified poorly correlated models, that is, new shapes for which a full CMB bispectrum analysis was warranted.   We also 
proposed a straightforward two-dimensional eigenmode analysis of shape functions, valid for 
nearly scale-invariant  models.   This allowed us to identify the shape correlations with  
products of eigenvalues of the non-orthogonal eigenmodes, immediately revealing, for example, why 
warm and local models are independent.  In principle, the analysis can guide the theoretical search 
for primordial models with distinctive non-Gaussian signatures.   It also revealed the constant eigenmode (or shape (\ref{eq:constmodel})) as the primary cause of the cross-correlation between many models, suggesting strategies for distinguishing, for example, between local and equilateral models. 

Finally, we proposed an alternative approach (\ref{eq:normalise}) to the normalisation of the non-Gaussianity parameter $f_{NL}$
using the shape autocorrelator. This new normalisation allows us to employ $f_{NL}$ to systematically compare the true level of non-Gaussianity in different models. In contrast, with current methods, the constraints for competing models can vary by a factor of four or more, with bounds varying significantly 
for models even in the same class of highly correlated shapes.  

A detection of non-Gaussianity would have profound consequences for our understanding of the 
early universe, uprooting the present simplest inflationary paradigm. 
The present work indicates that the next generation of CMB experiments (notably Planck) may be able
to distinguish between different classes of shapes for primordial non-Gaussianity.   Delineating 
the bispectrum shape would provide important clues about viable alternative theories for the origin of 
large-scale structure in the universe.

\section{Acknowledgements}

We are very grateful for many informative discussions with 
Michele Liguori and Kendrick Smith.  We also acknowledge useful
conversations with David Seery, Xingang Chen and Bartjan van Tent.
Simulations were performed on the COSMOS supercomputer (an Altix 4700) which is funded by 
STFC, HEFCE and SGI.   EPS was supported by STFC grant ST/F002998/1 and the 
Centre for Theoretical Cosmology.

\bibliographystyle{unsrt}
\bibliography{Paper}

\begin{thebibliography}{10}

\bibitem{0612713}
J.~R. Fergusson and Edward P.~S. Shellard.
\newblock {Primordial non-Gaussianity and the CMB bispectrum}.
\newblock {\em Phys. Rev.}, D76:083523, 2007.

\bibitem{09012572}
Kendrick~M. Smith, Leonardo Senatore, and Matias Zaldarriaga.
\newblock {Optimal limits on $f_{NL}^{local}$ from WMAP 5-year data}.
\newblock {\em ArXiv Astrophysics e-prints}, 2009.

\bibitem{07071647}
Ian~G Moss and Chris~M Graham.
\newblock {Testing models of inflation with CMB non-gaussianity}.
\newblock {\em JCAP}, 0711:004, 2007.

\bibitem{08030547}
E.~Komatsu et~al.
\newblock {Five-Year Wilkinson Microwave Anisotropy Probe (WMAP)
  Observations:Cosmological Interpretation}.
\newblock 2008.

\bibitem{07121148}
Amit P.~S. Yadav and Benjamin~D. Wandelt.
\newblock {Evidence of Primordial Non-Gaussianity $(f_{\rm NL})$ in the
  Wilkinson Microwave Anisotropy Probe 3-Year Data at 2.8$\sigma$}.
\newblock {\em Phys. Rev. Lett.}, 100:181301, 2008.

\bibitem{inprep}
James~R. Fergusson, Michele Liguori, and Edward P.~S. Shellard.
\newblock {CMB constraints on non-Gaussianity, in preparation}.
\newblock 2009.

\bibitem{0503375}
Daniel Babich.
\newblock {Optimal Estimation of Non-Gaussianity}.
\newblock {\em Phys. Rev.}, D72:043003, 2005.

\bibitem{08070231}
A.~Curto et~al.
\newblock {WMAP 5-year constraints on fnl with wavelets}.
\newblock 2008.

\bibitem{08023677}
Chiaki Hikage et~al.
\newblock {Limits on Primordial Non-Gaussianity from Minkowski Functionals of
  the WMAP Temperature Anisotropies}.
\newblock {\em Mon. Not. Roy. Astron. Soc.}, 389:1439--1446, 2008.

\bibitem{0005036}
Eiichiro Komatsu and David~N. Spergel.
\newblock {Acoustic signatures in the primary microwave background bispectrum}.
\newblock {\em Phys. Rev.}, D63:063002, 2001.

\bibitem{0405356}
Daniel Babich, Paolo Creminelli, and Matias Zaldarriaga.
\newblock {The shape of non-Gaussianities}.
\newblock {\em JCAP}, 0408:009, 2004.

\bibitem{0509029}
Paolo Creminelli, Alberto Nicolis, Leonardo Senatore, Max Tegmark, and Matias
  Zaldarriaga.
\newblock {Limits on non-Gaussianities from WMAP data}.
\newblock {\em JCAP}, 0605:004, 2006.

\bibitem{0612571}
Kendrick~M. Smith and Matias Zaldarriaga.
\newblock {Algorithms for bispectra: forecasting, optimal analysis, and
  simulation}.
\newblock {\em ArXiv Astrophysics e-prints}, 2006.

\bibitem{watson}
G.~N. Watson.
\newblock {\em A Treatise on the Theory of Bessel Functions}.
\newblock Cambridge University Press, 1966.

\bibitem{07114933}
Amit P.~S. Yadav et~al.
\newblock {Fast Estimator of Primordial Non-Gaussianity from Temperature and
  Polarization Anisotropies in the Cosmic Microwave Background II: Partial Sky
  Coverage and Inhomogeneous Noise}.
\newblock {\em Astrophys. J.}, 678:578, 2008.

\bibitem{0210603}
Juan~Martin Maldacena.
\newblock {Non-Gaussian features of primordial fluctuations in single field
  inflationary models}.
\newblock {\em JHEP}, 05:013, 2003.

\bibitem{08023218}
Neil Barnaby and James~M. Cline.
\newblock {Predictions for Nongaussianity from Nonlocal Inflation}.
\newblock {\em JCAP}, 0806:030, 2008.

\bibitem{0410486}
G.~I. Rigopoulos, E.~P.~S. Shellard, and B.~J.~W. van Tent.
\newblock {A simple route to non-Gaussianity in inflation}.
\newblock {\em Phys. Rev.}, D72:083507, 2005.

\bibitem{0404084}
Mohsen Alishahiha, Eva Silverstein, and David Tong.
\newblock {DBI in the sky}.
\newblock {\em Phys. Rev.}, D70:123505, 2004.

\bibitem{0605045}
Xingang Chen, Min-xin Huang, Shamit Kachru, and Gary Shiu.
\newblock {Observational signatures and non-Gaussianities of general single
  field inflation}.
\newblock {\em JCAP}, 0701:002, 2007.

\bibitem{0306122}
Paolo Creminelli.
\newblock {On non-gaussianities in single-field inflation}.
\newblock {\em JCAP}, 0310:003, 2003.

\bibitem{0312100}
Nima Arkani-Hamed, Paolo Creminelli, Shinji Mukohyama, and Matias Zaldarriaga.
\newblock {Ghost inflation}.
\newblock {\em JCAP}, 0404:001, 2004.

\bibitem{0503692}
David Seery and James~E. Lidsey.
\newblock {Primordial non-gaussianities in single field inflation}.
\newblock {\em JCAP}, 0506:003, 2005.

\bibitem{0406398}
N.~Bartolo, E.~Komatsu, Sabino Matarrese, and A.~Riotto.
\newblock {Non-Gaussianity from inflation: Theory and observations}.
\newblock {\em Phys. Rept.}, 402:103--266, 2004.

\bibitem{811001}
Andrei~D. Linde.
\newblock {A New Inflationary Universe Scenario: A Possible Solution of the
  Horizon, Flatness, Homogeneity, Isotropy and Primordial Monopole Problems}.
\newblock {\em Phys. Lett.}, B108:389--393, 1982.

\bibitem{820101}
Andreas Albrecht and Paul~J. Steinhardt.
\newblock {Cosmology for Grand Unified Theories with Radiatively Induced
  Symmetry Breaking}.
\newblock {\em Phys. Rev. Lett.}, 48:1220--1223, 1982.

\bibitem{0207295}
Francis Bernardeau and Jean-Philippe Uzan.
\newblock {Non-Gaussianity in multi-field inflation}.
\newblock {\em Phys. Rev.}, D66:103506, 2002.

\bibitem{0209330}
Francis Bernardeau and Jean-Philippe Uzan.
\newblock {Inflationary models inducing non-gaussian metric fluctuations}.
\newblock {\em Phys. Rev.}, D67:121301, 2003.

\bibitem{0504045}
David~H. Lyth and Yeinzon Rodriguez.
\newblock {The inflationary prediction for primordial non- gaussianity}.
\newblock {\em Phys. Rev. Lett.}, 95:121302, 2005.

\bibitem{0504508}
G.~I. Rigopoulos, E.~P.~S. Shellard, and B.~J.~W. van Tent.
\newblock {Non-linear perturbations in multiple-field inflation}.
\newblock {\em Phys. Rev.}, D73:083521, 2006.

\bibitem{0506056}
David Seery and James~E. Lidsey.
\newblock {Primordial non-gaussianities from multiple-field inflation}.
\newblock {\em JCAP}, 0509:011, 2005.

\bibitem{0511041}
G.~I. Rigopoulos, E.~P.~S. Shellard, and B.~J.~W. van Tent.
\newblock {Quantitative bispectra from multifield inflation}.
\newblock {\em Phys. Rev.}, D76:083512, 2007.

\bibitem{08071101}
Christian~T. Byrnes, Ki-Young Choi, and Lisa M.~H. Hall.
\newblock {Conditions for large non-Gaussianity in two-field slow- roll
  inflation}.
\newblock {\em JCAP}, 0810:008, 2008.

\bibitem{08070180}
Atsushi Naruko and Misao Sasaki.
\newblock {Large non-Gaussianity from multi-brid inflation}.
\newblock 2008.

\bibitem{0506704}
G.~I. Rigopoulos, E.~P.~S. Shellard, and B.~J.~W. van Tent.
\newblock {Large non-Gaussianity in multiple-field inflation}.
\newblock {\em Phys. Rev.}, D73:083522, 2006.

\bibitem{0603799}
F.~{Vernizzi} and D.~{Wands}.
\newblock {Non-Gaussianities in two-field inflation}.
\newblock {\em JCAP}, 5:19--+, May 2006.

\bibitem{0511736}
Andrei Linde and Viatcheslav Mukhanov.
\newblock {The curvaton web}.
\newblock {\em JCAP}, 0604:009, 2006.

\bibitem{0208055}
David~H. Lyth, Carlo Ungarelli, and David Wands.
\newblock {The primordial density perturbation in the curvaton scenario}.
\newblock {\em Phys. Rev.}, D67:023503, 2003.

\bibitem{0309033}
N.~Bartolo, S.~Matarrese, and A.~Riotto.
\newblock {On non-Gaussianity in the curvaton scenario}.
\newblock {\em Phys. Rev.}, D69:043503, 2004.

\bibitem{0411394}
Kari Enqvist, Asko Jokinen, Anupam Mazumdar, Tuomas Multamaki, and Antti
  Vaihkonen.
\newblock {Non-Gaussianity from Preheating}.
\newblock {\em Phys. Rev. Lett.}, 94:161301, 2005.

\bibitem{0501076}
Kari Enqvist, Asko Jokinen, Anupam Mazumdar, Tuomas Multamaki, and Antti
  Vaihkonen.
\newblock {Non-gaussianity from instant and tachyonic preheating}.
\newblock {\em JCAP}, 0503:010, 2005.

\bibitem{08054795}
Alex Chambers and Arttu Rajantie.
\newblock {Non-Gaussianity from massless preheating}.
\newblock {\em JCAP}, 0808:002, 2008.

\bibitem{0601481}
N.~Barnaby and J.~M.~Cline,
\newblock{Nongaussian and nonscale-invariant perturbations from tachyonic preheating in hybrid inflation}
\newblock {\em Phys. Rev.}, D73:106012, 2006.

\bibitem{0611750}
N.~Barnaby and J.~M.~Cline,
\newblock{Nongaussianity from Tachyonic Preheating in Hybrid Inflation}
\newblock {\em Phys. Rev.}, D75:086004, 2007.

\bibitem{08020588}
David Seery, Karim~A. Malik, and David~H. Lyth.
\newblock {Non-gaussianity of inflationary field perturbations from the field
  equation}.
\newblock {\em JCAP}, 0803:014, 2008.

\bibitem{0306006}
Matias Zaldarriaga.
\newblock {Non-Gaussianities in models with a varying inflaton decay rate}.
\newblock {\em Phys. Rev.}, D69:043508, 2004.

\bibitem{9208001}
Toby Falk, Raghavan Rangarajan, and Mark Srednicki.
\newblock {The Angular dependence of the three point correlation function of
  the cosmic microwave background radiation as predicted by inflationary
  cosmologies}.
\newblock {\em Astrophys. J.}, 403:L1, 1993.

\bibitem{0702165}
Paolo Creminelli and Leonardo Senatore.
\newblock {A smooth bouncing cosmology with scale invariant spectrum}.
\newblock {\em JCAP}, 0711:010, 2007.

\bibitem{07084321}
Kazuya Koyama, Shuntaro Mizuno, Filippo Vernizzi, and David Wands.
\newblock {Non-Gaussianities from ekpyrotic collapse with multiple fields}.
\newblock {\em JCAP}, 0711:024, 2007.

\bibitem{07105172}
Evgeny~I. Buchbinder, Justin Khoury, and Burt~A. Ovrut.
\newblock {Non-Gaussianities in New Ekpyrotic Cosmology}.
\newblock {\em Phys. Rev. Lett.}, 100:171302, 2008.

\bibitem{07123779}
Jean-Luc Lehners and Paul~J. Steinhardt.
\newblock {Non-Gaussian Density Fluctuations from Entropically Generated
  Curvature Perturbations in Ekpyrotic Models}.
\newblock {\em Phys. Rev.}, D77:063533, 2008.

\bibitem{08041293}
Jean-Luc Lehners and Paul~J. Steinhardt.
\newblock {Intuitive understanding of non-gaussianity in ekpyrotic and cyclic
  models}.
\newblock {\em Phys. Rev.}, D78:023506, 2008.

\bibitem{0205152}
Sujata Gupta, A.~Berera, A.~F. Heavens, and S.~Matarrese.
\newblock {Non-Gaussian signatures in the cosmic background radiation from warm
  inflation}.
\newblock {\em Phys. Rev.}, D66:043510, 2002.

\bibitem{9312033}
Alejandro Gangui, Francesco Lucchin, Sabino Matarrese, and Silvia Mollerach.
\newblock {The Three point correlation function of the cosmic microwave
  background in inflationary models}.
\newblock {\em Astrophys. J.}, 430:447--457, 1994.

\bibitem{0701302}
Ian~G Moss and Chun Xiong.
\newblock {Non-gaussianity in fluctuations from warm inflation}.
\newblock {\em JCAP}, 0704:007, 2007.

\bibitem{07101302}
R.~Holman and Andrew~J. Tolley.
\newblock {Enhanced Non-Gaussianity from Excited Initial States}.
\newblock {\em JCAP}, 0805:001, 2008.

\bibitem{0611645}
Xingang Chen, Richard Easther, and Eugene~A. Lim.
\newblock {Large non-Gaussianities in single field inflation}.
\newblock {\em JCAP}, 0706:023, 2007.

\bibitem{08020491}
Rachel Bean, Xingang Chen, Girma Hailu, S.~H.~Henry Tye, and Jiajun Xu.
\newblock {Duality Cascade in Brane Inflation}.
\newblock {\em JCAP}, 0803:026, 2008.

\bibitem{08013295}
Xingang Chen, Richard Easther, and Eugene~A. Lim.
\newblock {Generation and Characterization of Large Non-Gaussianities in Single
  Field Inflation}.
\newblock {\em JCAP}, 0804:010, 2008.

\end{thebibliography}

\end{document}